\documentclass[lettersize,journal]{IEEEtran}
\usepackage{amsmath,amsfonts}
\usepackage{algorithm}
\usepackage{algpseudocode}
\usepackage{array}
\usepackage{pifont}
\usepackage[caption=false,font=normalsize,labelfont=sf,textfont=sf]{subfig}
\usepackage{textcomp}
\usepackage{stfloats}

\usepackage{setspace}
\usepackage{url}
\usepackage{verbatim}
\usepackage{graphicx}
\usepackage{cite}

\begin{document}
\title{Logit Poisoning Attack in Distillation-based Federated Learning and its Countermeasures}

\author{Yonghao Yu, \IEEEmembership{Student Member, IEEE}, Shunan Zhu, \IEEEmembership{Student Member, IEEE}, and Jinglu Hu \IEEEmembership{Member, IEEE}
\thanks{Yonghao Yu, Shunan Zhu and Jinglu Hu are with Graduate school of Information, Production and Systems, Waseda University of Hibikino 2-7. Wakamatsu-Ku, Kitakyushu-shi, Fukuoka, Japan. 
(email: \mbox{yuyonghao@suou.waseda.jp}; shunan-zhu@ruri.waseda.jp; jinglu@waseda.jp).}% <-this % stops a space
\thanks{Yonghao Yu is the corresponding author.}
\thanks{This Manuscript has been submitted to IEEE Transactions on Big Data on 16-Nov-2023 with Manuscript ID: TBD-2023-11-0756}}

\markboth{Journal of \LaTeX\ Class Files,~Vol.~xx, No.~x, November~2023}%
{Shell \MakeLowercase{\textit{et al.}}: A Sample Article Using IEEEtran.cls for IEEE Journals}


\maketitle

\begin{abstract}
Distillation-based federated learning has emerged as a promising collaborative learning approach, where clients share the output logit vectors of a public dataset rather than their private model parameters. This practice reduces the risk of privacy invasion attacks and facilitates heterogeneous learning. The landscape of poisoning attacks within distillation-based federated learning is complex, with existing research employing traditional data poisoning strategies targeting the models' parameters. However, these attack schemes primarily have shortcomings rooted in their original designs, which target the model parameters rather than the logit vectors. Furthermore, they do not adequately consider the role of logit vectors in carrying information during the knowledge transfer process. This misalignment results in less efficiency in the context of distillation-based federated learning. Due to the limitations of existing methodologies, our research delves into the intrinsic properties of the logit vector, striving for a more nuanced understanding. We introduce a two-stage scheme for logit poisoning attacks, addressing previous shortcomings. Initially, we collect the local logits, generate the representative vectors, categorize the logit elements within the vector, and design a shuffling table to maximize information entropy. Then, we intentionally scale the shuffled logit vectors to enhance the magnitude of the target vectors. Concurrently, we propose an efficient defense algorithm to counter this new poisoning scheme by calculating the distance between estimated benign vectors and vectors uploaded by users. Through extensive experiments, our study illustrates the significant threat posed by the proposed logit poisoning attack and highlights the effectiveness of our defense algorithm. %This research opens new avenues for understanding and mitigating risks in the evolving field of distillation-based federated learning.
\end{abstract}

\begin{IEEEkeywords}
Federated learning, knowledge distillation, poisoning attack, poisoning detection.
\end{IEEEkeywords}

\section{Introduction}
\IEEEPARstart{F}{ederated} learning (FL) \cite{mcmahan2017communication}, \cite{park2019wireless}, \cite{lim2020federated}, \cite{kairouz2021advances}, \cite{zhou2022communication}, \cite{zeng2023hfedms} represents a paradigm shift in collaborative machine learning, enabling parallel and distributed training across multiple clients. In this approach, individual devices or nodes undertake the training process locally. Only the relevant updates, rather than raw training data, are shared with a central server for aggregation. This methodology attempts to balance efficiency and privacy; however, traditional FL still faces risks. Privacy leakage from uploading local model parameters has not only been identified but has also proven to be a tangible threat to user confidentiality. Distillation-based federated learning \cite{chang2019cronus}, \cite{li2019fedmd}, \cite{lin2020ensemble}, \cite{itahara2021distillation}, \cite{cheng2021fedgems}, \cite{fang2022robust} has emerged as a significant area of research to address these security and privacy challenges, inspired by the knowledge transfer algorithms \cite{hinton2015distilling}, \cite{ba2014deep}, \cite{papernot2016semi}, \cite{papernot2018scalable}, \cite{wang2018kdgan}, \cite{anil2018large}. By exchanging local model prediction results on public datasets rather than plain model parameters or gradients, distillation-based federated learning promises to provide enhanced protection against potential breaches. It has become a subject of extensive academic investigation. In this approach, the local model output of prediction is mainly in the form of logit vectors, which are the unnormalized outputs of the model's final feature encoding layer before applying a softmax function. Distillation-based federated learning offers an innovative approach to enhancing security in collaborative machine learning. Moreover, distillation-based federated learning introduces the capability for heterogeneous FL, allowing users to maintain their own unique model structure while still acquiring knowledge from others\cite{ejigu2023robust}.

Despite the abundance of research analyzing poisoning attacks in the context of distillation-based federated learning, prevalent studies often employ antiquated methods that fail to effectively address the unique challenges posed by this learning approach.
For example, Cronus \cite{chang2019cronus} is an FL framework that employs model distillation for collaborative learning. Local nodes use private data to train the model and further generate soft labels from public datasets. These soft labels are aggregated at a central server and then distributed back to nodes. The authors provide an in-depth examination of both targeted and untargeted poisoning attacks employed in the study, an approach intended to evaluate the effectiveness of their proposed scheme robustly. They tested four different poisoning attack schemes, label flip poisoning \cite{jagielski2018manipulating}, \cite{hayes2018contamination}, \cite{munoz2017towards}, naive poisoning (PAF) \cite{chang2019cronus}, little is enough attack (LIE) \cite{baruch2019little}, and one far one mean (OFOM) \cite{chang2019cronus}. The experimental results affirm that a robust FL scheme is assured under these attacks. 
Distillation-Based Semi-Supervised Federated Learning (DS-FL) \cite{itahara2021distillation} algorithm enhances model performance by sharing local model outputs. It uses labeled open datasets as shared datasets for further local model training. These shared datasets can not only expand the datasets but also can boost local models' performance. In order to examine the robustness of their scheme, the authors adapt the model poisoning attack \cite{bagdasaryan2020backdoor} to DS-FL, with malicious clients transmitting logit vectors produced by an arbitrary model. However, it should be noted that the model poisoning attack was initially designed to attack FL. Consequently, implementing this particular poisoning attack strategy in the given context fails to yield optimal performance.
Cheng \textit{et al.} \cite{cheng2021fedgems} introduced a framework called Federated Learning of Larger Server Models via Selective Knowledge Fusion (FedGEMS). Within this framework, the server model strategically gathers insights from its own analysis and teacher clients, enhancing its comprehension of the learning process. The author also uses the previously proposed poisoning attack to evaluate the robustness of the proposed scheme, including PAF, LIE, and OFOM. Similarly, these antiquated attack strategies fail to exert any significant influence on their FL framework.
While most of these attack strategies aim to poison the dataset or model parameters to alter the distribution of logit vectors, they often fail to account for the primary factor encapsulating knowledge within the logit vectors. In other words, these poisoning strategies are designed based on the distribution of the model parameters other than the underlying logic of the logit vectors' ability to transfer knowledge. Therefore, the poisoning attack designed to disrupt the expression of the logit vectors based on their knowledge transfer principle can significantly threaten the distillation-based federated learning schemes, highlighting an urgent need to address these vulnerabilities.
\begin{figure*}
    \centering
    \includegraphics[width=.75\textwidth]{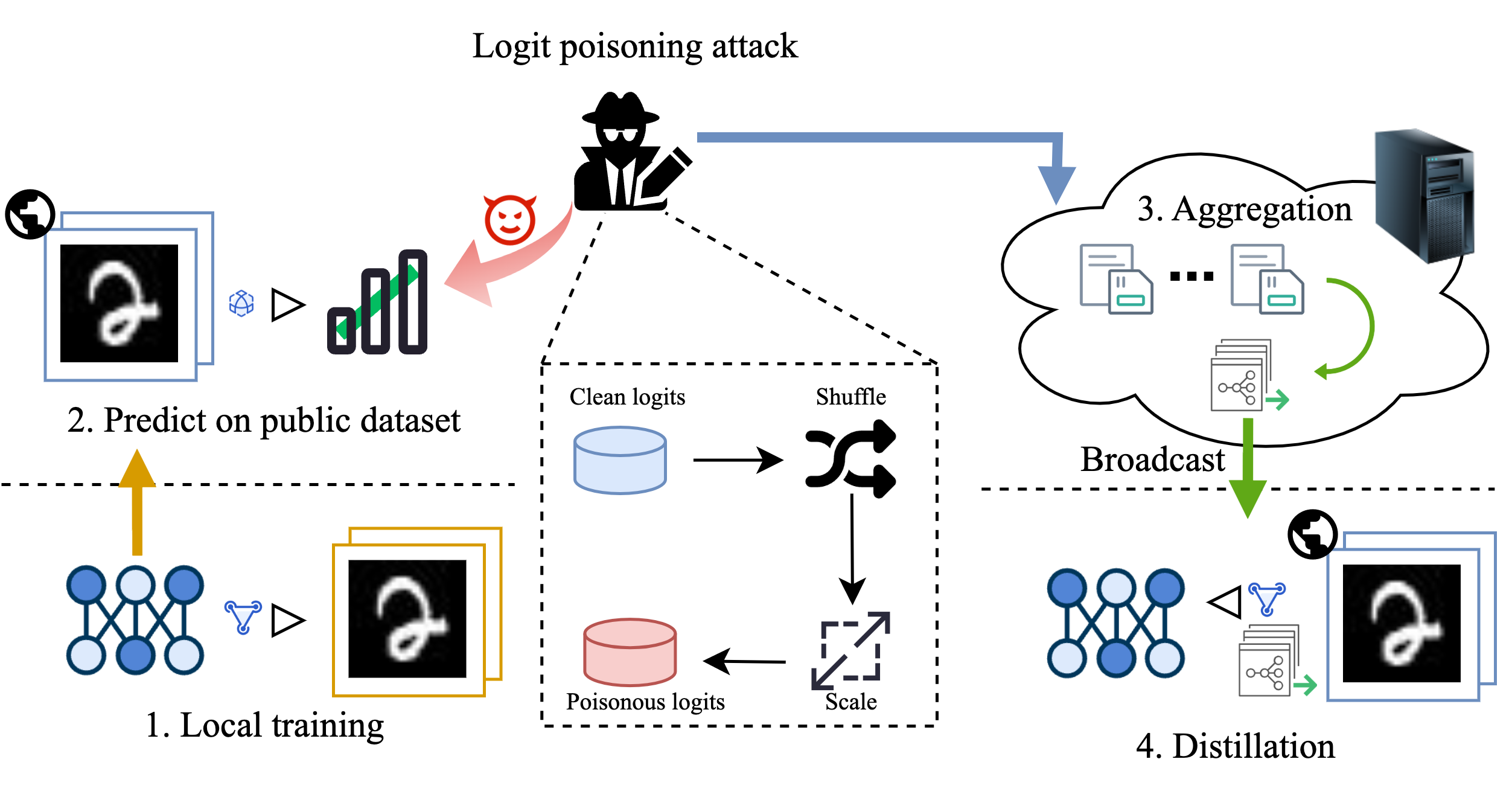}
    \caption{Overview of proposed logit poisoning attack} 
    \label{fig:over}
\end{figure*}

To address the above challenge, in this work, we delve into an analysis of the knowledge encapsulated in logit vectors, exploring it from two critical perspectives: the inter-class similarity and the position of maximum value. The former assesses the local model's judgment on how closely related a sample is to other classes. The latter indicates the certainty with which a sample is assigned to a particular class. Building on this foundation, we design targeted attack schemes designed to dismantle the information within the logit vectors by well-designed shuffling and scaling. By doing so, we develop schemes to destroy the information carried in these logit vectors and mislead models to the greatest extent possible, thereby maximizing the poisoning effect. The overview of our poisoning process is depicted in Fig. \ref{fig:over}. We further design a robust aggregation algorithm to defend the federated learning scheme against poisoning attacks. This defense strategy is founded on estimating the distribution of benign vectors and calculating the distance between user-uploaded vectors and these benign representations. Following this, we can assign an aggregation weight for each user. Our main contributions can be summarized as follows,

% \begin{table}[t]
%   \centering
%   \begin{tabular}{lllll}
%   \hline\hline
%     Name & FedMd & Cronus & FedGEMS & DS-FL \\\hline
%     Label-flipping poisoning & \xmark & \cmark & ? & ?\\
%     Naive poisoning & \xmark & \cmark & \cmark & ?\\
%     Little is enough attack & \xmark & \cmark & \cmark & ?\\
%     OFOM & \xmark & \cmark & \cmark & ?\\
%     Model poisoning & \xmark & ? & ? & \cmark\\
%     Our scheme & \xmark & \xmark & \xmark & \xmark \\\hline
%   \end{tabular}
%   \caption{Comparison}
%   \label{tab:comparison}
% \end{table}
\begin{itemize}
    \item We perform the first systematic examination of logit poisoning attacks in the context of distillation-based federated learning. Our research involves an in-depth analysis of the underlying principle governing knowledge transfer through logit vectors.
    \item  We propose a customized attack scheme for distillation-based federated learning involving the manipulation of local logits. This scheme employs shuffling and scaling techniques to craft malicious local logits. Furthermore, we design a robust aggregation algorithm to identify malicious logit vectors, thereby protecting distillation-based federated learning from various poisoning attacks.
    \item We conduct extensive experiments using the MNIST dataset to evaluate the effectiveness of our proposed poisoning scheme and the defense mechanism. The experimental results reveal that our poisoning scheme can substantially impair the performance of the models. Conversely, our defense scheme demonstrates robust resistance to various poisoning attacks.
\end{itemize}

The rest of this paper is structured as follows. Section II provides an introduction to the related literature. Then, in Section III, we overview the preliminaries and introduce the threat model in this work. Section IV elaborates on the attack method design. Section V gives the introduction of our defense scheme. Section VI experimentally evaluates the performance of the proposed schemes. Finally, Section VII concludes this paper. 

\section{Related Work}
In this section, we briefly review the state-of-the-art research works focusing on distillation-based federated learning and poisoning attacks in federated learning.

\subsection{Distillation-based Federated Learning}
% \vspace{-5mm}
Distillation-based federated learning utilizes model distillation techniques, also known as knowledge transfer methods, to share knowledge. Such an approach exchanges model output logit vectors instead of model parameters, providing an alternative approach to centralizing local models. 
Cronus \cite{chang2019cronus} is an FL framework proposed by Chang \textit{et al.} that uses the model distillation technique to achieve collaborative learning. During the training process, each local node trains its model on private data and then makes predictions on the public dataset, producing soft labels for server aggregation. Finally, the server aggregates the collected soft labels and sends the resulting information back to the local nodes, allowing them to update their models with the latest global aggregation update. 
Li \textit{et al.} \cite{li2019fedmd} proposed Federated Learning with Model Distillation (FedMD). Each participating party uses private data and a public dataset for local training in this approach. Afterward, each party transmits its predicted logit vectors on the public dataset to the server. The server aggregates these logit vectors and returns them to each party for the following process iteration.
Lin \textit{et al.} \cite{lin2020ensemble} proposed an innovative method involving ensemble distillation for model fusion. In this technique, the central model is trained using data sourced from the outputs of various client models. This knowledge distillation strategy offers the adaptability required for aggregating heterogeneous client models, which may differ in size, numerical precision, or architecture design.
Itahara \textit{et al.} \cite{itahara2021distillation} proposed an innovative algorithm called DS-FL. In DS-FL, the outputs of local models are shared, and the communication cost is solely dependent on the output dimensions, not the model size. The shared model outputs a label for each instance in the open dataset, generating a new dataset. Leveraging this expanded dataset for further training local models can improve their performance.
Cheng \textit{et al.} \cite{cheng2021fedgems} introduced an approach in FL known as FedGEMS. In this framework, the large deeper server model strategically acquires knowledge from numerous teacher clients and itself to efficiently fuse and accumulate knowledge. This process fosters an accumulation of enriched knowledge, thereby significantly boosting the model performances on both server and client aspects.
% The global consensus in heterogeneous federated learning has been limited due to the need for clients to adjust their learning direction to account for differences among them and the assumption that each client has a clean dataset. 
Fang \textit{et al.} \cite{fang2022robust} proposed Robust Heterogeneous Federated Learning (RHFL) to handle label noise while conducting heterogeneous federated learning. RHFL employs public data to align feedback from heterogeneous models without requiring additional shared global models for collaboration. It also uses a robust, noise-tolerant loss function to minimize the negative impact of internal label noise\cite{wang2019symmetric}. Furthermore, in situations where feedback from other participants is challenging, a novel client confidence re-weighting scheme has been designed to adaptively assign corresponding weights to each client during collaborative learning.

\subsection{Poisoning Attacks in FL}
Poisoning attacks pose a significant security threat in FL. A malicious participant can intentionally introduce tainted data or updates into the shared model \cite{cao2019understanding}, undermining its accuracy or compromising its integrity. Such attacks can lead to decreased accuracy and the implementation of hidden backdoors within the model. Adversaries can exploit these vulnerabilities to degrade the model's performance or extract sensitive data \cite{lyu2023poisoning}.
Works  have demonstrated that a single Byzantine user could manipulate the parameter server to select any vector. As a result, the Byzantine user could impact the convergence of any conventional averaging-based FL methodolog \cite{blanchard2017machine}\cite{shejwalkar2021manipulating}.
Label-flipping attack \cite{fung2020limitations} was proposed by Fung \textit{et al.} to reduce the performance of the global model. The attacker intentionally modifies the labels of adversary examples to mislead the model during training, causing it to learn incorrect associations between features and labels. 
A backdoor attack is another type of attack that can be achieved by poisoning \cite{sundar2022distributed}. In such an attack, the adversary modifies specific features or small regions of the training data to implant a backdoor trigger into the model \cite{lyu2022privacy}. When the model is supplied with untainted data, it performs as expected. However, the presence of the backdoor trigger in the input causes the model to predict the class predetermined by the attacker invariably. This attack carries significant risk, allowing an attacker to manipulate the model's behavior in a specific context while leaving its overall performance on unrelated tasks unaffected.
Bagdasaryan \textit{et al.} \cite{bagdasaryan2020backdoor} carried out a backdoor attack against the FL system using a model replacement technique, which one or more malicious participants can carry out. By replacing the model with a modified version with backdoor functionality, the attacker could manipulate the model's behavior to their advantage.
Fang \textit{et al.} \cite{fang2020local} explored local model poisoning threats in Byzantine-robust FL. The authors comprehensively analyzed local model poisoning attacks on Byzantine-robust federated learning systems. Their study formalized the adversarial behavior as an optimization problem to decrease the global model accuracy, assuming that attackers had infiltrated certain clients, who then manipulated the local model parameters on these compromised devices during the learning stage. 
Yang \textit{et al.} \cite{yang2023model} introduced a new type of model poisoning attack called Model Shuffle Attack (MSA). While differential privacy protects sensitive data by adding calculated noise, this could allow attackers to avoid being detected \cite{wei2020federated}. Unlike other model poisoning attacks, the compromised model performs as expected during testing, which obscures the malevolent alterations. Nonetheless, the manipulated model detrimentally impacts the overall model's learning pace.
\section{Proposed Formulation}
\begin{table}[t]
    \caption{Notation Description}
    \centering
    \begin{tabular}{c  c}
    \hline\hline
    Notations & Descriptions\\ [0.5ex]
    \hline
    $\theta$ & Model parameter\\
    $\mathcal{L}$ & Loss function\\
    $C$ & Total classes\\
    $K$ & Total clients number\\
    $N$ & Total samples number in dataset\\
    $X_k$ & Samples in private dataset\\
    $Y_k$ & Labels in private dataset\\
    $X_0$ & Sampels in public dataset\\
    $Y_0$ & Labels in public dataset\\
    $\hat{Y}_k$ & local logits\\
    $\hat{Y}$ & global logits \\
    $\hat{Y}^\prime$ & Poisonous global logits\\
    $\alpha$ & Auxiliary information\\
    $\eta$ & Scaling factor \\[1ex]
    \hline\hline
    \end{tabular}
    \label{tab:notation}
\end{table}
In this section, we commence by offering an introduction to distillation-based federated learning. This unique FL paradigm prioritizes the exchange of predictive outputs over the direct transmission of model parameters, aiming to uphold privacy and facilitate diverse or heterogeneous settings. Following this, we outline the objectives of the poisoning attack as formulated in this study. A comprehensive list of notations can be found in the Table \ref{tab:notation}.

\subsection{Distillation-based Federated Learning}
% Public dataset $\mathrm{D}_0$, private dataset $\mathrm{D}_k$, private model $f_k, k = 1 ... m$, labels $T_k$, labeled private dataset $(d_{i,k},t_{i,k})_{i=1}^{I_k}$, $d_{i,k}$ represents the vectorized input samples. Moreover, $I_k$ denotes the number of samples in the labeled dataset. Considering $N_L$ as the number of objective class, the term $t_{i,k} = [t_{1,i,k},...,t_{N_L,i,k}]^T$ is the vectorized form of the label attached to the sample $d_{i,k}$ and is in the one-hot representation, wherein the element $t_{n,i,k}$ equals 1 if the $n$th label is the ground truth and 0 otherwise. Public logit $t^t_i$. $N_L$ is the number of objective classes.
The initial phase involves local training within the scope of distillation-based federated learning. This process is done independently by each client and does not require any data exchange or communication between the clients. During this phase, each client strives to minimize the loss on their private dataset. It is worth noting that, in certain existing schemes \cite{li2019fedmd}, clients utilize the public dataset to train the models. The formulation of this training process is presented as Eq. \ref{eq:update}.
\begin{equation}
    \label{eq:update}
        \arg \min_{\theta} \mathcal{L}\left(X_k,Y_k;\theta\right) 
\end{equation}
The term $\mathcal{L}$ is the loss function used in the training. The loss function adopted varies depending on the training scenarios or schemes. The commonly used loss functions are cross-entropy loss, mean absolute error (MAE) loss, and the Kullback-Leibler (KL) divergence loss \cite{kullback1951information}. 
% \begin{equation}
% \label{eq:1}
%     w_k^{t} \xleftarrow{} w_k^{t-1} - \eta\nabla\phi\left(\mathrm{D}_k,\mathrm{T}_k. \mid w_k^{t-1}\right)
% \end{equation}
 In the case of corss-entropy, the $\mathcal{L} \left( X_k,Y_k;\theta \right)$ is given as Eq. \ref{eq:crossentropy}. In most instances, the cross-entropy loss function is employed in scenarios that involve hard label training, where each sample is associated with a definite categorical label.
\begin{equation}
    \label{eq:crossentropy}
    \mathcal{L}\left( X_k, Y_k; \theta \right) = - \frac{1}{N} \sum_{i=1}^N\sum_{c=1}^C y_{ic} \log f(x_{ic},\theta)
\end{equation}

Subsequently, the clients pass the public dataset to the trained model, generating local logits, as depicted in Eq \ref{eq:predict}. 
\begin{equation}
    \label{eq:predict}
    \hat{Y_k} = \left\{ f\left( x_i,\theta \right) \mid x_i \in X_0 \right\}
\end{equation}
These local logits are then exchanged and aggregated into global ones, as outlined in Eq. \ref{eq:aggregate}. It should be noted that various strategies employ distinct aggregation methods during the aggregation process. 
\begin{equation}
    \label{eq:aggregate}
    \hat{Y} = Aggregate \left(\sum_{k=1}^{K} \hat{Y_k}\right)
\end{equation}

Finally, these global logits serve a critical role in the iterative refinement of the models' training. To put it more precisely, utilizing aggregated global logits from public datasets fulfills the aim of facilitating knowledge transfer. This operation, commonly termed digest, is detailed in Eq. \ref{eq:digest}. 
\begin{equation}
    \label{eq:digest}
    \arg \min_{\theta} \mathcal{L}\left(X_0,\hat{Y};\theta\right)
\end{equation}
Under this circumstance, the MAE loss and KL divergence loss are frequently adopted, as these facilitate training on logit vectors, which can be considered as the form of soft labels. The computation methods for MAE loss and KL divergence loss are outlined in Eq. \ref{eq:meanabsoluteerror} and Eq. \ref{eq:KL}, respectively.
\begin{equation}
    \label{eq:meanabsoluteerror}
    \mathcal{L}\left(X_0,\hat{Y};\theta\right) = \frac{1}{N} \sum_{i=1}^N \left| y_i - f(x_i,\theta) \right|
\end{equation}
\begin{equation}
    \label{eq:KL}
    \mathcal{L}\left(X_0,\hat{Y};\theta\right) = \sum_{i=1}^N y_i\cdot \log \left(\frac{y_i}{f\left(x_i,\theta\right)}\right)
\end{equation}
% \begin{equation}
% \label{eq:4}
% \phi\left(\mathrm{D}_{k}^{p}, \mathrm{T}_k\mid w_k^t\right)=\sum_{i=1}^{I_k}  t_{k,i}^t \cdot \log\left(\frac{t_{k,i}^t}{F\left(d_{i,k}\mid w_k\right)}\right)
% \end{equation}
% In \ref{eq:1}, the $\eta$ represents the learning rate. 

These procedures are repeated for several rounds until a stopping criterion is met, such as a maximum number of rounds, a minimum improvement in the global model's performance, or a threshold on the divergence between local and global models.

\subsection{logit poisoning attacks on distillation-based federated learning}
\textbf{Attacker's goal}
As with many previous studies on poisoning attacks, we consider untargeted attack \cite{cao2022mpaf}, \cite{xie2020fall}, \cite{li2022learning}, \cite{yu2023untargeted} in this work. In contrast to targeted attack \cite{bhagoji2019analyzing}, \cite{xie2019dba}, \cite{suciu2018does},\cite{biggio2013poisoning} untargeted attack aims to degrade the usability and maximize the error rate of the resulting model rather than generating specific outcomes for certain instances. Specifically, the untargeted attacker's objective is to corrupt the global logits by maliciously modifying the logit vectors generated by the local model, ultimately aiming to decrease the accuracy of the resulting model. In the conventional poisoning attacks against the FL system, the execution of the poisoning attack is modeled as an optimization problem, with the attacker's objective defined as inducing directed deviation. Specifically, the attacker aims to maximize the divergence of the global model's parameter update direction from its correct path. However, this approach is not applicable to our scenario. In the distillation-based federated learning paradigm, the client cannot upload model parameters. Knowledge transfer is facilitated solely through the logit vectors generated by the local model. Consequently, an attacker cannot directly affect the global model's deviation by manipulating the local model's parameters. In our work, we approach this challenge from a distinct perspective. Instead of controlling the update direction of the model parameters, we aim to distort the information contained within the local output logit vectors to accomplish the poisoning objective.

\textbf{Attacker's capability}
In this work, we assume that the attacker has control over numerous malicious clients and can manipulate the uploaded predictive outputs yet lacks knowledge about the benign clients and the ability to modify them. Importantly, we stipulate that the number of devices under the attacker's control should not exceed 50\%, as surpassing this limit would render the tampering of the model a straightforward task. We formulate the poisoning process with Eq. \ref{eq:poison1}, where the $\alpha$ is auxiliary information the attacker possesses. More precisely, the auxiliary information in our proposed attack scheme is a shuffle table built on the output statistics derived from a pre-trained classifier. This specific process is described in detail in Section \ref{sec:attack}. The manipulated local logits $\hat{Y}^\prime_k$ are further aggregated, culminating in corrupted global logits $\hat{Y}^\prime$. These corrupted global logits $\hat{Y}^\prime$ then serve as the data for subsequent training, ultimately yielding a maliciously altered model $\theta^\prime$. This process is formulated as Eq. \ref{eq:ptrain}.

\begin{equation}
    \label{eq:poison1}
    \hat{Y}^\prime_k = Poison \left(\hat{Y}_k,\alpha\right)
\end{equation}
\begin{equation}
    \label{eq:ptrain} 
    \arg \min_{\theta^\prime} \mathcal{L}\left(X_0, \hat{Y}^\prime \right)
\end{equation}

\textbf{Attacker's background knowledge}
First of all, the attacker possesses knowledge of the code, the private datasets used for local training, and the model parameters on the compromised devices. Additionally, given the transparent nature of the public dataset, the attacker also has access to it. Moreover, the attacker is informed about the aggregation rules the central server applies.

\section{Attack Method Design}\label{sec:attack}
In this section, we propose a novel logit poisoning attack in distillation-based federated learning. This method addresses the shortcomings of traditional poisoning attacks, significantly enhancing the efficiency of the poisoning process. We begin with a detailed introduction of the proposed scheme's workflow. Subsequently, we present the two principal components of our poisoning mechanism.
\subsection{Overview of Logit Poisoning Attack}
\begin{figure}
    \centering
    \includegraphics[width=0.45\textwidth]{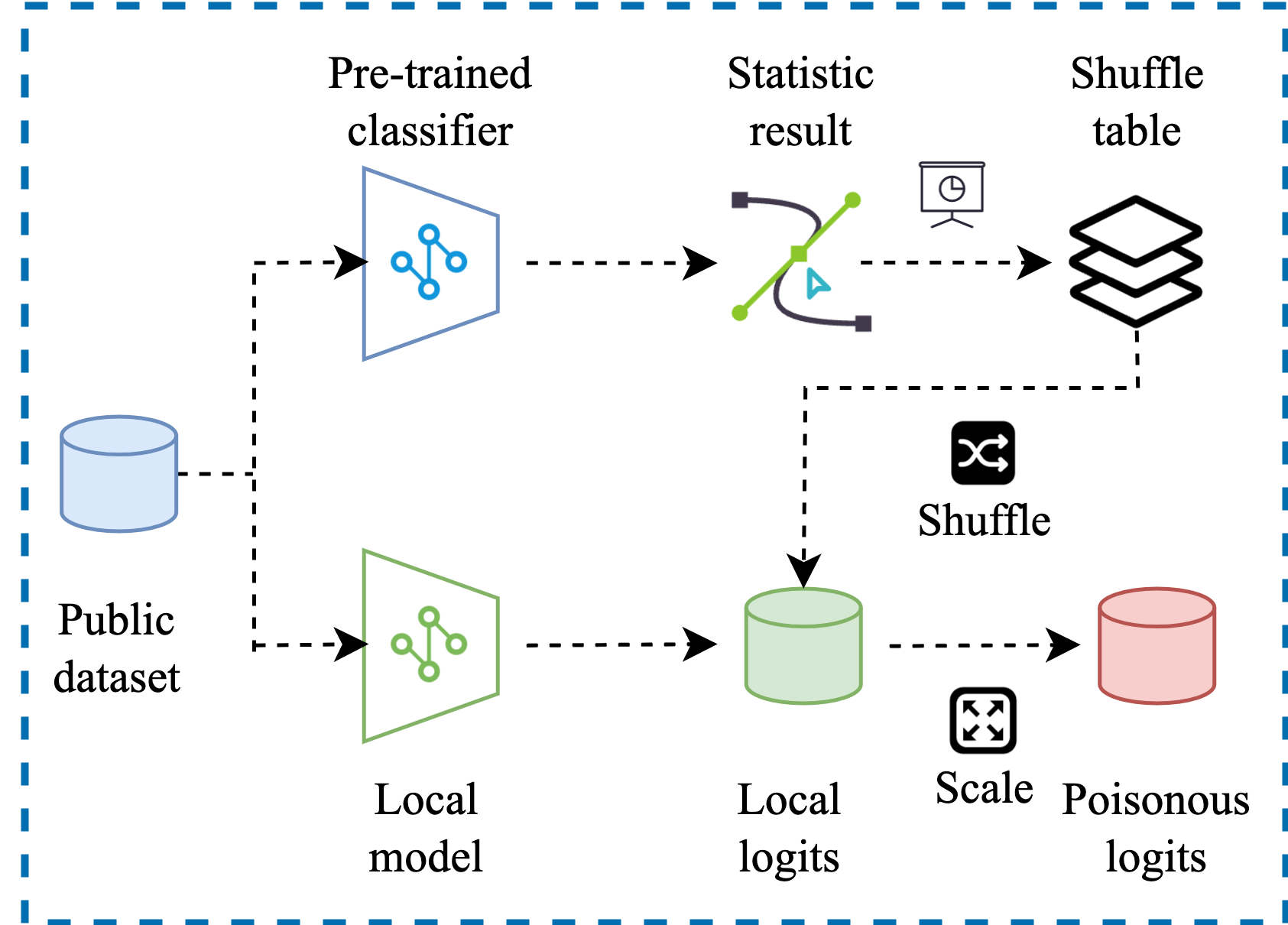}
    \caption{Illustration of the logit poisoning process}
    \label{fig:attack}  
\end{figure}

Fig. \ref{fig:attack} provides an illustration of our attack strategy, which encompasses the following two stages.
First of all, the attacker needs to create a shuffle table to distort the correlation among the various logit elements in logic vectors. Specifically, the attacker first trains a classifier using a public dataset. This pre-trained classifier's output logit vectors then undergo statistical analysis, leading to the generation of unique shuffle tables for each class. In the ensuing poisoning attack, the attacker utilizes this shuffle table to shuffle the logit elements within the logit vectors. Subsequently, the attacker scales the permuted logit elements to intensify the poisoning impact.

Overall, our poisoning attack aims to undermine the principle of knowledge representation in logit vectors. 
Each logit element in the logit vectors corresponds to the raw prediction score for a particular class. If the logit for a class is relatively high, it signifies that the model is highly confident that the given input corresponds to that class. The relative magnitudes of logit elements for different classes reflect the model's confidence level in assigning each class to a given input. In cases where the logit elements for several classes are similar, the model still determines which class to assign to the given input. This can also signify where the model struggles to distinguish between certain classes. Specifically due to these principles, distillation-based federated learning can extract knowledge. Therefore, in our poisoning attack, the attacker must disrupt these principles to amplify the error.

\subsection{Shuffling Strategy}
In our attack strategy, we first initiate the disruption of the logit vectors through a shuffling procedure. The following section explains the reason and procedure for this strategy's adoption. As mentioned earlier, one of our poisoning attack's core principles is to disrupt the associativity of the logit elements in logit vectors. The relative discrepancies between the logit elements of distinct classes signify the model's preference among those classes for a certain input. For similar classes, the model might generate comparable logit elements for these classes in a logit vector. For example, consider an image from the MNIST \cite{deng2012mnist} dataset; If the target label is '9', both '9' and '7' will register high logit values in the model-generated logit vector due to the shared characteristics between '9' and '7'. As such, these logit values reflect the model's judgment regarding the degree of similarity in the data attributes. 

An additional goal of implementing the shuffling strategy is to misdirect the trained model towards the target label, typically represented by the maximum logit value within the logit vector. The maximum logit value holds the highest significance in the logit vector. It does two things: first, it points to the target label of the provided input data, and second, it mirrors the degree of certainty the model holds about this classification. As a result of these factors, the maximum logit value captures and holds the most critical information the model provides. Therefore, in the context of a logit vector, swapping the highest and lowest values can result in severe damage. This action not only misdirects the model's data interpretation but also exacerbates the issue by steering the target label towards the most dissimilar category. Essentially, it inaccurately aligns the input data with the least probable class, thereby significantly distorting the model's comprehension and subsequent training. 
% \begin{algorithm}
%     \caption{Entropy Calculation}
%     \textbf{Input:} The sorted labels $A^*$ of group $A$;
    
%     \textbf{Output:} The information entropy $E$ of group $A$;
%     \begin{algorithmic}[1]
%         %\State $values, counts \gets Unique(Labels, return\_counts = True)$
%         \State $\mathbf{u}\gets \phi$ 
%         \State $\mathbf{c}\gets \phi$ 
%         \For{each $l$ $\in$ $A^*$} \Comment{Calculate the number of times each unique item appears in $A^*$}
%         \If{$l$ not in $\mathbf{u}$}
%                 \State $\mathbf{u} \gets \mathbf{u} \oplus l$ 
%                 \State $\mathbf{c[l]}  \leftarrow  1 $
%         \Else
%                 \State $\mathbf{c[l]} \gets \left\{ c_1,c_2,\ldots,c_n+1 \right\}$ 
%         \EndIf
%         \EndFor
%         \State $p_i\gets \frac{c_i}{\sum_{i=1}^n c_i}$ \Comment{Calculate the frequencies}
%         %\State $Entropy \gets - \sum (probability \times log_2(probability))$
%         \State $E \gets - \sum_{i=1}^n (p_i \times \log_2^{p_i})$ 
%         \State \Return{$E$}
%     \end{algorithmic}
%     \label{alg:entro}
%     \end{algorithm}
\begin{algorithm}
    \caption{Entropy Calculation}
    \textbf{Input:} The labels $A^*$ of each element in group $A$;
    
    \textbf{Output:} The information entropy $E$ of group $A$;
    \begin{algorithmic}[1]
        %\State $values, counts \gets Unique(Labels, return\_counts = True)$
        % \State $\mathbf{c}\gets \phi$ 
        \State Initialize an empty dictionary ${D}$
        \State ${c}\gets \phi$ 
        
        \For{each $l$ $\in$ $A^*$} \Comment{Calculate the number of times each unique item appears in $A^*$}
        \If{$l$ is a key in dictionary $D$}
              \State $D[l] \gets D[l] + 1$
            \Else
        \State $D[l] \gets 1$
        \EndIf
        \EndFor
        \For {each key $k$ in dictionary $D$}
        % \State $c \gets c.append D[k]$
        \State $ c$.insert$(D[k])$
        \EndFor
        \State $p_i\gets \frac{c_i}{\sum_{i=1}^n c_i}$ \Comment{Calculate the frequencies}
        %\State $Entropy \gets - \sum (probability \times log_2(probability))$
        \State $E \gets - \sum_{i=1}^n (p_i \times \log_2 {p_i})$ 
        \State \Return{$E$}
    \end{algorithmic}
    \label{alg:entro}
    \end{algorithm}
    
Also, given the varying characteristics of data within each class and the disparate levels of inter-class similarity, it is essential to develop class-specific shuffle tables. This strategy maximizes the efficacy of the poisoning attack. Initially, the attacker needs to employ the known dataset to train a model until convergence as a pre-trained classifier. After training the classifier, the attacker sends the images in the test dataset with a specific chosen class $c$ into the classifier and records the output logit vectors. In our work, we assume that the training and testing data used by the attacker are from public datasets $\{X_0, Y_0\}$.
Since a model's classification prediction for specific data may not always be accurate, our proposed method involves averaging the logit vectors that belong to the same class formulated as Eq. \ref{eq:classaverage}, where the $n_c$ is the total number of samples in the specific class $c$ and $y_c$ is the average logit vector in terms of class $c$. The averaged logit vector for each class can be seen as the representative vector for this class. This approach not only enables the creation of an optimally disruptive shuffle table but also allows for a comprehensive, statistically grounded interpretation of the model's judgment on a particular classification.
\begin{equation}
    \label{eq:classaverage}
    % y_c = \frac{\sum_{i=1}^{n_c} y_{i}}{n_c} = 
    % \frac{1}{n_c} \times
    % \begin{pmatrix}
    %     \sum_{i=1}^{n_c} e_i1 \\
    %     \sum_{i=1}^{n_c} e_i2 \\
    %     \vdots\\
    %     \sum_{i=1}^{n_c} e_in
    % \end{pmatrix}
    %\begin{aligned}
        y_c 
        %& = \frac{\sum_{i=1}^{n_c} y_{i}}{n_c} \\
         = \frac{1}{n_c} \times
        \begin{pmatrix}
            \sum_{i=1}^{n_c} e_{i1} \\
            \sum_{i=1}^{n_c} e_{i2} \\
            \vdots\\
            \sum_{i=1}^{n_c} e_{in}
        \end{pmatrix}\\
         = \begin{pmatrix}
            e_{c1} \\
            e_{c2} \\
            \vdots\\
            e_{cn}
        \end{pmatrix}
    %\end{aligned}
\end{equation}

We categorize the logit elements within the logit vector into three distinct groups according to their values, namely, 'the most likely', 'likely', and 'the least likely'. This classification facilitates more precise manipulation of the logit elements. The application of k-means clustering \cite{lloyd1982least}, chosen for its efficiency in handling this relatively low-dimensional space, aids this process. This choice not only simplifies the computational task but also contributes to reducing the overall time consumption.

Let $y = (e_1, e_2, \ldots, e_n)$ be a vector. We deploy k-means algorithm to classify the logit elements of $y$ into three groups:
\begin{itemize}
    \item Group $A = (e_1^a, e_2^a, \ldots, e_{n_a}^a)$: Logit elements with higher scores represent their associated labels most akin to the target label.
    \item Group $B = (e_1^b, e_2^b, \ldots, e_{n_b}^b)$: Logit elements with intermediate scores represent their associated labels less akin to the target label.
    \item Group $C = (e_1^c, e_2^c, \ldots, e_{n_c}^c)$: Logit elements with lower scores represent their associated labels least akin to the target label.
\end{itemize}

To maximize the disruption to the original vector, we leverage the notion of information entropy as a measure of its disorder level. Specifically, we mark the original logit elements in a logit vector into three groups, with logit elements within the same group belonging to the same class. The group's information entropy rises as it contains a greater variety of logit element types. The entropy calculation is formalized in Algorithm \ref{alg:entro}. While computing entropy, our first step is to tally the occurrences of each category $\mathbf{c}=\left\{c_1,c_2,c_3\right\}$. Subsequently, we calculate the entropy based on these tallied frequencies. While maintaining the size of each group, we further proceed with shuffling. The whole shuffling process is trying to maximize the information entropy of the three groups. Throughout the shuffling process, we continuously evaluate the information entropy of the logit elements within the group. The maximization entropy process is formulated in Algorithm \ref{alg:max}. Following the shuffling process, we undertake the fine-tuning of the result vector. As previously highlighted, the swapping of the largest logit element in a vector with the smallest one can cause significant misdirection. However, given that the entire shuffling process is random, we can't guarantee such an occurrence. Consequently, in the concluding phase, we manually adjust the vector to ensure that the position originally occupied by the smallest logit element now hosts the largest one.

% \begin{algorithm}
% \caption{Numpy Unique Function}
% \begin{algorithmic}[1]
% \Procedure{Unique}{$array, return\_counts$}
%     \State $unique\_elements \gets empty\_list$
%     \State $counts \gets empty\_list$
%     \State $sorted\_array \gets sort(array)$
%     \For{each $element$ in $sorted\_array$}
%         \If{$element$ not in $unique\_elements$}
%             \State Append $element$ to $unique\_elements$
%             \State Append 1 to $counts$
%         \Else
%             \State Increment the last element in $counts$
%         \EndIf
%     \EndFor
%     \State \Return{$unique\_elements, counts$}
% \EndProcedure
% \end{algorithmic}
% \end{algorithm}

\begin{algorithm}
\caption{Entropy Maximization Algorithm}
\textbf{Input:} Three groups $A$, $B$, and $C$, and their respective labels $A^*$, $B^*$, and $C^*$;

\textbf{Output:} Shuffled groups $A$, $B$, and $C$;
\begin{algorithmic}[1]
% \Ensure The modified $set1$, $set2$, and $set3$ that potentially have a higher collective entropy
    \State $\mathcal{S} \gets [A,B,C]$
    \State $\mathcal{S}^* \gets [A^*,B^*,C^*]$
    \State $N_r \gets 500$ \Comment{Define the total shuffling rounds}
    \State $E \gets \sum_{s^* \in \mathcal{S}^*} \operatorname{Entropy}(s^*)$  \Comment{Calculate the entropy using Algorithm \ref{alg:entro}}
    \For{$i \gets 1$ to $N_r$}
        \State Randomly select two distinct sets from $\mathcal{S}$ as $s_a,s_b$
        \State Randomly select two indexes from $s_a$ and $s_b$ as $i_a,i_b$
        %\State $i_a, i_b \gets \text{select two indexes from $s_a$ and $s_b$ at random}$
        \State $\operatorname{Swap}\left(s_a\left[i_a\right], s_b\left[i_b\right]\right)$
        \State $\mathcal{S}^* \gets \operatorname{getLabel}(\mathcal{S})$
        \State $E^{\prime} \gets \sum_{s^* \in \mathcal{S}^*}\operatorname{Entropy}(s^*)$
        \If{$E^{\prime} \leq E$}
            \State $\operatorname{Swap}\left(s_a\left[i_a\right], s_b\left[i_b\right]\right)$
            %\State Revert $set\_a[index\_a]$ and $set\_b[index\_b]$ \Comment{Undo the swap}
        \Else
            \State $E \gets E^{\prime}$
        \EndIf
    \EndFor
    \State \Return $\mathcal{S} = \left[A, B, C\right]$ \Comment{Return the shuffled groups}
\end{algorithmic}
\label{alg:max}
\end{algorithm}

% \begin{equation}
%     \label{eq:cf}
%     c_i = \sum_{j=1}^n 1_{u_i}\left(e_j\right)
% \end{equation}

%  \begin{equation}
%     \label{eq:cf}
%     c_i = \sum_{j=1}^n
%     \begin{cases}
%         1, & \text{if $e_j = u_i$}\\
%         0, & \text{otherwise}
%      \end{cases}
% \end{equation}

\begin{figure*}
    \centering
    \includegraphics[width=\textwidth]{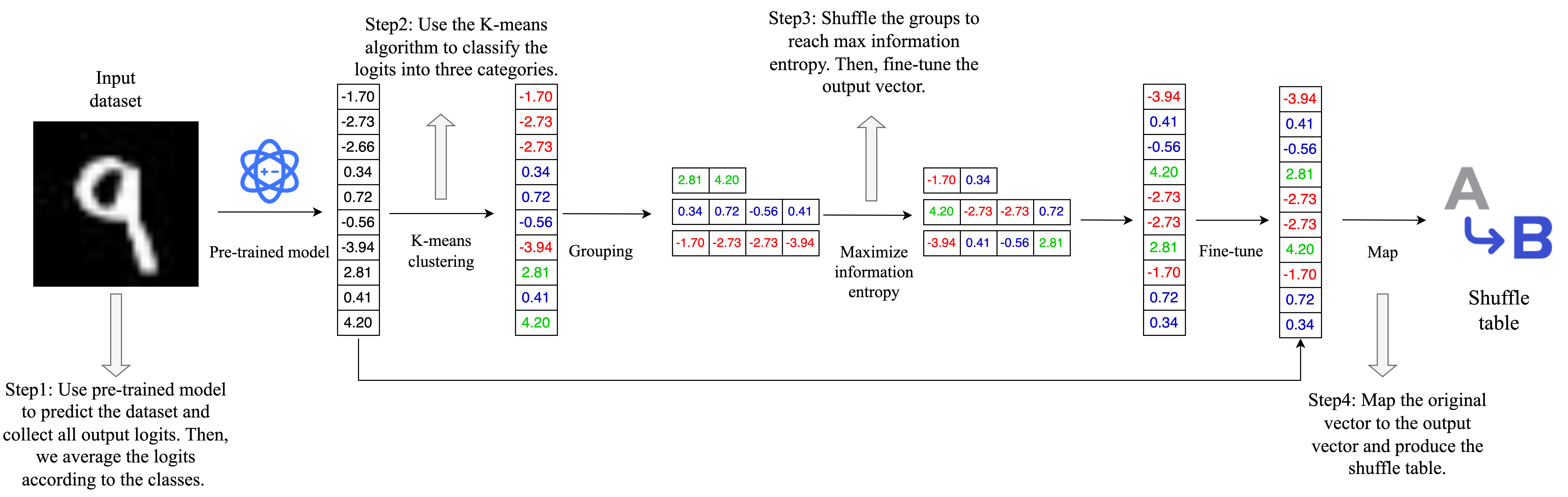}
    \caption{The generation process of shuffling table}
    \label{fig:st} 
\end{figure*}

Finally, we collect the mapping of original vectors to resulting vectors as the final shuffled table. Formulated in Eq. \ref{eq:shuffletable}, the shuffle table for each class can be defined as a function $P$, where $\pi$  is a permutation of the indices $\{1,2,\ldots,n\}$.
The whole shuffle table generation process is shown in Fig. \ref{fig:st}.
\begin{equation}
    \label{eq:shuffletable} 
    \begin{gathered}
        P: y \rightarrow y^\prime \\
        \text{s.t.} \quad y^\prime = P(y) = \begin{pmatrix}e_{\pi(1)} & e_{\pi(2)} & \ldots & e_{\pi(n)}\end{pmatrix}^\mathsf{T}
    \end{gathered}
\end{equation}

\subsection{Scaling Strategy}
Scaling is a common technique used in conventional model poisoning attacks \cite{yang2023model}. In model poisoning attacks, scaling the model parameters can effectively enhance the effects of poisonous models during the aggregation process. Employing the scaling operations can easily alter the parameter distribution of the final model. In our approach, we conduct the scaling operation by multiplying the shuffled vector by a predetermined scaling factor as $y^\prime = \eta * y$, where $\eta$ is the scaling factor. The scaling operation also holds significant utility in the context of logit poisoning, given that the knowledge carried by the logit vectors plays a pivotal role during the knowledge transfer phase. Since the shuffled logit vectors are already capable of misguiding the model that is trained with it, our goal is merely to magnify the error. In order to accomplish this, we typically employ a scaling factor greater than 1. By scaling, the poisonous logit vectors can maximize their poisoning effect during the aggregation process and subsequent training process. 
The whole poisoning process can be formulated as Eq. \ref{eq:poison2}.

\begin{equation}
    \label{eq:poison2}
    \begin{gathered}
        \begin{aligned}
            Y^\prime & = Poison\left(Y, \pi, \eta\right) \\
            & = \left\{ \eta \cdot P\left(y_i\right) \mid y_i \in Y \right\}, 
        \end{aligned}\\
        \text{ where } P(y_i) = \begin{pmatrix}e_{\pi(1)} & e_{\pi(2)} & \ldots & e_{\pi(n)}\end{pmatrix}^\mathsf{T}
    \end{gathered}
    % \begin{aligned}
    %     Y^\prime & = Poison\left(Y, \pi, \eta\right) \\
    %     & = \left\{ \eta \cdot P\left(y_i\right) \mid y_i \in Y \right\}, \\
    %     \text{ where } P(y_i) = \begin{pmatrix}e_{\pi(1)} & e_{\pi(2)} & \ldots & e_{\pi(n)}\end{pmatrix}^\mathsf{T}\\
    %     %& =\left\{ \begin{pmatrix} \eta \cdot e_{\pi(1)} & \eta \cdot e_{\pi(2)} & \ldots & \eta \cdot e_{\pi(n)} \end{pmatrix}^\mathsf{T}_i \mid y_i \in Y \right\} 
    % \end{aligned}
\end{equation}

\subsection{Discussion}
As previously discussed, we can't optimize the poisoning effect using the offset direction due to the differing nature of our poisoning targets. In many conventional FL poisoning schemes, the effectiveness of the poisoning attack is typically optimized by either maximizing the loss function or the deviation of the vector. Such methods aim to inflict maximum damage to the model or to distort the vector direction away from the normal direction.
However, such approaches can't be applied to our scenario to optimize the poisoning effect, as they do not completely capture the impact of poisoning. For instance, if cosine similarity is used to measure the poisoning effect, the aim would be to minimize the similarity between the poisonous and original vectors. Therefore, negating the original vector would accomplish this. However, this doesn't disrupt the knowledge embedded in the logit vectors to the maximum extent. For example, consider a model that produces a logit vector of $\begin{pmatrix} 0.9 & 0.8 & -0.9 & -0.8 \end{pmatrix}^\mathsf{T}$, where each logit element represents the confidence that the input data belongs to a particular class. Let's assume that the four classes are represented as \textit{classA}, \textit{classB}, \textit{classC}, and \textit{classD}, respectively, and the negated logit vector is $\begin{pmatrix} -0.9 & -0.8 & 0.9 & 0.8 \end{pmatrix}^\mathsf{T}$. These two logit vectors convey two pieces of crucial information. First, it represents the model's current judgment, implying that the input data most likely belongs to the class with the highest output in the vector. Secondly, it signifies the model's judgment about the similarity between different classes. For instance, it suggests that \textit{classA} is very similar to \textit{classB}, and \textit{classC} is very similar to \textit{classD}, as their respective logit values are alike. Conversely, this can also be interpreted as \textit{classA} and \textit{classB} being markedly different from \textit{classC} and \textit{classD}. However, although the vector after negating has the smallest cosine similarity compared with the original vector, it still can express the similarity between \textit{classA} and \textit{classB}, \textit{classC} and \textit{classD}.

For this reason, in this work, we delve into methods to optimize the effects of logit poisoning. We assess the effectiveness of logit poisoning from two distinct perspectives. First, we use inter-class relevance as one perspective to evaluate the impact of the poisoning attack. For each output logit vector, we calculate the distance of the logit element $i$ from $j$ as Eq. \ref{eq:relevance}. Also, we define the similarity score of a logit element to itself as zero. Therefore, we can calculate one relevance vector for each logit element. This vector signifies the level of similarity between this logit element and the other logit elements within the logit vector. The higher the distance score of logit element $i$ in terms of logit element $j$, the less similar they are.

% \begin{figure*}
%     \centering
%     \includegraphics[width=\textwidth]{pictures/distribution.png}
%     \caption{Distribution of logit vectors}
%     \label{fig:dist}
% \end{figure*}

\begin{equation}
    \label{eq:relevance}
    s_{ij} = \frac{\left|e_i - e_j\right|}{\sum_{k=0}^n \left| e_i - e_k \right|}
\end{equation}

% \begin{equation}
%     \label{eq:relevanceVector}
%     d_i =
%     \begin{bmatrix}
%         s_{i1}\\
%         s_{i2}\\
%         \vdots \\
%         s_{in}
%     \end{bmatrix}
% \end{equation}

We subsequently gather all the calculated relevance vectors within the logit vector and assemble a distance matrix as Eq. \ref{eq:distanceMatrix}. This distance matrix houses the relational data of all logit elements within the logit vector with respect to others. Therefore, we can employ this matrix to evaluate the inter-class relevance present within the logit vector quantitatively. Specifically, the higher the score in the matrix, the greater the distance between the class represented by its row and the corresponding other class, and thus, the larger the gap.
\begin{equation}
    \label{eq:distanceMatrix}
    \mathbf{M} = 
    \begin{pmatrix}
        {d_{1}} \\
        {d_{2}} \\
        \vdots \\
        {d_{n}}
    \end{pmatrix}
    % ^\mathsf{T}
    = 
    \begin{pmatrix}
        s_{11} & s_{12} & \cdots & s_{1n} \\
        s_{21} & s_{22} & \cdots & s_{2n} \\
        \vdots  & \vdots  & \ddots & \vdots  \\
        s_{n1} & s_{n2} & \cdots & s_{nn} 
    \end{pmatrix}
\end{equation}

\begin{figure*}
    \centering
    \includegraphics[width=\textwidth]{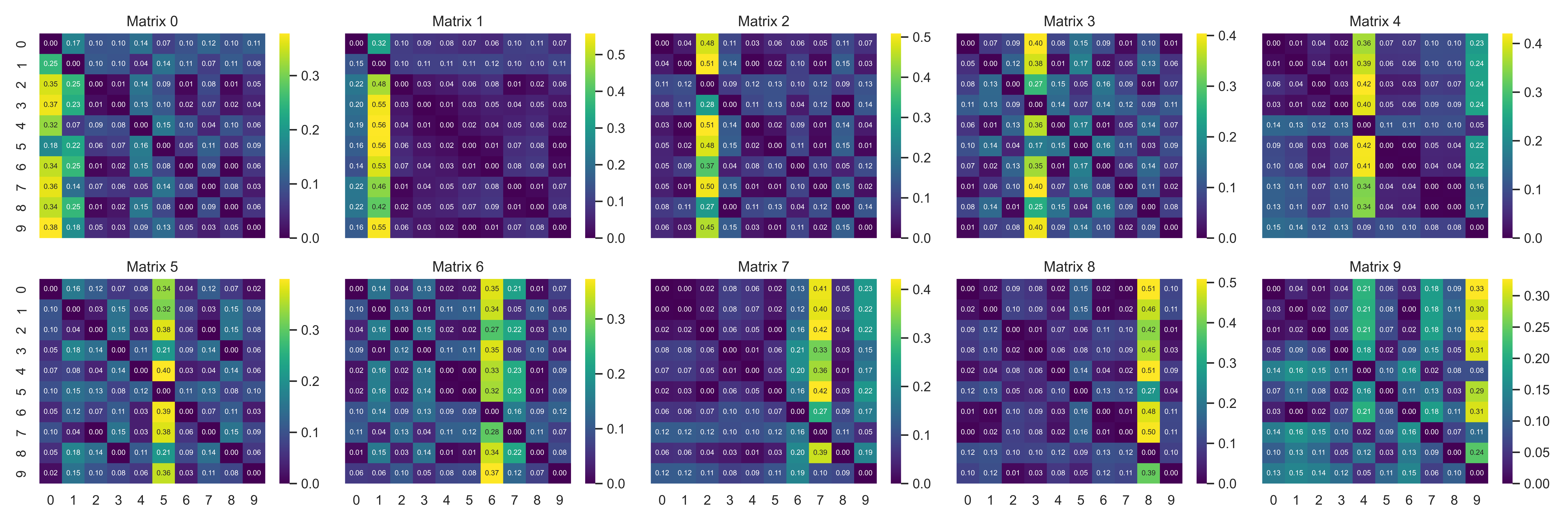}
    \caption{Distance matrix of logit vectors before poisoning}
    \label{fig:d1}
\end{figure*}

\begin{figure*}
    \centering
    \includegraphics[width=\textwidth]{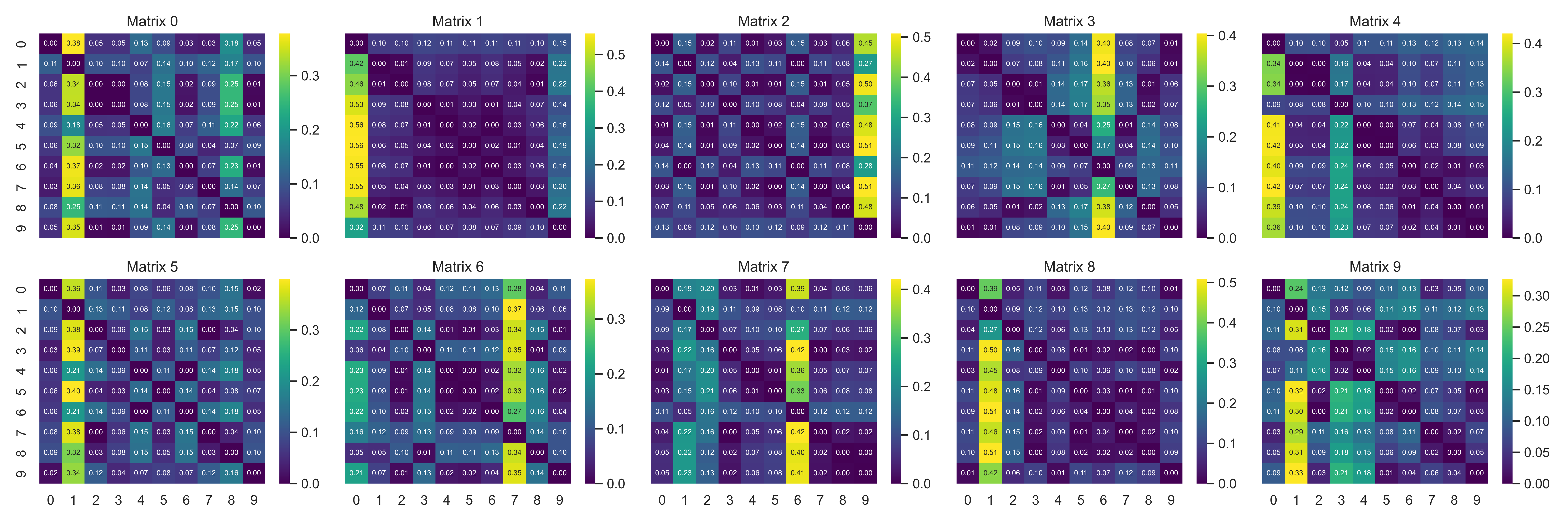}
    \caption{Poisoning effect from the perspective of distance matrix}
    \label{fig:d2}
\end{figure*}

In our work, we recognize that the original logit vector contains the accurate judgment of the model on the input data. Consequently, we use the distance matrix corresponding to the original vector as a reference. We then apply Eq. \ref{eq:matrix1norm} to assess the impact of the poisoning on the original. Specifically, we subtract the distance matrix of the poisonous vectors $\mathbf{M}^\prime$ from the distance matrix of the original logit vectors $\mathbf{M}$ and calculate the L1 norm of the resulting matrix $\left|\left| \mathbf{M}-\mathbf{M}^\prime \right|\right|$, which involves taking the absolute values of all elements in the resulting matrix and summing them up. The distance matrix of the original logit vectors and the distance matrix of the poisonous vectors are depicted in Fig. \ref{fig:d1} and Fig. \ref{fig:d2}, respectively. It's evident that the two sets of distance matrices differ significantly, with each element having been shuffled along with its associated elements.

\begin{equation}
    \label{eq:matrix1norm}
    \begin{aligned}
        \mathcal{S}_1 
        & = \left|\left| \begin{pmatrix}
            s_{11} -s_{11}^\prime & s_{12}-s_{12}^\prime & \cdots & s_{1n}-s_{1n}^\prime \\
            s_{21} -s_{21}^\prime & s_{22}-s_{22}^\prime & \cdots & s_{2n}-s_{2n}^\prime \\
            \vdots  & \vdots  & \ddots & \vdots  \\
            s_{n1}-s_{n1}^\prime & s_{n2} -s_{n2}^\prime& \cdots & s_{nn} - s_{nn}^\prime
        \end{pmatrix} \right|\right| \\
        & = \sum_{i=0}^n\sum_{j=0}^n \left| s_{ij} - s_{ij}^\prime \right|
    \end{aligned}
\end{equation}

\begin{figure}
    \centering
    \includegraphics[width=.48\textwidth]{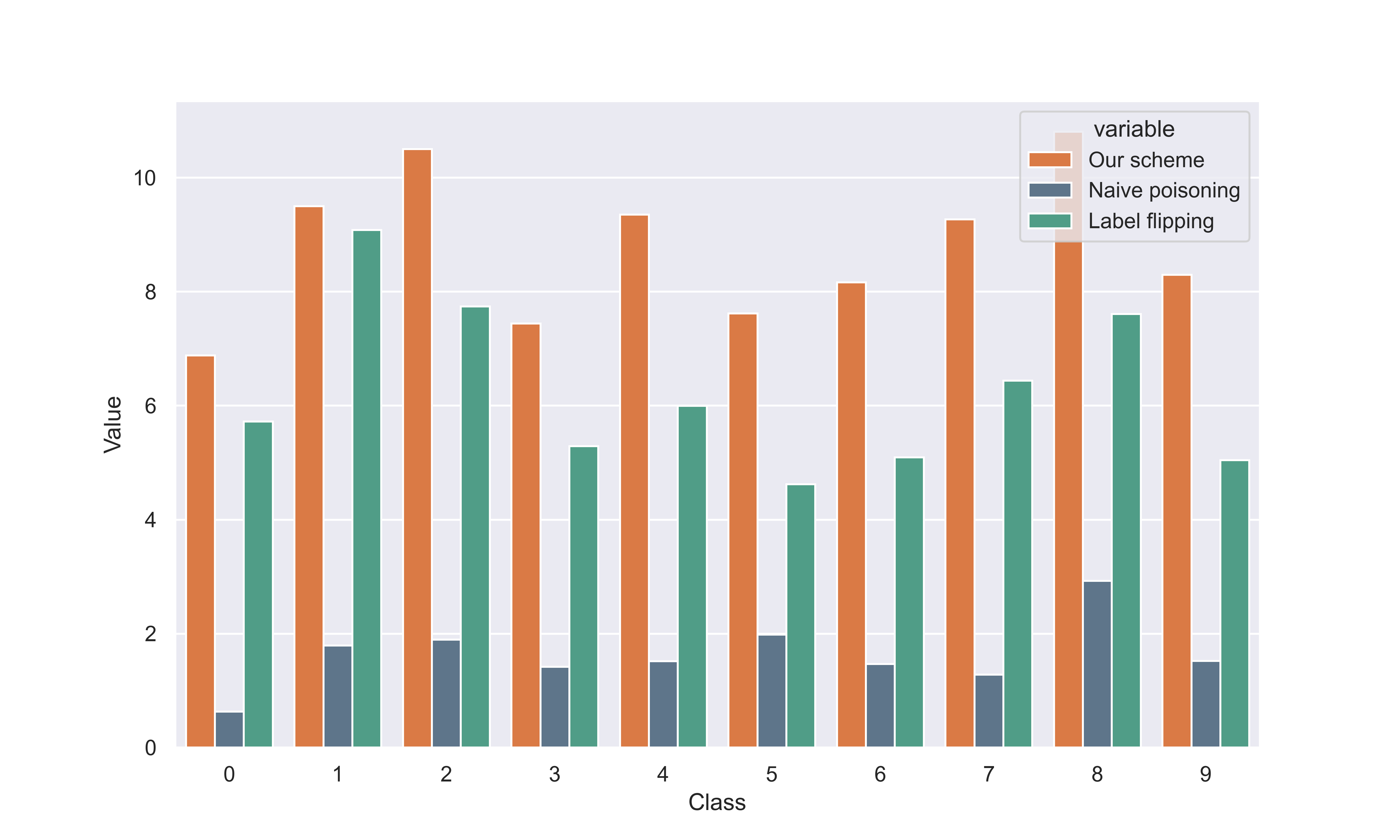}
    \caption{Comparsion of distance matrixes under different poisoning schemes}
    \label{fig:dmcs}
\end{figure}

In order to compare our poisoning scheme with existing ones, we used this calculation to derive the poisoning scores and showcased the results in Fig. \ref{fig:dmcs}. The figure clearly indicates that the poisoning score for vectors tainted using our strategy is significantly higher than that of the existing schemes. Notably, the shuffling strategy in our plan is custom-tailored for each class based on statistical data. Moreover, the label-flipping attack essentially substitutes elements in the vector, which is why it can also cause a considerable poisoning effect on the similarity matrix. Nonetheless, our attack method utilizes a meticulously designed scheme to maximize this poisoning effect.

% \begin{figure*}
%     \centering
%     \includegraphics[width=\textwidth]{pictures/distanceMatrix1.png}
%     \caption{Distance matrix of logit vectors before poisoning}
%     \label{fig:d1}
% \end{figure*}
% \begin{figure*}
%     \centering
%     \includegraphics[width=\textwidth]{pictures/distanceMatrix2.png}
%     \caption{Poisoning effect from the perspective of distance matrix}
%     \label{fig:d2}
% \end{figure*}
Another angle from which we optimize the poisoning effect concerns the position of the maximum value in the logit vector. As we know, the maximum value in the model's predictions signifies the predicted result. The gap between the data class symbolized by the position of the maximum value post-shuffling and the actual class profoundly affects the poisoning effect. Furthermore, the magnitude of the maximum value represents the model's confidence in the outcome. Thus, the poisonous vector can not only misdirect the model trained with the data but also augment the influence on the model due to the value.
% The distance between the original largest logit element and the poisonous one.
Let $y = (e_1, e_2, \ldots, e_n)$ be a vector. The index of the largest logit element in $y$ can be denoted by Eq. \ref{eq:largestElement}.

\begin{equation}
    \label{eq:largestElement}
    i_{m} = \underset{i}{\text{arg max}} \ e_i 
\end{equation}

\begin{figure}
    \centering
    \includegraphics[width=.48\textwidth]{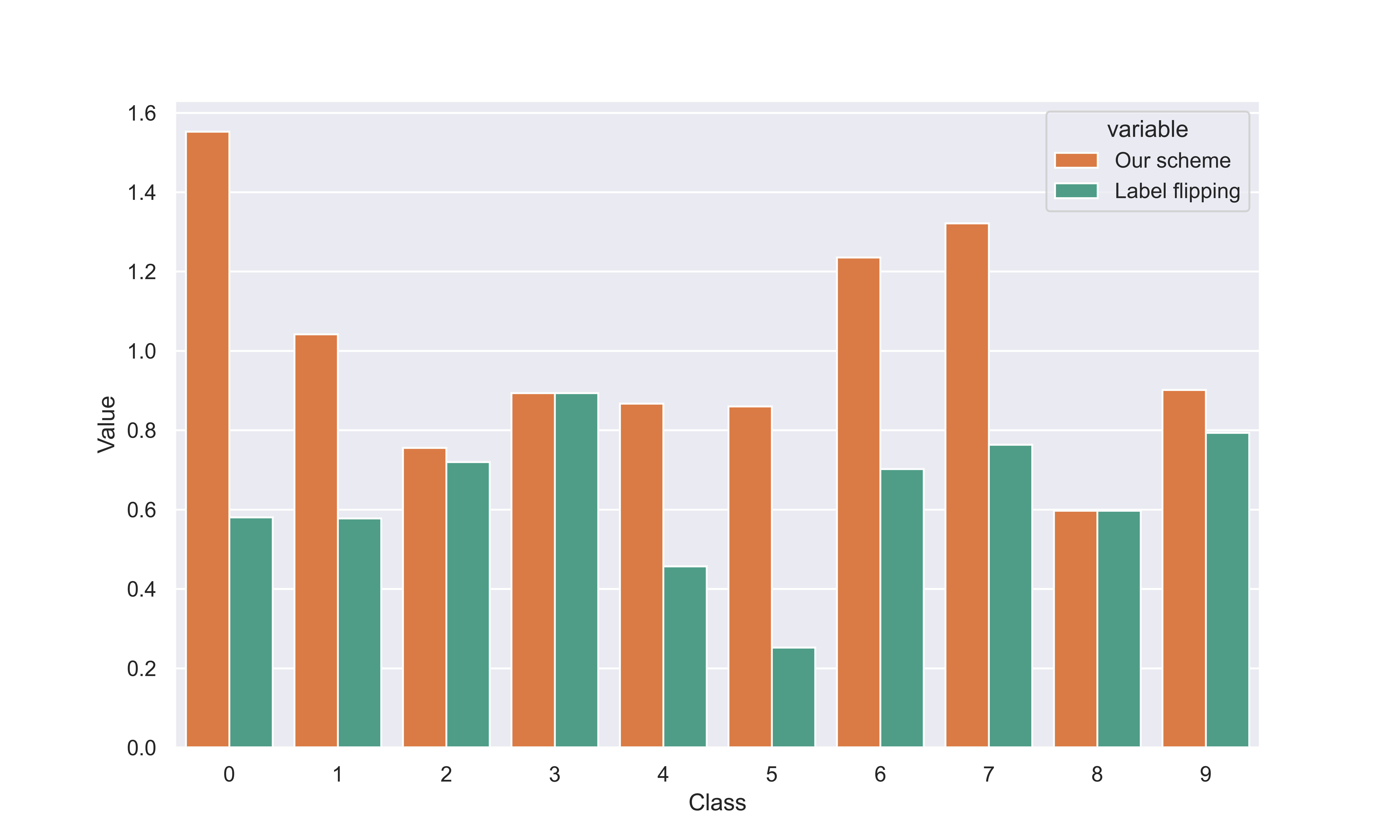}
    \caption{Poisoning effect from the perspective of max value position} 
    \label{fig:poisonScore2}
\end{figure}

Therefore, we further define the second poisoning score as Eq. \ref{eq:sps}. We measure the impact of poisoning by calculating the product of the difference in values at the maximum position and the distance. Similarly, the higher the poisoning score, the better the poisoning effect of the solution. We compare the poisoning score of our approach with that of the label-flipping attack and showcase the comparison in Fig. \ref{fig:poisonScore2}. At this stage, we do not collect the poisoning score of the naive poisoning, as this technique rarely alters the position of the maximum value within the logit vector. Consequently, it fails to gain any advantage using this poisoning evaluation method. In essence, a naive attack usually scores zero in this particular context.

\begin{equation}
    \label{eq:sps}
    \begin{aligned}
        \mathcal{S}_2 &= \left| e_{i_m^\prime} - e^\prime_{i_m^\prime}\right| \times s_{i_m i_m^{\prime}}\\ &=  \frac{\left| e_{i_m^\prime} - e^\prime_{i_m^\prime} \right|\times \left|e_{i_m} - e_{i_m^{\prime}}\right|}{\sum_{k=0}^n \left| e_{i_m} - e_k \right|} 
    \end{aligned}
\end{equation}

% The objective is to:
% \begin{equation}
%     \max \left(\mathcal{S}_1 + \mathcal{S}_2\right)
% \end{equation}

\section{Toward Secure Distillation-based Federated Learning}
In this section, we propose a robust aggregation algorithm to defend against the aforementioned logit poisoning attack. This enables secure distillation-based learning in an adversarial environment.
\subsection{Proposed Defense Scheme}
We propose a robust aggregation algorithm specifically designed to defend distillation-based federated learning, aiming to guard against logit poisoning attacks. This novel aggregation method functions by identifying and subsequently discarding local logits that may be potentially malicious before computing the global logits in each iterative process. The general procedure of the proposed robust logit aggregation algorithm is shown in Alg. \ref{alg:defense}.

In the initial phase, it computes the average logit vectors uploaded by each user and associated with the same class, thus calculating a representative vector for each class. The algorithm then needs to generate the corresponding aggregation weights tailored to each class. The algorithm employs spectral clustering \cite{von2007tutorial} to categorize the representative vectors that belong to identical classes. It is worth noting that our threat model assumes that the number of attackers constitutes less than 50\% of the total. Hence, we operate under the assumption that the cluster with the largest size encompasses all benign logit vectors. Furthermore, we posit that the centroid of this dominant cluster represents the benign distribution. This clustering technique plays a crucial role in the algorithm's efficacy, aiding in identifying potentially harmful entities and ensuring the integrity of the federated learning process.
Subsequently, the algorithm computes the average of all logit vectors within a cluster determined to be benign according to predefined criteria. This calculation then yields the distribution of benign logit vectors.

\begin{algorithm}
\begin{spacing}{1.1}
    \caption{Robust Logit Aggregation Algorithm for Distillation-based Federated Learning Framework}
    \textbf{Input}: Private datasets $\left\{X_k, Y_k\right\}_{k=1}^K$, public dataset $\{X_0,Y_0\}$, local models $\left\{\theta_k\right\}_{k=1}^K$, number of communications $T$, $F_s\left(\cdot \mid \mathcal{T}^e\right)$ softmax function with respect to temperature $\mathcal{T}^e$.
    \begin{algorithmic}[1]
        \State Initialize the training iteration index by $t = 1$
        \For {$t \leq T$}
            \State Each client minimize loss on $\{X_k,Y_k\}$
            \State Each client computes the local logits and sends them to the server:
            \State \indent $Y^t_k = 
            \{ f\left(x_i, \theta_k^t\right) | x_i \in X_0 \}$
            \For {$c \in C$}
                \State The server generates corresponding representative vectors of logit vectors uploaded by each user:
                \State \indent $Y^t_{kc} = \frac{\sum_{i=1}^{n_c} y_{i}}{n_c}$ 
                \State For representative vectors, the server applies spectral clustering and obtain labels $L = \{l_1, l_2, ..., l_n\}$, and finds the label $l_{benign}$ with the highest number of occurrences $F$ in $L$
                \State Compute the benign vector distribution:
                \State \indent $y_{benign} = \frac{1}{F} \sum_{i : l_i = l_{benign}} y_i$
                \State Compute cosine similarity:
                \State \indent $S_{kc} = \left\{\frac{y_i \cdot y_{benign}}{\lVert y_i \rVert \lVert y_{benign} \rVert} \mid y_i \in Y^t_{kc} \right\}$
                \State Assign weight according to the representative vector of class $c$:
                \State \indent $w_{kc} = F_s \left( S_{kc} \mid \mathcal{T}^e \right)$
            \EndFor
            \State Calculate the weights for individual users by averaging the class representative weights:
            \State \indent $w_k = \frac{1}{||C||}\sum_{c=0}^{||C||} w_{kc}$
            \State The server aggregates the logit vectors based on the weights:
            \State \indent $\hat{Y}^t = \frac{\sum_{k=1}^K w_k \times \hat{Y}_k^t}{K}$
            %\State \indent $\hat{Y}^t = F_s \left( \frac{\sum_{k=1}^K Y_k^t}{K} \mid \mathcal{T}^e \right)$
            \State Each client minimize the loss on $\{X_0, \hat{Y}^t\}$
        \EndFor
    \end{algorithmic}
    \label{alg:defense}
\end{spacing}
\end{algorithm}

Following this, we calculate the cosine similarity between the representative vectors and the previously computed distribution of benign vectors. This similarity is then processed through the softmax function, incorporating a temperature parameter \cite{hinton2015distilling}, in order to derive the aggregated weights. 
In the final stage of our algorithm, each user is assigned $|C|$ aggregation weights representing distinct classes. When all logit vectors are aggregated at the end of the process, the average weight corresponding to each user is utilized. This structure ensures a balanced and robust aggregation, in alignment with our defensive approach against potential poisoning attacks. We utilize the weights computed to perform a weighted average of the logit vectors from all users. This weighted aggregation results in the derivation of the global logits, embodying collective intelligence and ensuring the integrity of the federated learning process.
In our approach, the incorporation of the softmax function with a temperature parameter serves as a mechanism to finely tune the sensitivity of the malicious logit vectors detection algorithm. Specifically, when the temperature is set to a value less than 1, the mathematical effect is to magnify the gap between the benign vector and the other logit vector. This enlargement of the gap translates to the algorithm assigning lower weights to vectors that are likely to be malicious. Conversely, when the temperature is set to a value greater than 1, the algorithm exhibits a higher tolerance for vectors that deviate from the benign distribution. This means that vectors differing from the benign distribution are assigned greater weights, reflecting a more permissive stance. Consequently, the temperature parameter provides a valuable tool for calibrating the detection threshold, allowing for nuanced control over the robustness of the federated learning process.

\subsection{Discussion}
To prove the robustness of our aggregation scheme within the context of distillation-based federated learning, we delineate the security properties inherent to our approach here.
Here, we define the logit vectors uploaded by the honest users as $\{Y_i,i\in H\}$ and the logit vectors of the adversary as $\{Y_i^\prime,i\in A\}$. Considering $E[Y] = \mathbf{y}$ as an unbiased estimator of the benign logit vectors, we have $E\left[Y_i\right] = E[Y] = \mathbf{y}$. Furthermore, we assume that the difference between the benign and poisonous logit vectors as $\delta = \frac{\sum_{i\in A} Y_i^\prime}{|A|}-\frac{\sum_{i \in H} Y_i}{|H|}$. As a result, we have:
\begin{equation}
    \begin{aligned}
        \sum_{i \in A} Y_{\mathbf{i}}^\prime & =|A| \delta+\frac{\sum_{i \in H} Y_i }{|H|}|A| \\
        & =|A| \delta+\left(\frac{|A|}{|H|}-1\right) \sum_{i \in H} Y_i +\sum_{i \in H} Y_i \\
        & =|A| \delta+\left(\frac{|A|}{|H|}-1\right)|H| \mathbf{y}+\sum_{i \in H} Y_i
    \end{aligned}
    \label{eq:proof1}
\end{equation}
We have thus established that an error term exists between the malicious and benign logit vectors, denoted by $\Delta = |A| \delta+\left(\frac{|A|}{|H|}-1\right)|H|\mathbf{y}$. If the poisonous logit vectors are sent directly to the server. This error term will paralysis other models during the subsequent process.

Since poisonous logit vectors follow a different distribution compared to benign ones, they will generally be situated farther away from the majority of the logit vectors. As a vector becomes more effective at poisoning, its position in the space shifts farther away from the benign distribution. This movement exemplifies the increased divergence and assists in characterizing the malicious behavior within the data. Given that an attacker aims to maximize the poisoning effect produced by the poisonous logit vectors, it follows that:
\begin{equation}
    S_{kc} = \frac{y_i^\prime\cdot y_{benign}}{||y_i^\prime||||y_{benign}||}\to {-1}
\end{equation}
As a consequence, the weights assigned to the poisonous logit vectors exhibit a tendency to approach zero $w_{kc} = F_s \left(S_{kc}\ |\ \mathcal{T}^e \right) \rightarrow 0$ while the benign logit vectors still hold greater weights. The error terms can also be written as $\Delta = \sum_{i \in A} Y_{\mathbf{i}}^\prime - \sum_{i \in H} Y_i$. From this perspective,  our robust aggregation algorithm can effectively eliminate the effect of the error terms $\Delta \rightarrow 0$ by assigning very low weights for poisonous logit vectors.

\section{Experiments Analysis}
In this section, we begin by executing the proposed logit poisoning attack within the context of distillation-based federated learning. Subsequently, we conduct experiments to evaluate the effectiveness of our proposed defense algorithm. All experiments are run in a high-performance server with the configuration of Ubuntu 22.04, Intel Xeon E5-2680 2.40GHz CPU, RTX 3060, and 128GB RAM.
\subsection{Experimental Setup}
We employed three distillation-based federated learning schemes to evaluate our schemes: FedMD \cite{li2019fedmd}, DS-FL \cite{itahara2021distillation}, and FedDF \cite{lin2020ensemble}. The pseudocode for these three schemes is presented in Alg. \ref{alg:fedmd}, \ref{alg:dsfl}, and \ref{alg:feddf}, respectively. The dataset we use is MNIST \cite{deng2012mnist}. The MNIST dataset is a large database of handwritten digits that is commonly used in the field of machine learning. The dataset contains 60,000 training images and 10,000 testing images, each of which is a grayscale image of 28x28 pixels. These images represent handwritten digits from 0 to 9 and are labeled with the corresponding integer. We partition the dataset, allocating 65,000 samples as private user data and dividing them equally among the users. Out of these, 4,000 samples are designated as public datasets for knowledge transfer, while 1,000 samples are reserved as the test dataset to evaluate the model's accuracy and loss.

\begin{algorithm}
  \caption{FedMD}
  \textbf{Input}: Private datasets $\{X_k, Y_k\}_{k=1}^K$, public dataset $\{X_0,Y_0\}$, local models $\left\{\theta_k\right\}_{k=1}^K$, number of communications $T$.
  
  \begin{algorithmic}[1]
    \State Initialize the training iteration index by $t$ = 1.
    \State Each client minimize the loss on $\{X_0, Y_0\}$
    \State Each client minimize the loss on $\{X_k, Y_k\}$
    \For {$t \leq T$}
    \State Each client computes the local logits and sends them to the server:
    \State \indent $\hat{Y}_k^t = 
    \{f\left(x_i; \theta_k^t\right) | x_i \in X_0\}$
    \State The server computes the global logits:
    \State \indent $\hat{Y}^t = \frac{\sum_{k=1}^K \hat{Y}_k^t}{K}$
    \State Each client receives $\hat{Y}^t$ and minimize the loss on $\left\{X_0, \hat{Y}^t\right\}$
    \State Each client minimize the loss on $\left\{X_k, Y_k\right\}$
    \EndFor
  \end{algorithmic}
  \label{alg:fedmd}
\end{algorithm}

\begin{algorithm}
    \caption{DS-FL}
    \textbf{Input}: Private datasets $\left\{X_k, Y_k\right\}_{k=1}^K$, public dataset $\{X_0,Y_0\}$, local models $\left\{\theta_k\right\}_{k=1}^K$, number of communications $T$, $F_s\left(\cdot \mid \mathcal{T}^e\right)$ softmax function with respect to temperature $\mathcal{T}^e$.
    \begin{algorithmic}[1]
    \State Initialize the training iteration index by $t$ = 1.
    \For {$t \leq T$}
    \State Each client minimize loss on $\{X_k,Y_k\}$
    \State Each client computes the local logits and sends them to the server:
    \State \indent $Y^t_k = 
    \{f\left(x_i, \theta_k^t\right) | x_i \in X_0\}$
    \State The server computes the global logits:
    % \State \indent $L^t=F_{\mathrm{s}}\left(\frac{1}{K} \sum_{k=1}^{K} \hat{T}_{k} \mid T^e\right)$
    \State \indent $\hat{Y}^t = F_s \left( \frac{\sum_{k=1}^K Y_k^t}{K} \mid \mathcal{T}^e \right)$
    %\State \indent $L^t = \frac{\sum_{k=1}^K L_k^t}{K}$
    \State Each client minimize the loss on $\{X_0, \hat{Y}^t\}$
    \EndFor
    \end{algorithmic}
    \label{alg:dsfl}
\end{algorithm}

\begin{algorithm}
    \caption{FedDF}
    \textbf{Input}: Private datasets $\left\{X_k, Y_k\right\}_{k=1}^K$, unlabeled public dataset $\{X_0\}$, local models $\left\{\theta_k\right\}_{k=1}^K$, global model $\theta$, number of communications $T$, the softmax function $F_s$.
    \begin{algorithmic}[1]
    \State Initialize the training iteration index by $t$ = 1.
    \For {$t \leq T$}
    \State Each client minimize the loss on $\{X_k, Y_k\}$ and sends the model parameters $\theta_k^t$ to the server.
    \State The server aggregates the local models to generate the student model:
    \State \indent $\hat{\theta}^t = \frac{\sum_{k=1}^K}{K}\theta_k^t$
    \State The server computes the logit vectors using local models:
    \State \indent $Y^t_k = 
    \{f\left(x_i, \theta_k^t\right) | x_i \in X_0\}$
    \State The server aggregate logit vectors:
    \State \indent $\hat{Y}^t = \frac{\sum_{k=1}^K Y_k^t}{K}$
    \State The server trains the global model on $\{X_0, F_s\left(\hat{Y}^t\right)\}$ and produces the latest global model $\theta^t$.
    %\State Each client receives $L^t$ and trains $w_k$ on $\left(D_0, L^t\right)$
    \State The server distributes the global model $\theta^t$ as the latest model
    \EndFor
    \end{algorithmic}
    \label{alg:feddf}
\end{algorithm}

In our experiments, we investigate three distinct poisoning attack schemes: naive poisoning, label flipping poisoning, and our proposed logit poisoning attack. In the case of the label-flipping attack, the adversary alters the labels of the local training data in a specific manner to poison it. This label-flipping strategy is uniformly executed across all the malicious parties under the adversary's control. Subsequently, these malicious parties employ the poisoned data to train their local models and share the corresponding updates with the central server. For the naive poisoning attack, the adversary is aware of the distribution of benign updates. The adversary then can introduce a malicious update by adding a significantly large vector to the benign distribution. It is essential to highlight that the naive poisoning attack necessitates the construction of a large vector, whereas, in our poisoning scheme, the logit vector must be scaled. To ensure a fair comparison, we constrain the magnitude of the large vector in the naive attack to align with the value range of the scaled vector in our logit vectors poisoning attack. This alignment is significant because our scheme only requires a relatively modest scaling factor, rendering the impact of the naive poisoning attack comparatively minimal. 

We adopt a 3-layer Convolutional Neural Network \cite{lecun1998gradient} for image classification.
The first convolutional layer accepts an input shape of (28,28), utilizing filters with a kernel size of (3,3) and a stride of 1. A dropout rate \cite{hinton2012improving} of 0.2 is applied following this layer. The second convolutional layer employs filters with a kernel size of (2,2) and a stride of 2, again followed by a dropout rate of 0.2. The third convolutional layer uses filters with a kernel size of (3,3), a stride of 2, and the same dropout rate of 0.2. The Dense output layer consists of ten units and applies L2 regularization with a factor of 1e-3. The model is compiled using the Adam optimizer \cite{kingma2014adam}.

For the FedMD and DS-FL schemes, the configuration is as follows: the number of local training epochs is 1, the number of transfer epochs is 3, the learning rate for local training is 2e-6, and the transfer learning rate is 1e-5. For DS-FL specifically, the number of epochs for local updates and distillation is 2, while the number of server-side distillation epochs is 1.
For the FedDF scheme, the client-side epoch count is 2, the number of server-side epochs is 3, the client-side learning rate is 5e-6, and the server-side learning rate is 1e-5. In all schemes, the number of communication rounds is set to 10. For the FedMD and DS-FL schemes, the loss function utilized in knowledge transfer is the MAE loss. Conversely, for the FedDF scheme, the KL divergence loss is employed in the knowledge transfer process.

For the FedMD and DS-FL schemes, the results are represented by the average of the accuracy and loss of all local models on the test dataset. In contrast, for the FedDF scheme, the results are determined by the accuracy and loss of the global model on the test dataset.
% \begin{equation}
%     \theta_m=\frac{\sum_{i=1}^n \theta_i}{(1-\epsilon) n}+\theta^{\prime}
% \end{equation}

\subsection{Logit Poisoning Attack}
The experiments detailed here are conducted to validate the efficacy of our logit poisoning attack in compromising distillation-based federated learning. Each experiment is performed multiple times to acquire a robust set of trial findings, from which an average is calculated. The comparative results demonstrate that our approach exhibits superior performance to conventional poisoning schemes under equivalent value magnitudes. To more effectively illustrate the efficacy of our approach, we initially examined the influence of varying attacker proportions on model accuracy across three different distillation-based federated learning schemes. Subsequently, we fixed the proportion of attackers at 30\% to evaluate the comparative performance of our attack scheme against both naive poisoning and label-flipping attacks under this condition.

\begin{figure}
    \centering
        \subfloat[]{\includegraphics[width=0.24\textwidth]{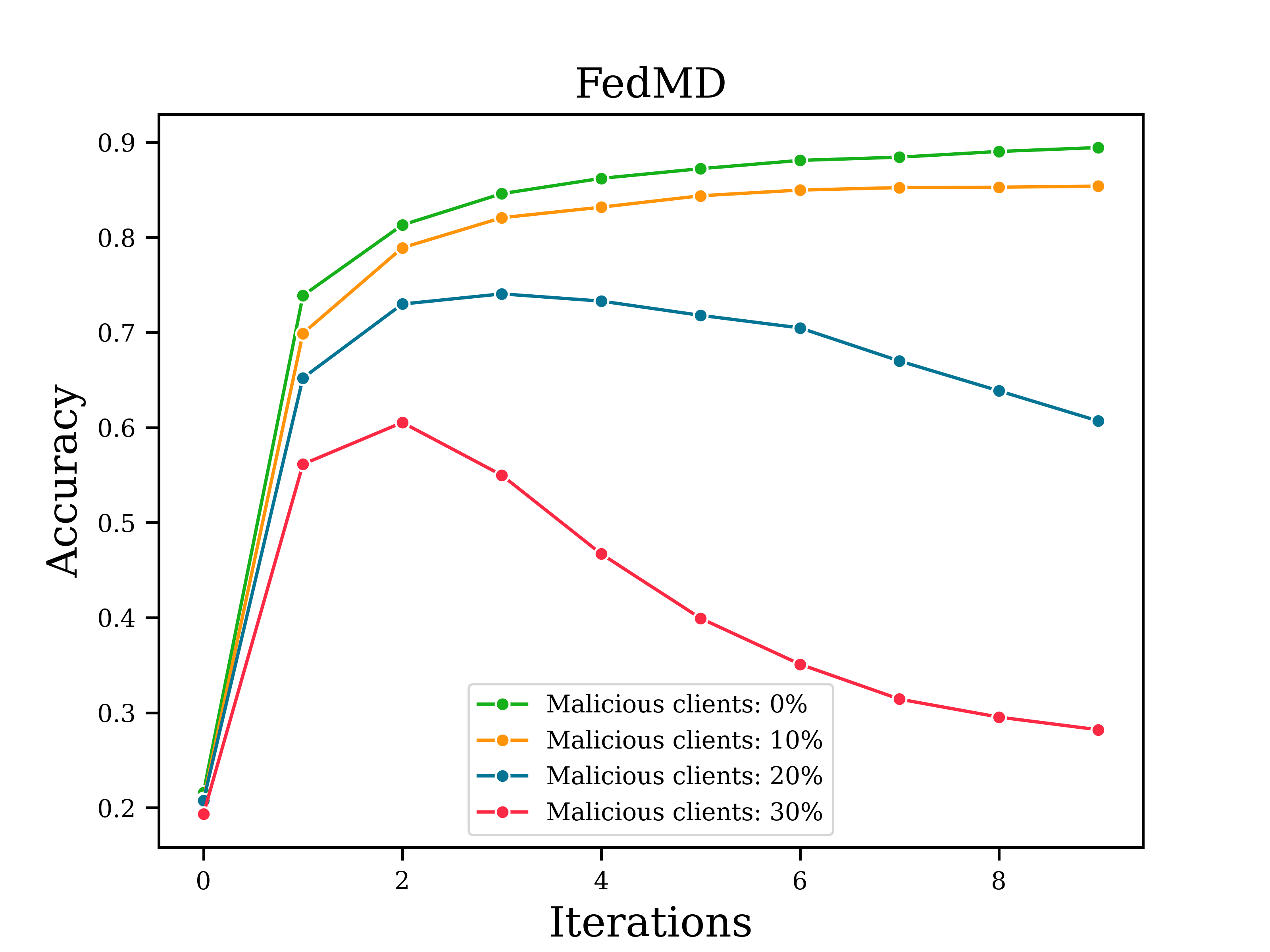}%
        \label{subfig:1}}
        \subfloat[]{\includegraphics[width=0.24\textwidth]{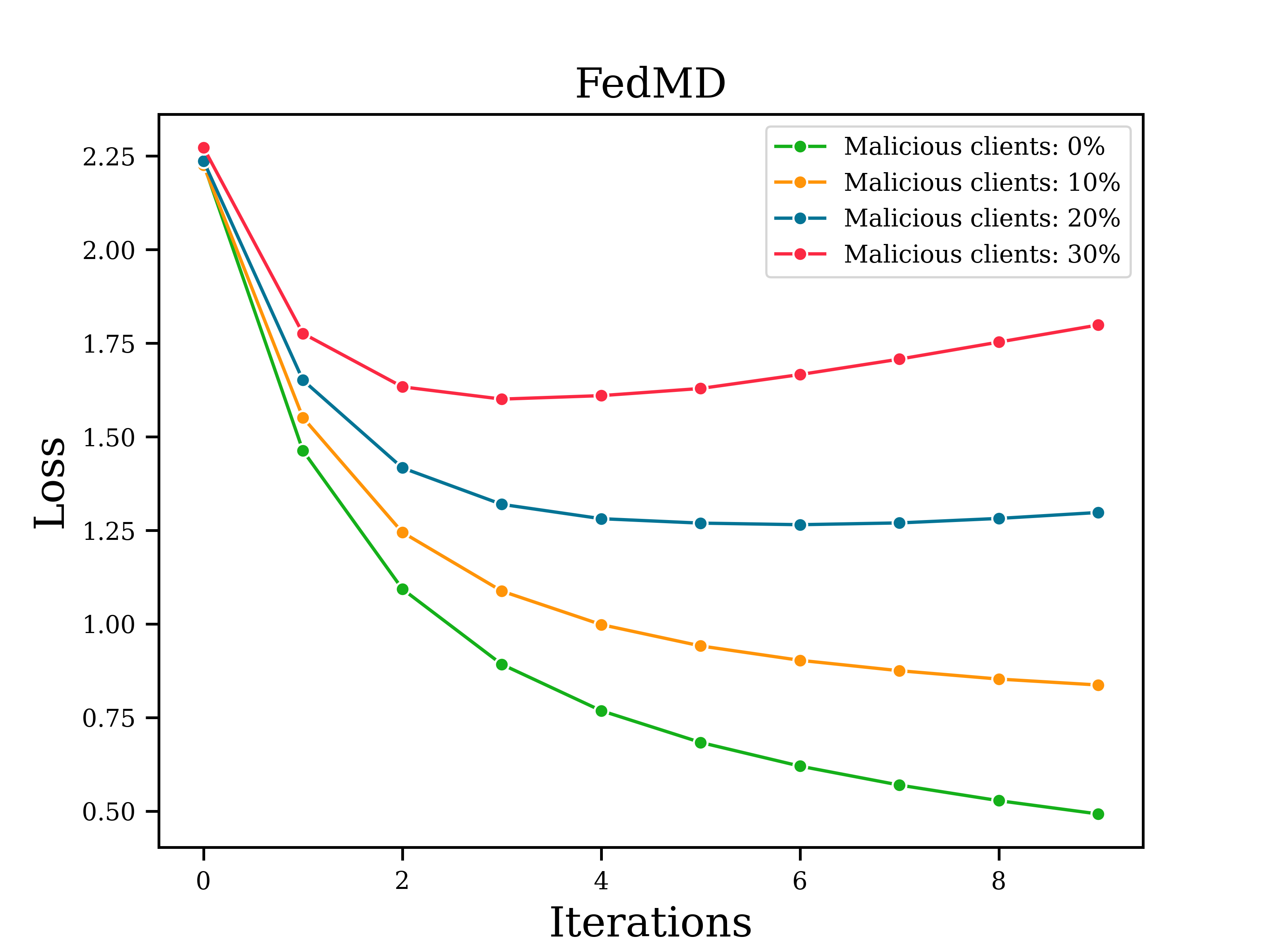}%
        \label{subfig:2}}
        \quad
        \subfloat[]{\includegraphics[width=0.24\textwidth]{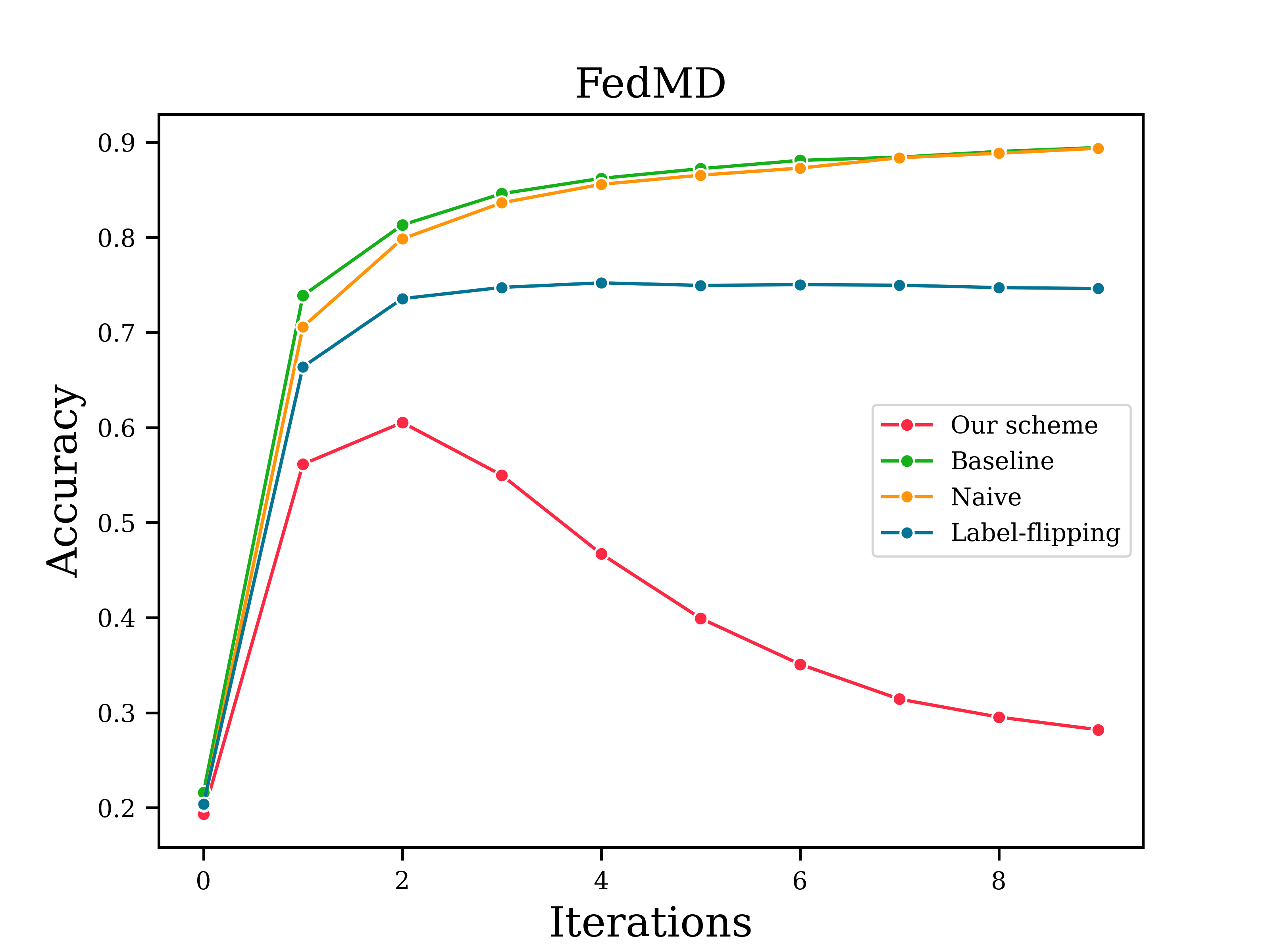}%
        \label{subfig:3}}
        \subfloat[]{\includegraphics[width=0.24\textwidth]{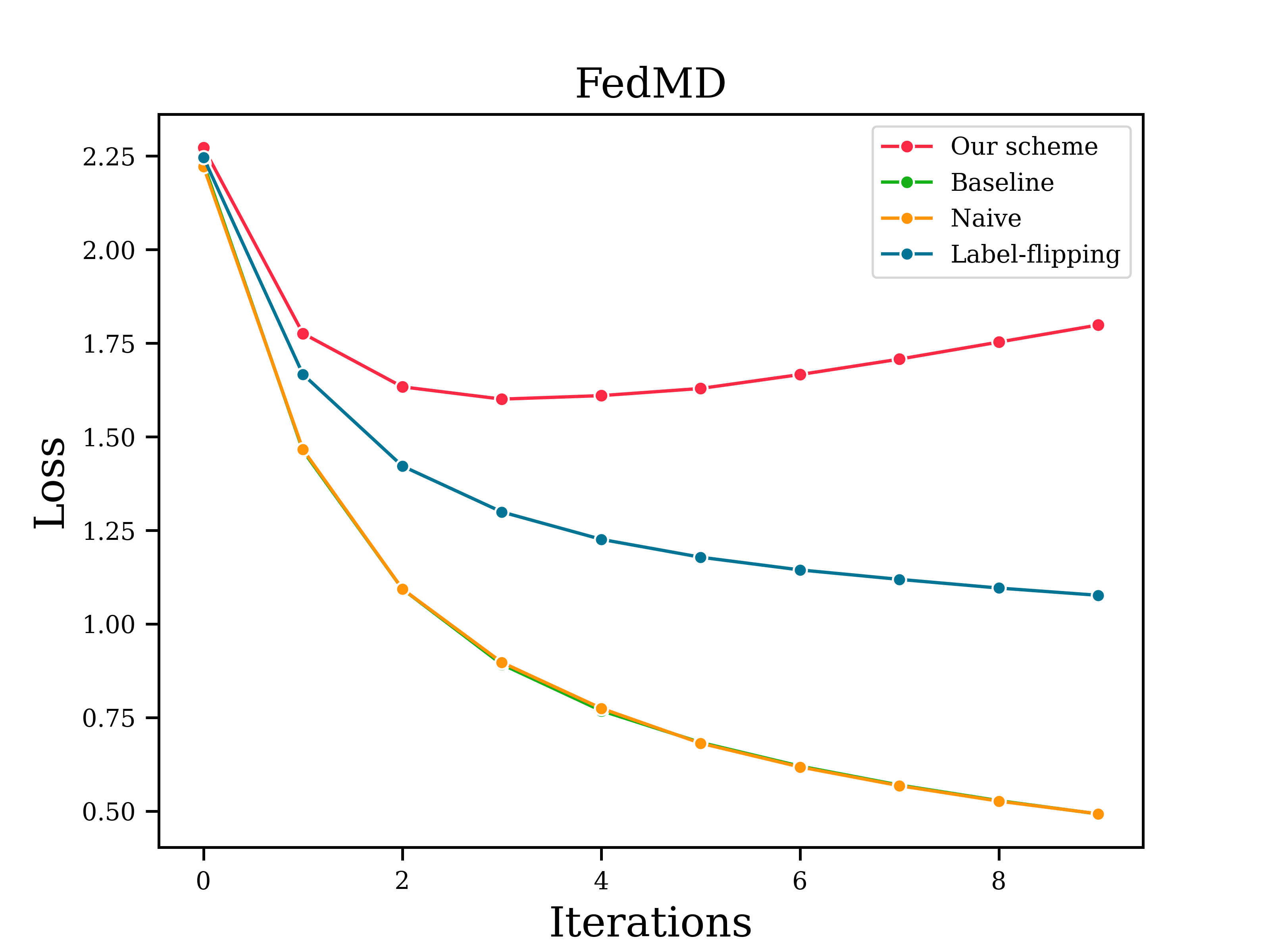}%
        \label{subfig:4}}   
    \caption{Poisoning effect in FedMD}
    \label{fig:attfedmd}
\end{figure}

\begin{figure}
    \centering
        \subfloat[]{\includegraphics[width=0.24\textwidth]{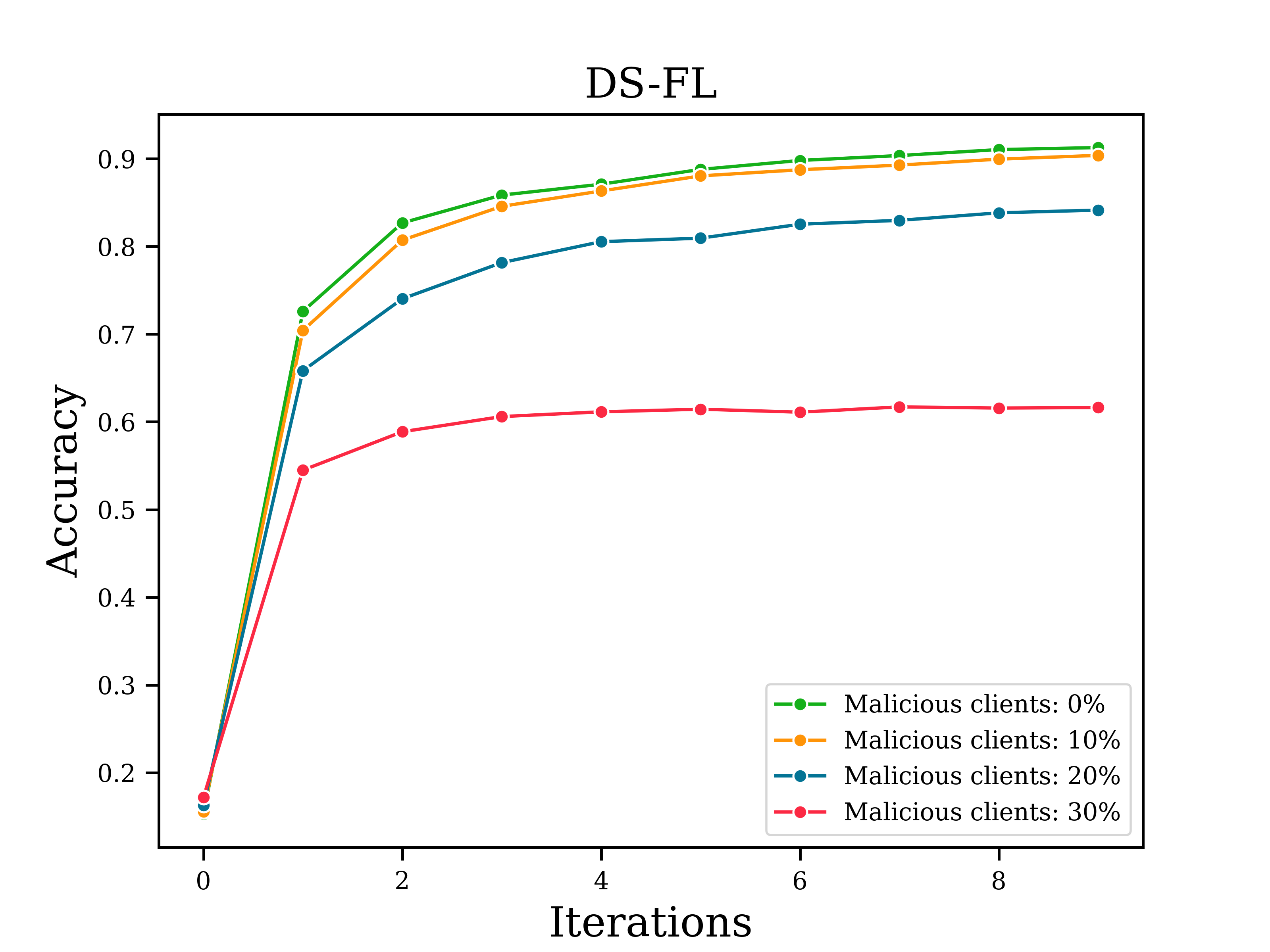}%
        \label{subfig:5}}
        \subfloat[]{\includegraphics[width=0.24\textwidth]{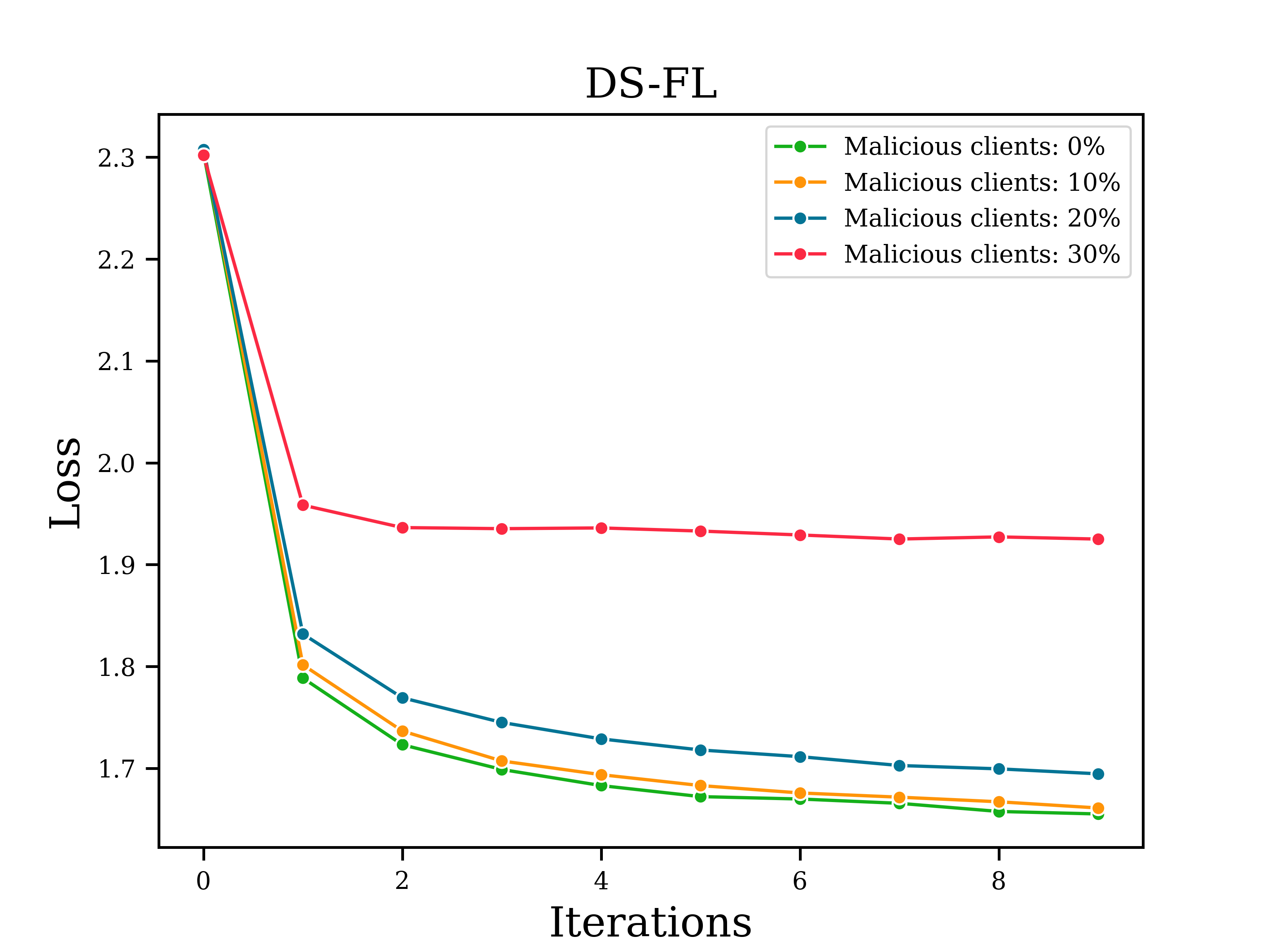}%
        \label{subfig:6}}
        \quad
        \subfloat[]{\includegraphics[width=0.24\textwidth]{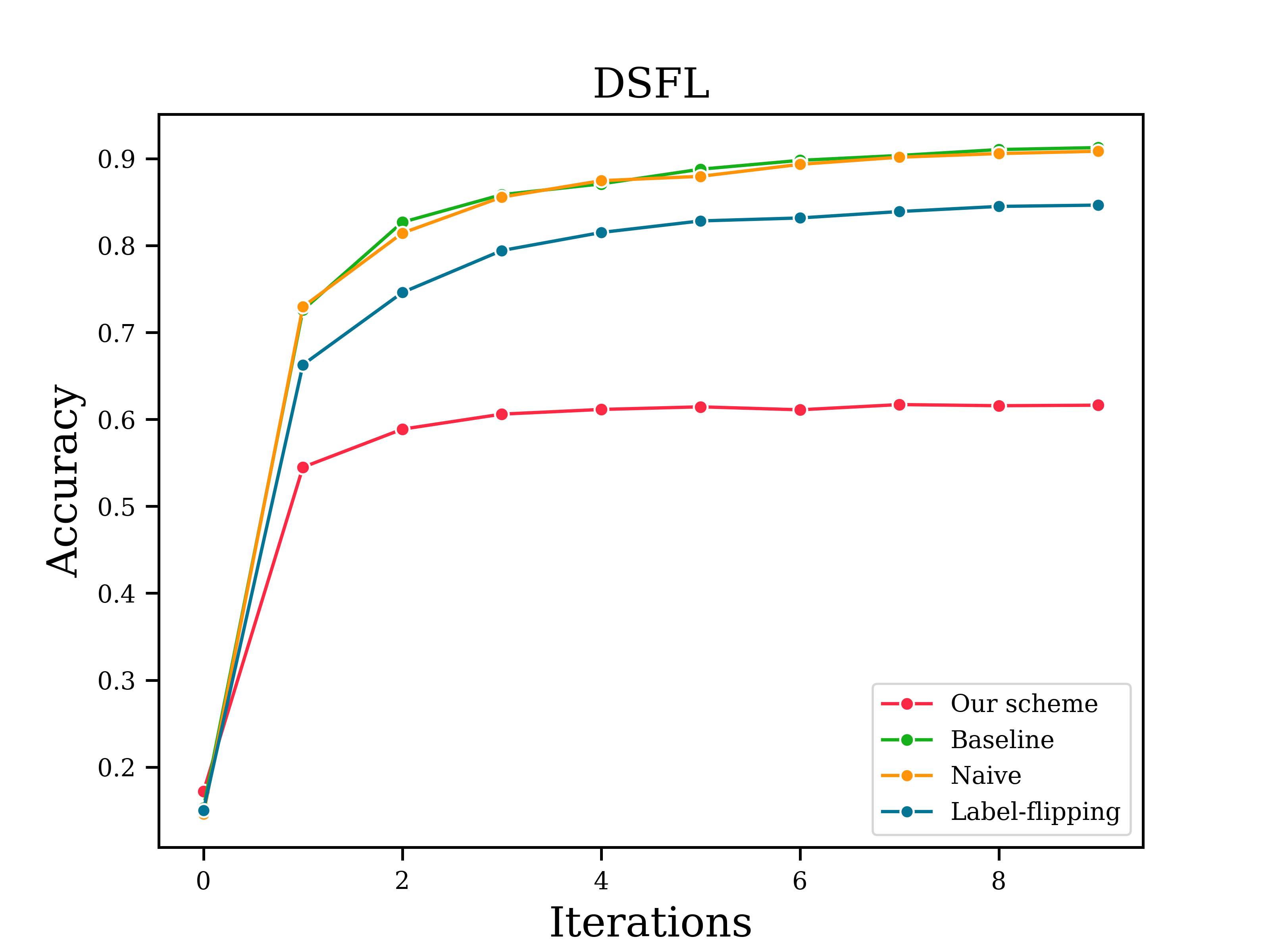}%
        \label{subfig:7}}
        \subfloat[]{\includegraphics[width=0.24\textwidth]{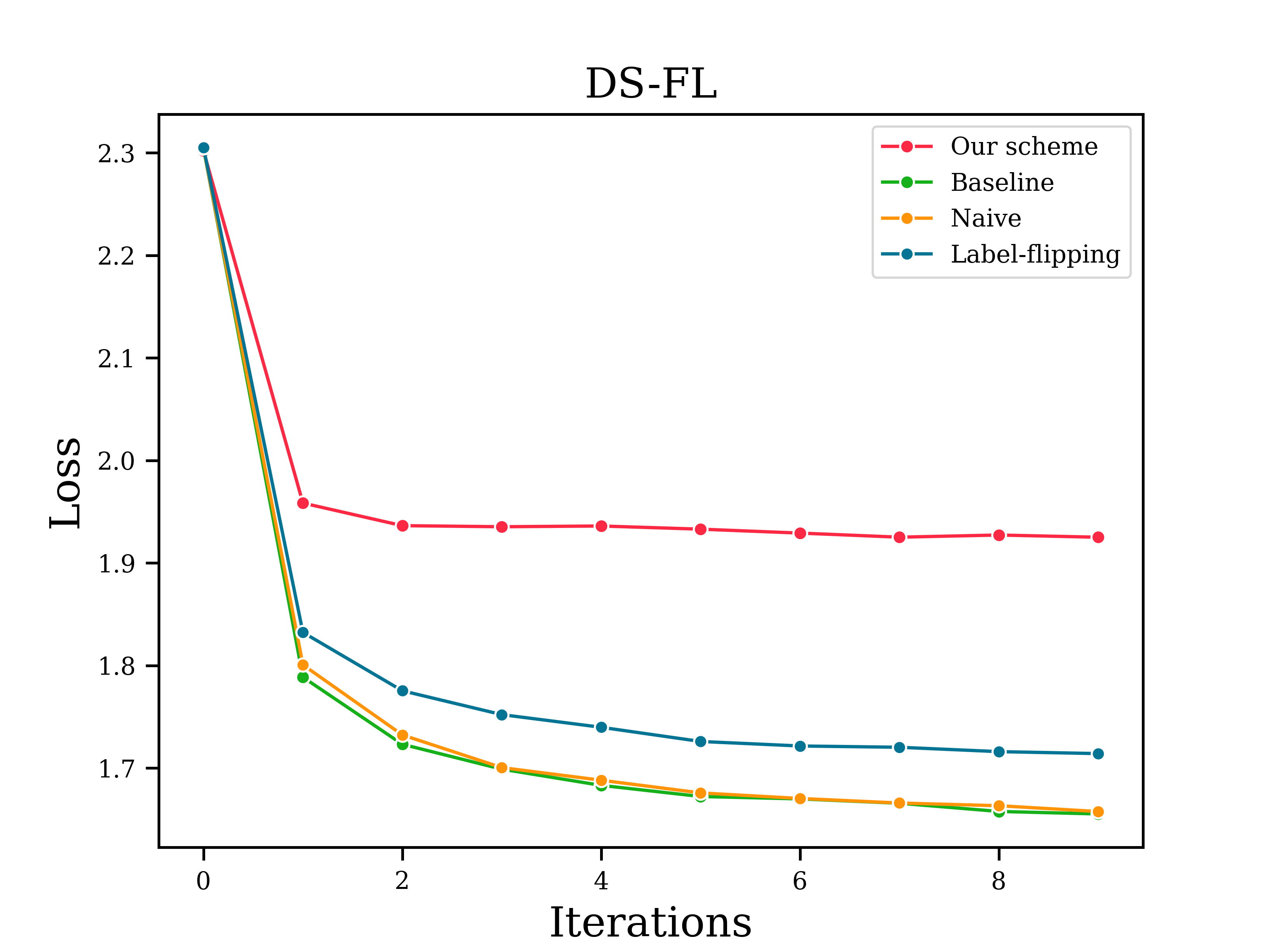}%
        \label{subfig:8}}  
    \caption{Poisoning effect in DS-FL}
    \label{fig:attdsfl}
\end{figure}

\begin{figure}
    \centering
        \subfloat[]{\includegraphics[width=0.24\textwidth]{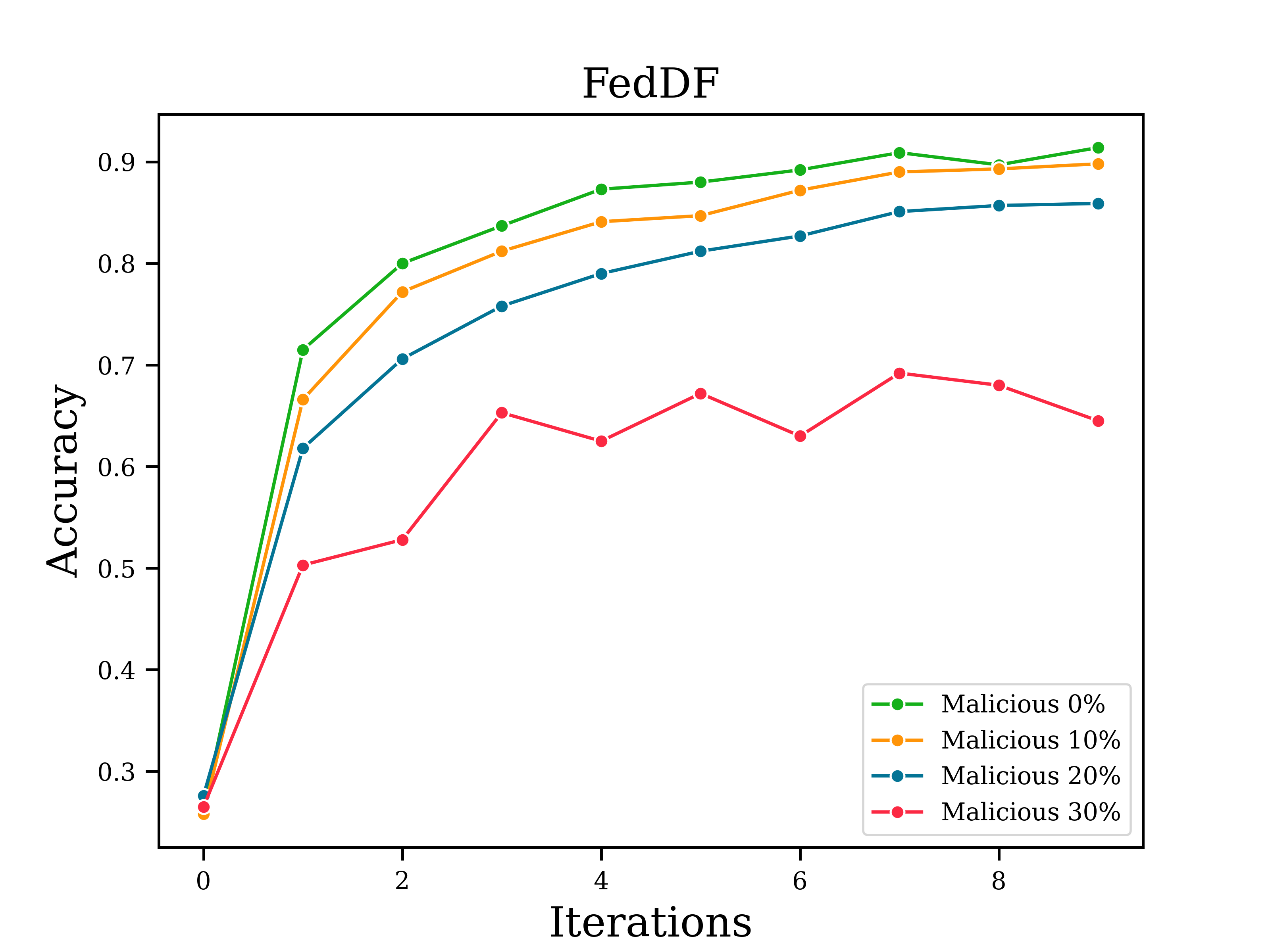}%
        \label{subfig:9}}
        \subfloat[]{\includegraphics[width=0.24\textwidth]{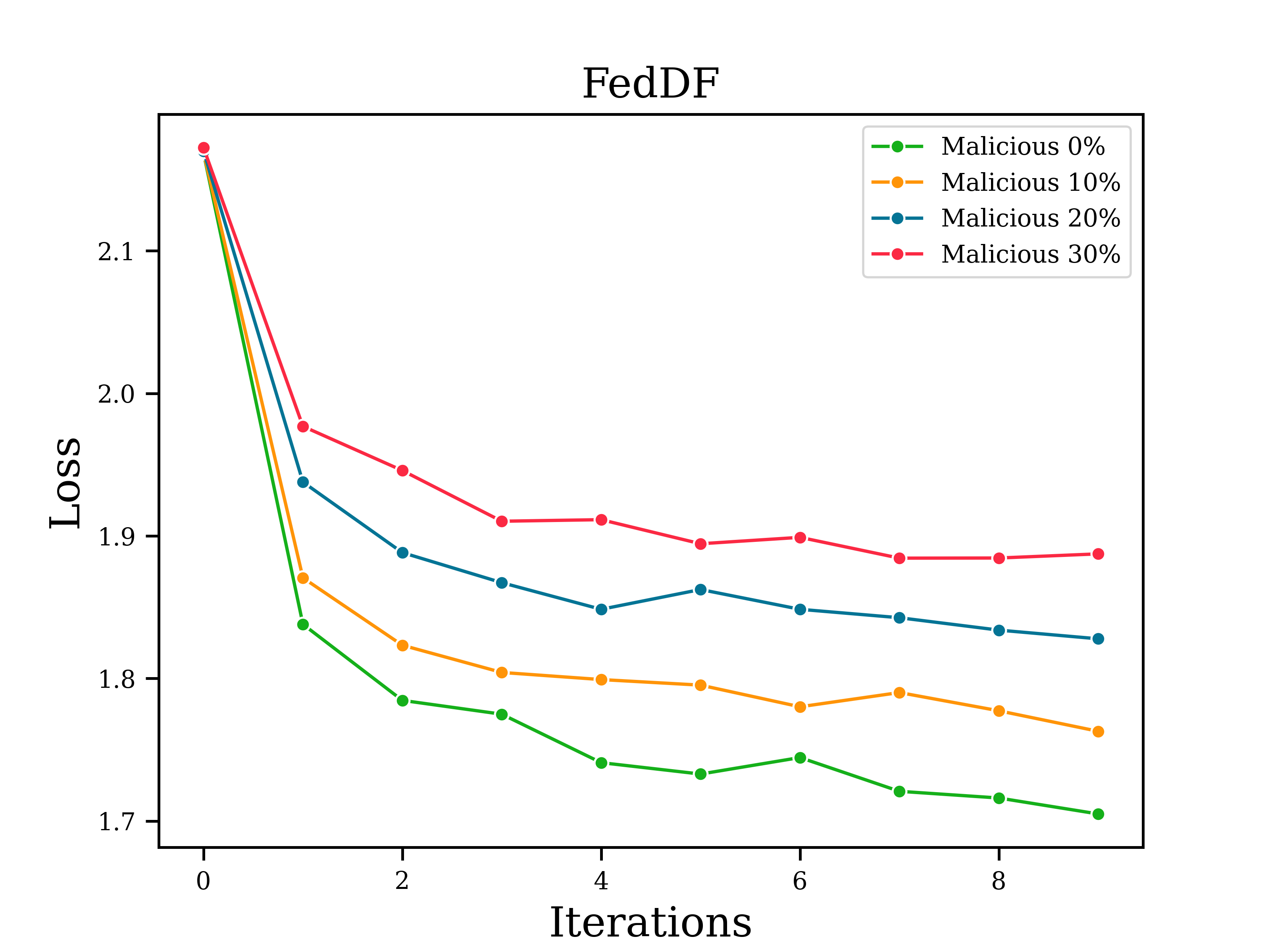}%
        \label{subfig:10}}
        \quad
        \subfloat[]{\includegraphics[width=0.24\textwidth]{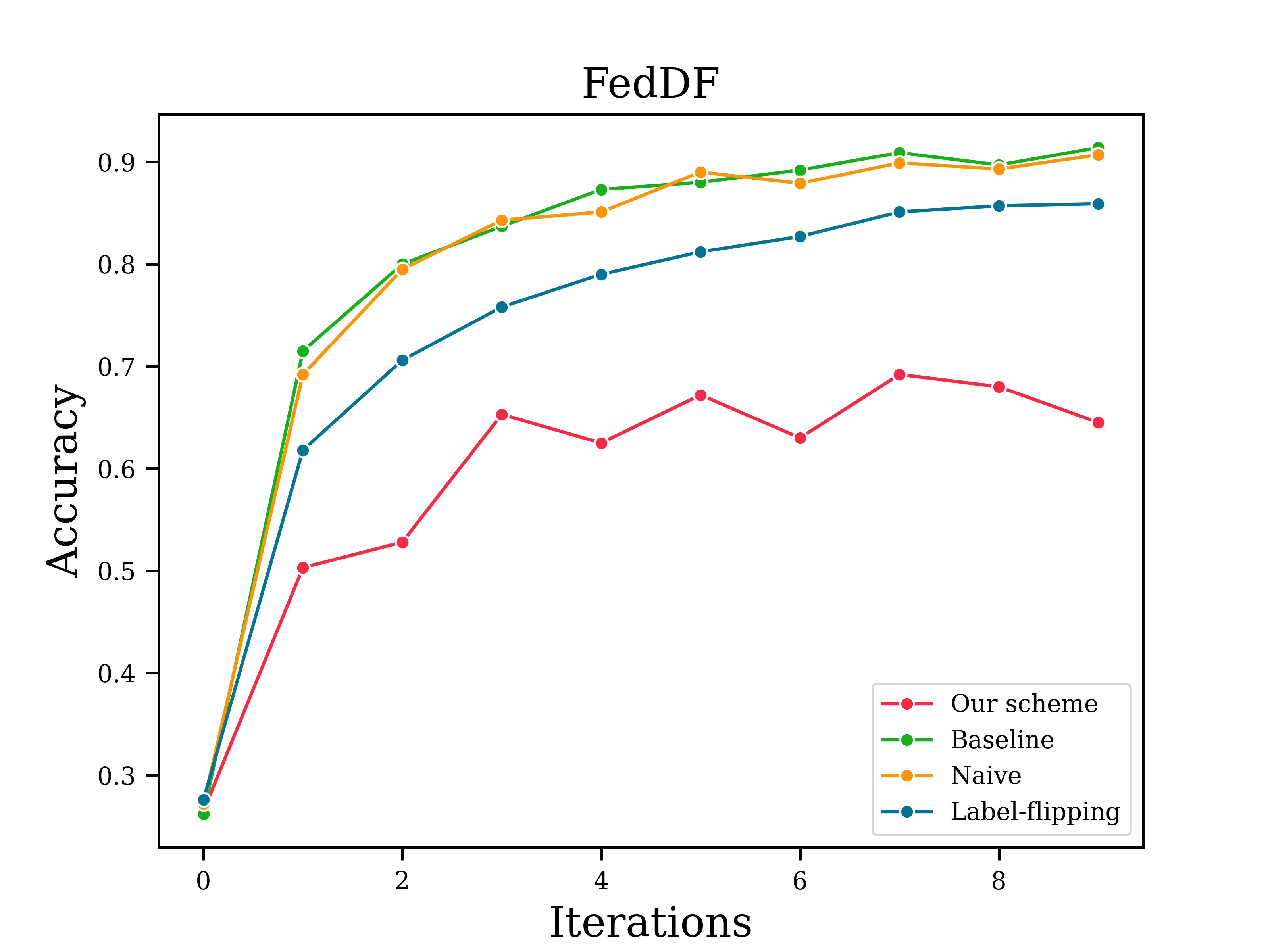}%
        \label{subfig:11}}
        \subfloat[]{\includegraphics[width=0.24\textwidth]{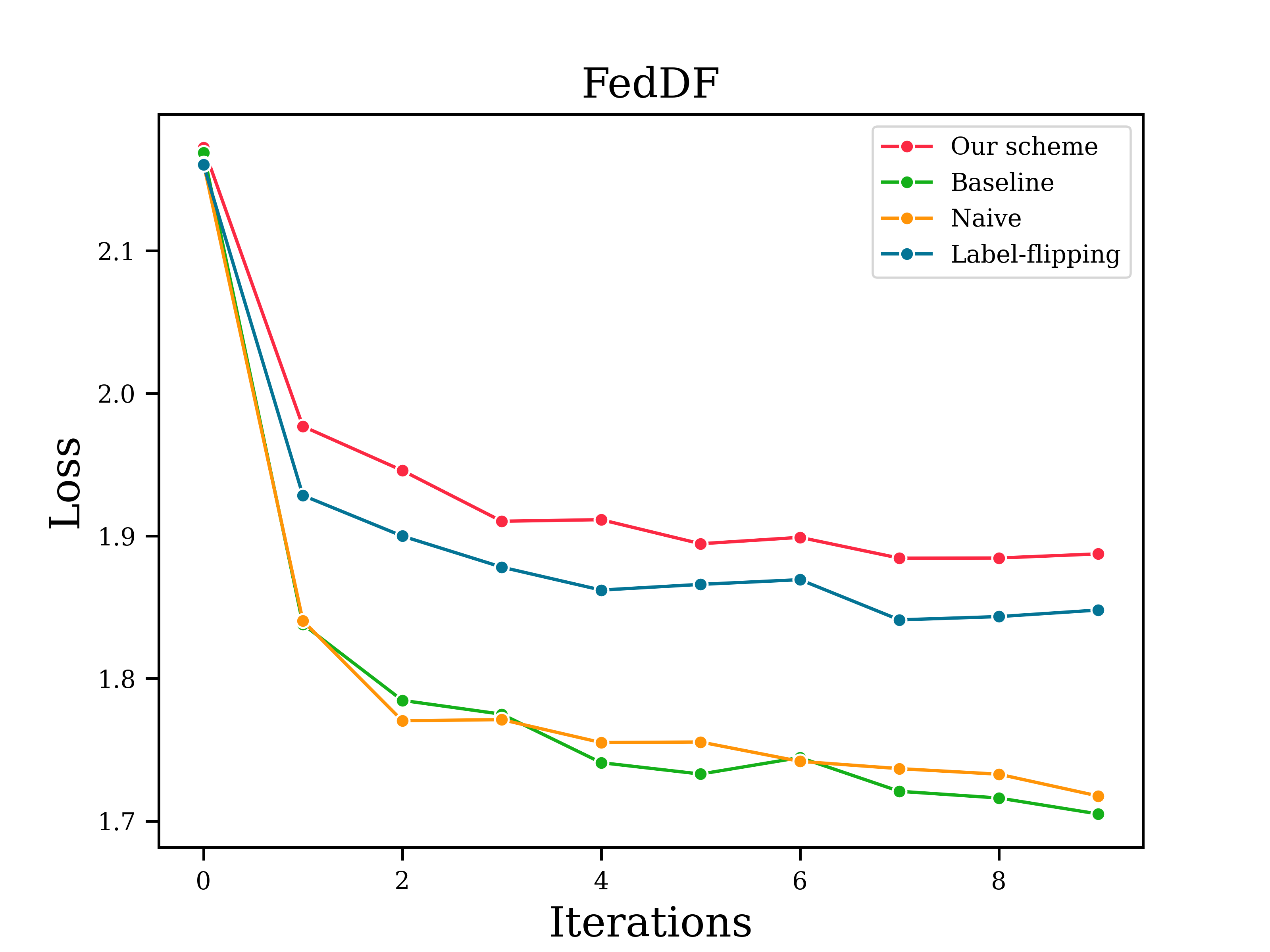}%
        \label{subfig:12}}
    \caption{Poisoning effect in FedDF}
    \label{fig:attfeddf}
\end{figure}

The outcomes of the experiments conducted under these three distinct settings are depicted in Figures \ref{fig:attfedmd}, \ref{fig:attdsfl}, and \ref{fig:attfeddf}, respectively.
As illustrated in the corresponding figures, it is evident that, for the three distinct federated learning schemes, our poisoning attack scheme achieves a more pronounced poisoning effect when the proportion of attackers is set at 30\%. Under these conditions, the accuracy of the three federated learning schemes on the test dataset substantially declines, while the loss concomitantly experiences a significant increase.
For the DS-FL scheme, its inherent entropy reduction algorithm mitigates the impact when the attackers constitute 10\% and 20\% of the total. However, our scheme proves successful when the attackers account for 30\%. For the FedDF method, it is observed that the resistance to any attack scheme is notably robust. Specifically, all poisoning schemes inflict less damage to this approach. This resilience can be attributed to the scheme's utilization of the softmax function on the logit vector and the execution of the knowledge transfer process on the server side. Nevertheless, our poisoning scheme is still able to achieve a satisfactory poisoning effect, leading to a reduction in the accuracy of the global model. Furthermore, the efficacy of our approach surpasses that of other poisoning schemes. Another noteworthy observation is the substantially reduced poisoning effects when implementing naive poisoning attacks compared to results in analogous studies \cite{cheng2021fedgems}. This discrepancy arises from our deliberate constraint on the vectors employed in naive poisoning attacks, aligning them within a similar value range to our poisoning schemes. This methodological decision ensures a fair and consistent comparison across the different approaches. Within this controlled value range, our scheme demonstrates a more effective poisoning effect, further emphasizing the superiority and efficacy of our proposed method.

\subsection{Defense Scheme for Distillation-based Federated Learning}
Having illustrated the efficacy of the logit poisoning attack on distillation-based federated learning, we proceed to evaluate the performance of our proposed defense framework in this subsection.

We enhance the three implemented distillation-based federated learning schemes with our proposed robust aggregation algorithm and subject them to evaluation under three distinct poisoning methods. The sole distinction lies in the substitution of the aggregation algorithm within the three FL schemes with our newly proposed robust aggregation algorithm. By integrating the robust aggregation algorithm, we intend to forge a solution that exhibits enhanced security and resilience against poisoning attacks. We utilize the running results obtained with zero percent attackers as the baseline for comparison. The efficacy of the defense mechanism can be gauged by how closely the accuracy and loss metrics when under attack, align with those of the baseline group. In this instance, the settings and parameters for the attackers remain consistent with those delineated in the previous subsection. The percentage of attackers is set to 30\%.

\begin{figure*}
    \centering
        \subfloat[]{\includegraphics[width=0.3\textwidth]{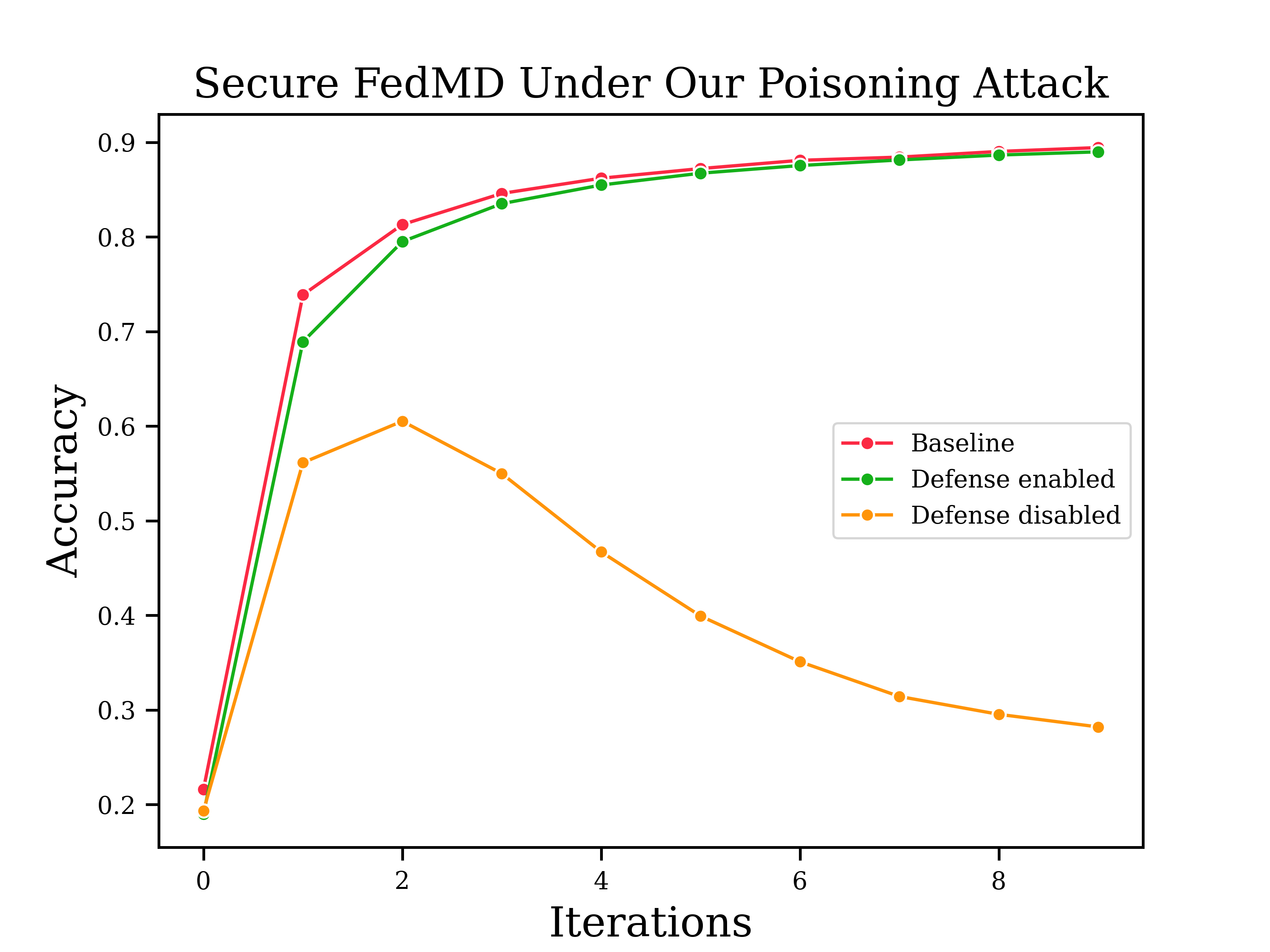}%
        \label{subfig:13}}
        \subfloat[]{\includegraphics[width=0.3\textwidth]{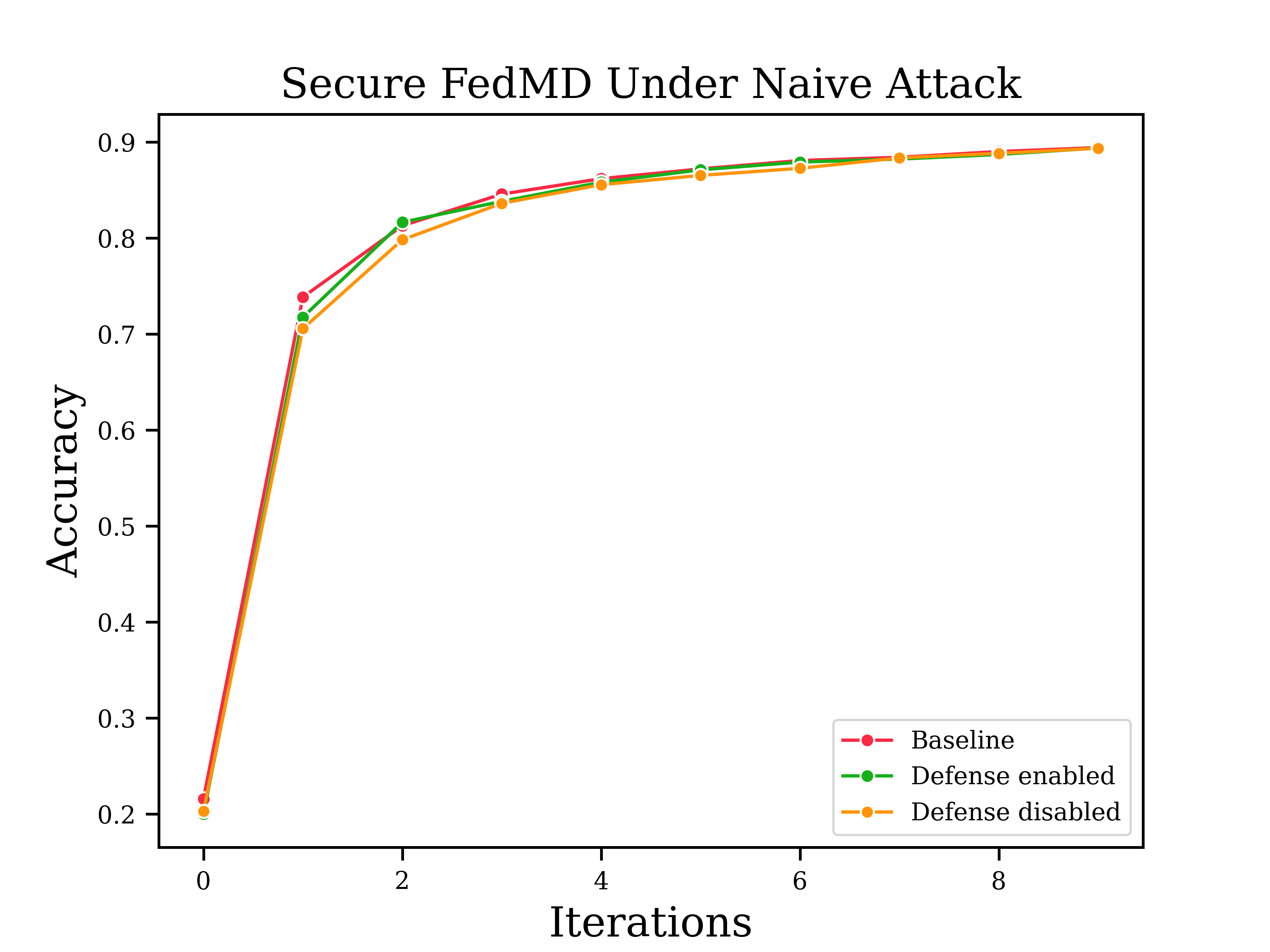}%
        \label{subfig:14}}
        \subfloat[]{\includegraphics[width=0.3\textwidth]{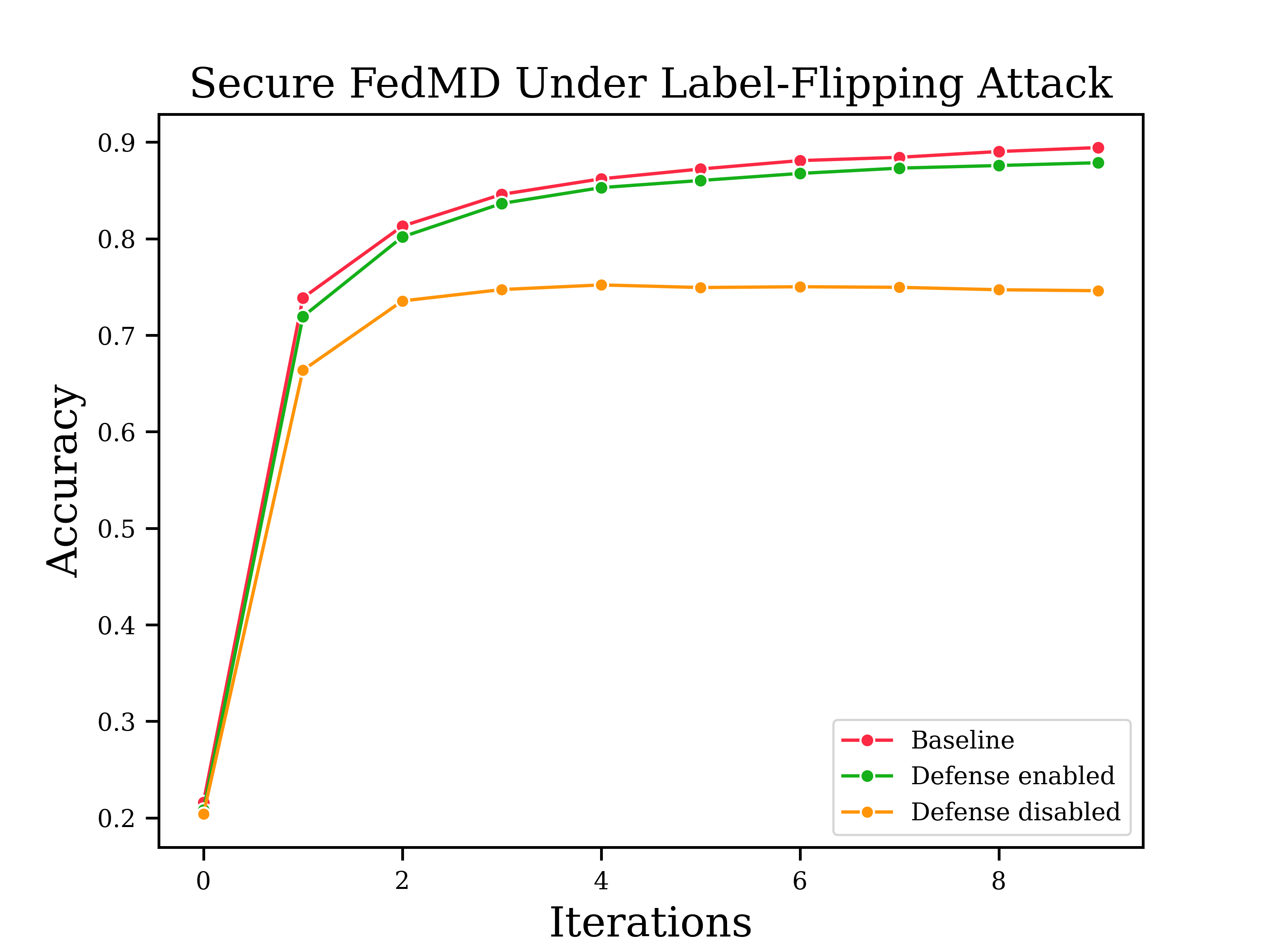}%
        \label{subfig:15}}
        \quad
        \subfloat[]{\includegraphics[width=0.3\textwidth]{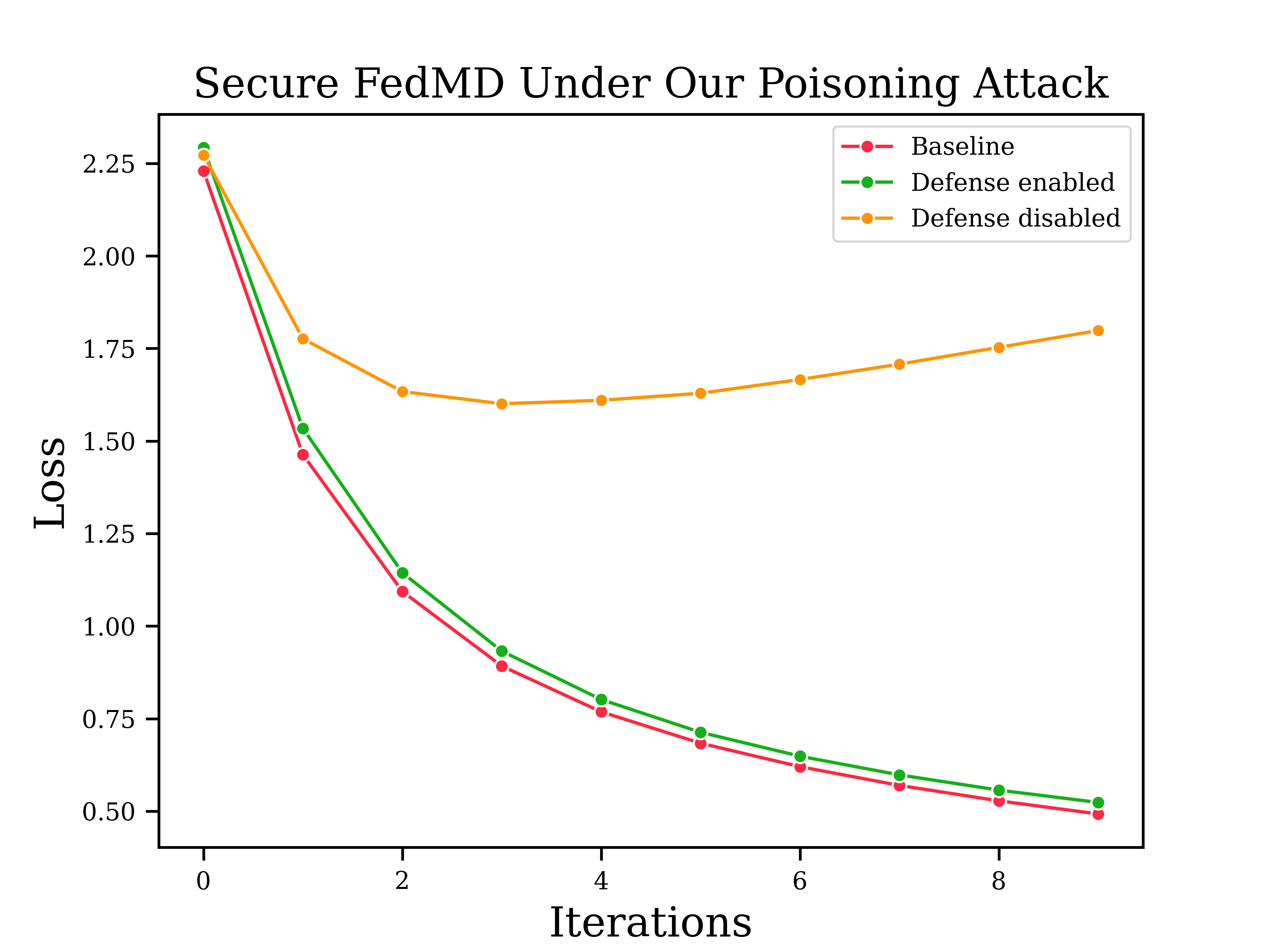}%
        \label{subfig:16}}
        \subfloat[]{\includegraphics[width=0.3\textwidth]{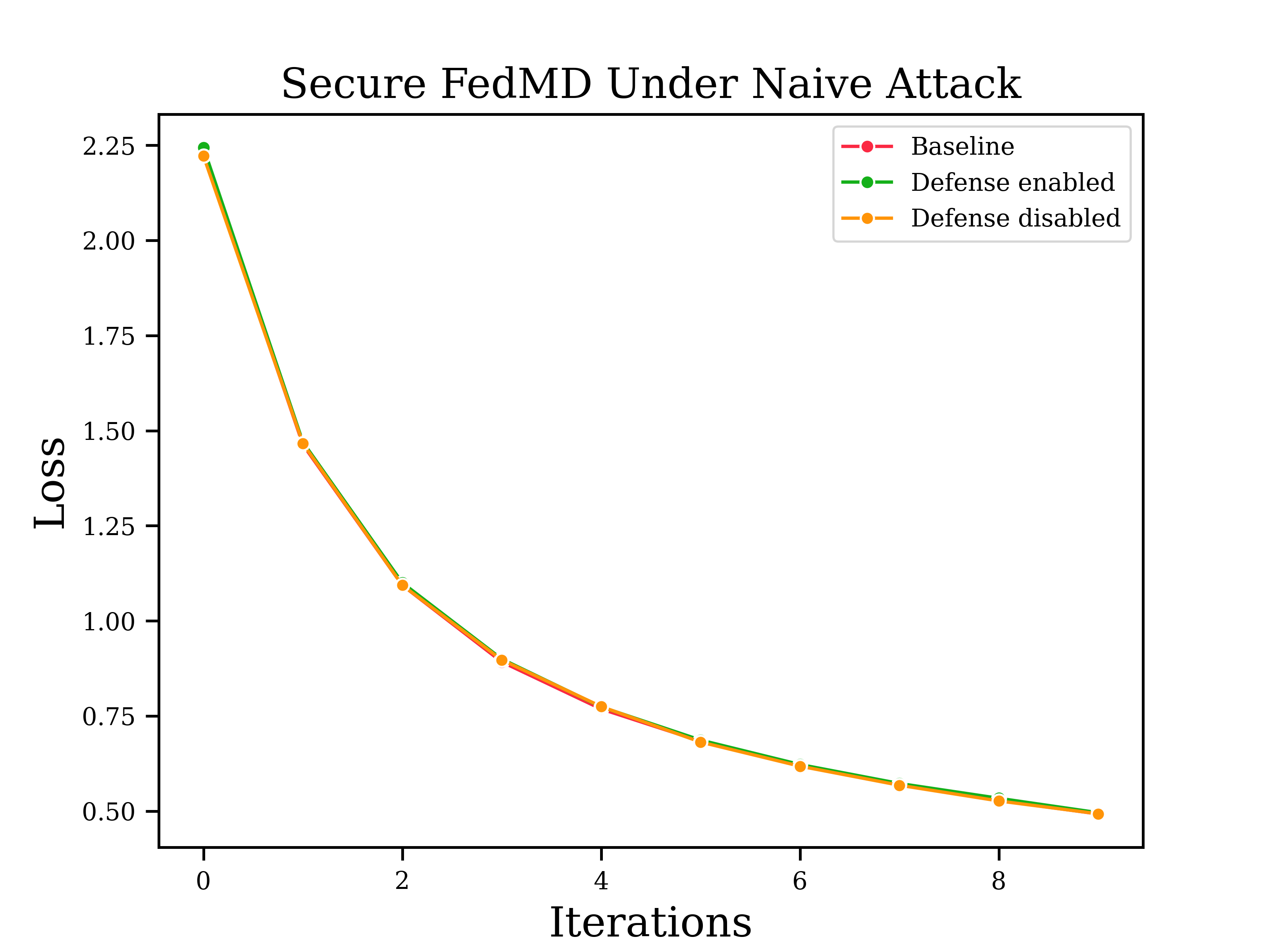}%
        \label{subfig:17}}   
        \subfloat[]{\includegraphics[width=0.3\textwidth]{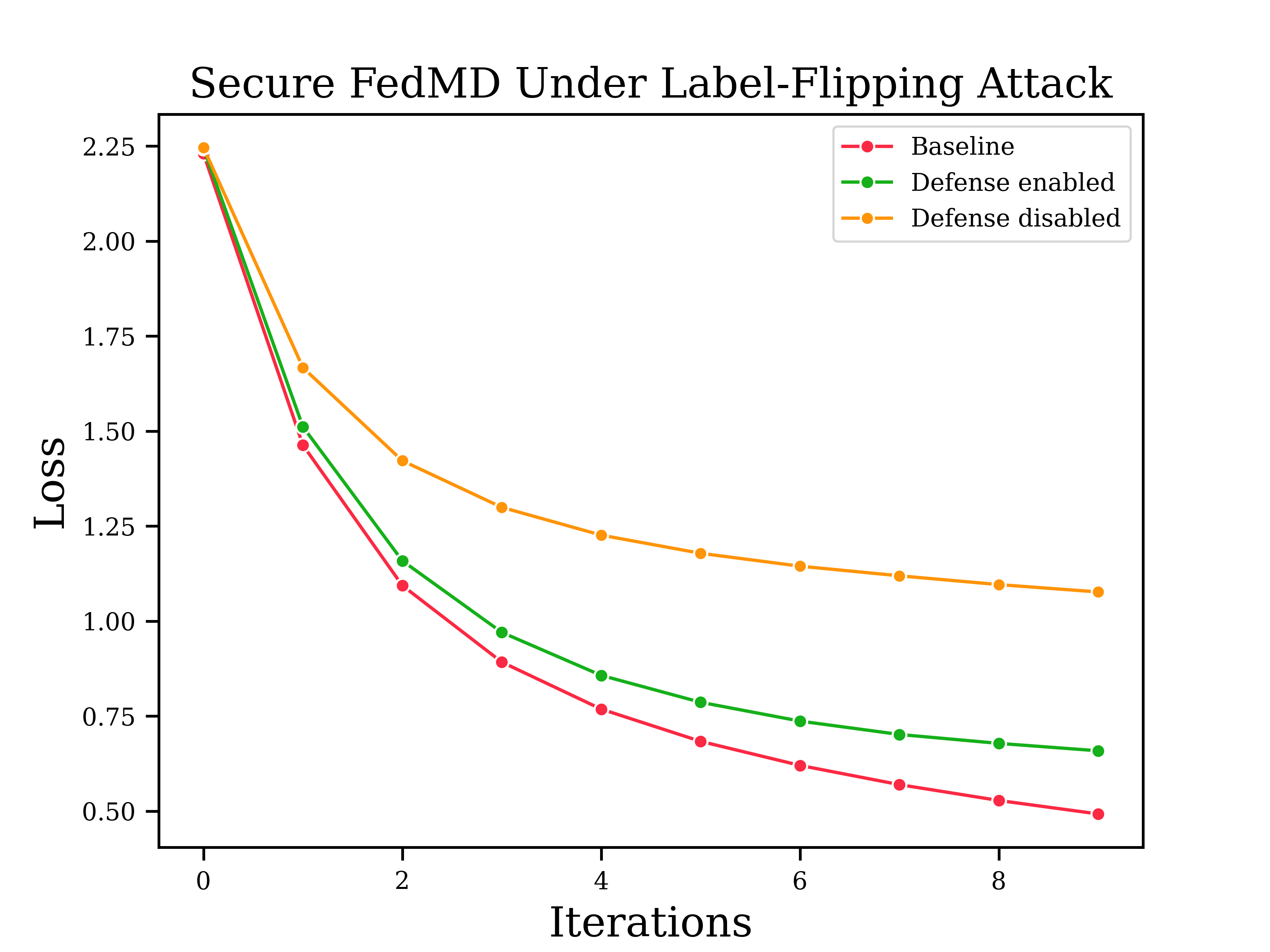}%
        \label{subfig:18}}           
        \caption{FedMD enhanced by proposed robust aggregation algorithm}
        \label{fig:defedmd}
\end{figure*}

\begin{figure*}
    \centering
        \subfloat[]{\includegraphics[width=0.3\textwidth]{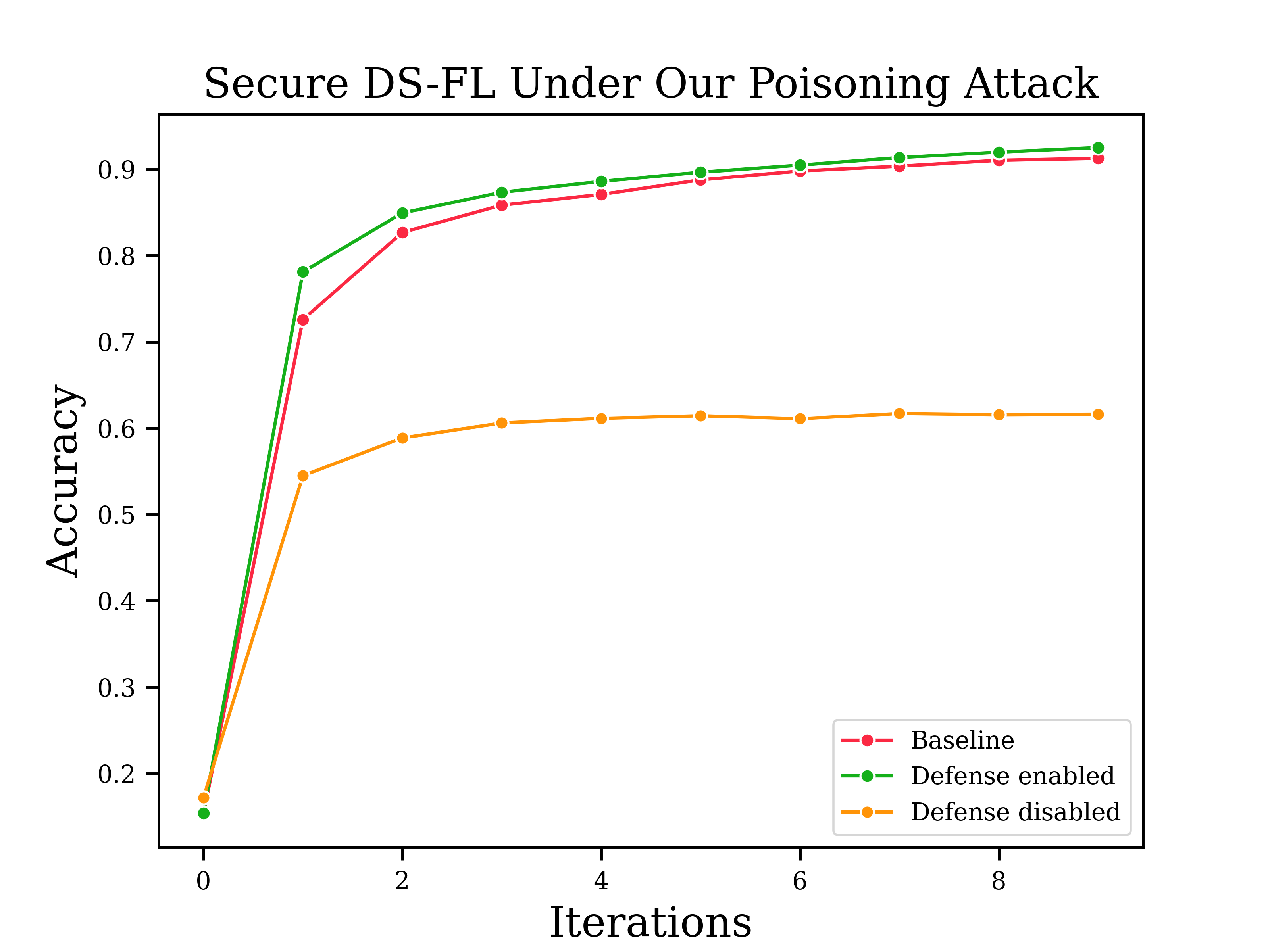}%
        \label{subfig:19}}
        \subfloat[]{\includegraphics[width=0.3\textwidth]{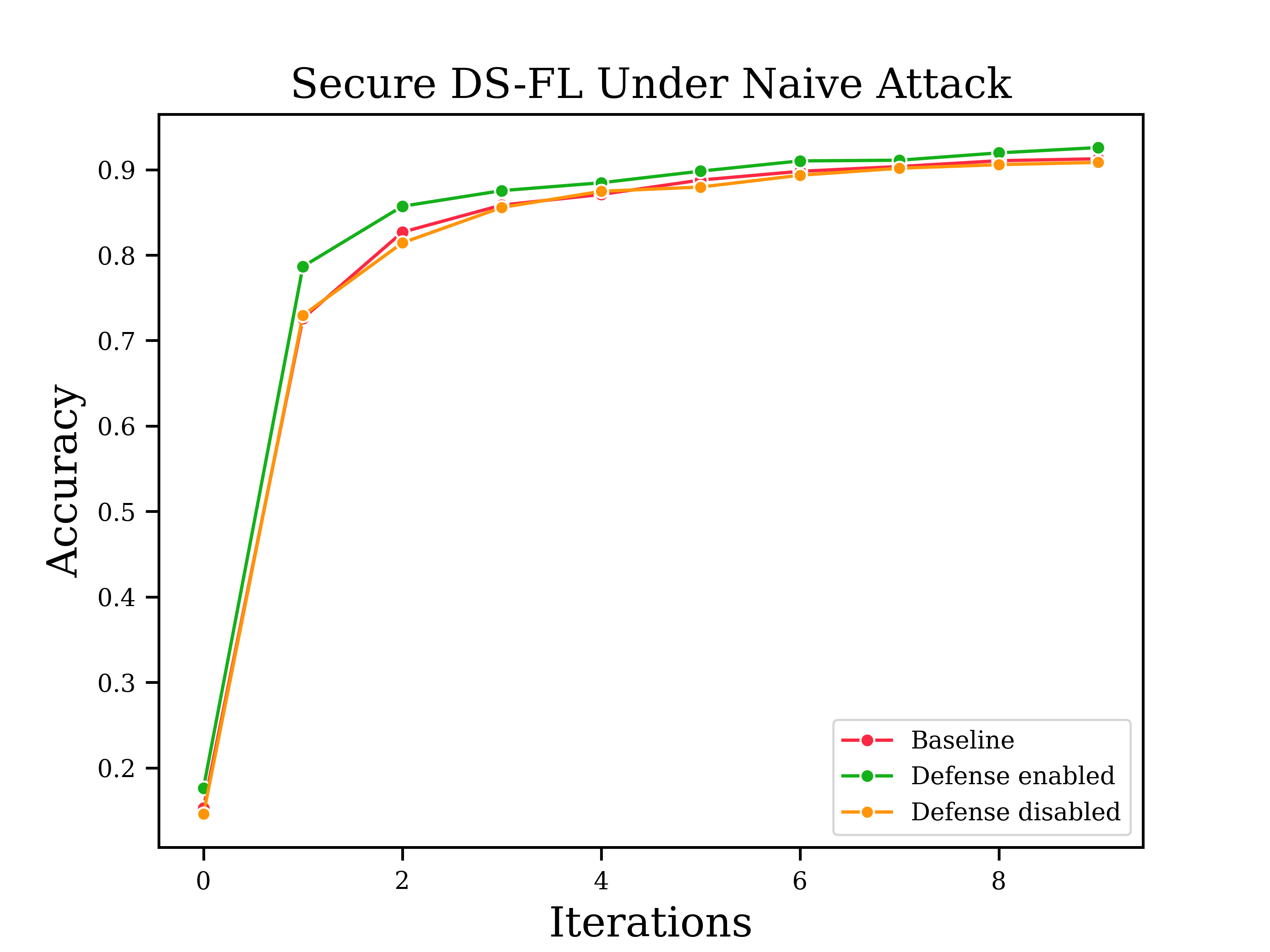}%
        \label{subfig:20}}
        \subfloat[]{\includegraphics[width=0.3\textwidth]{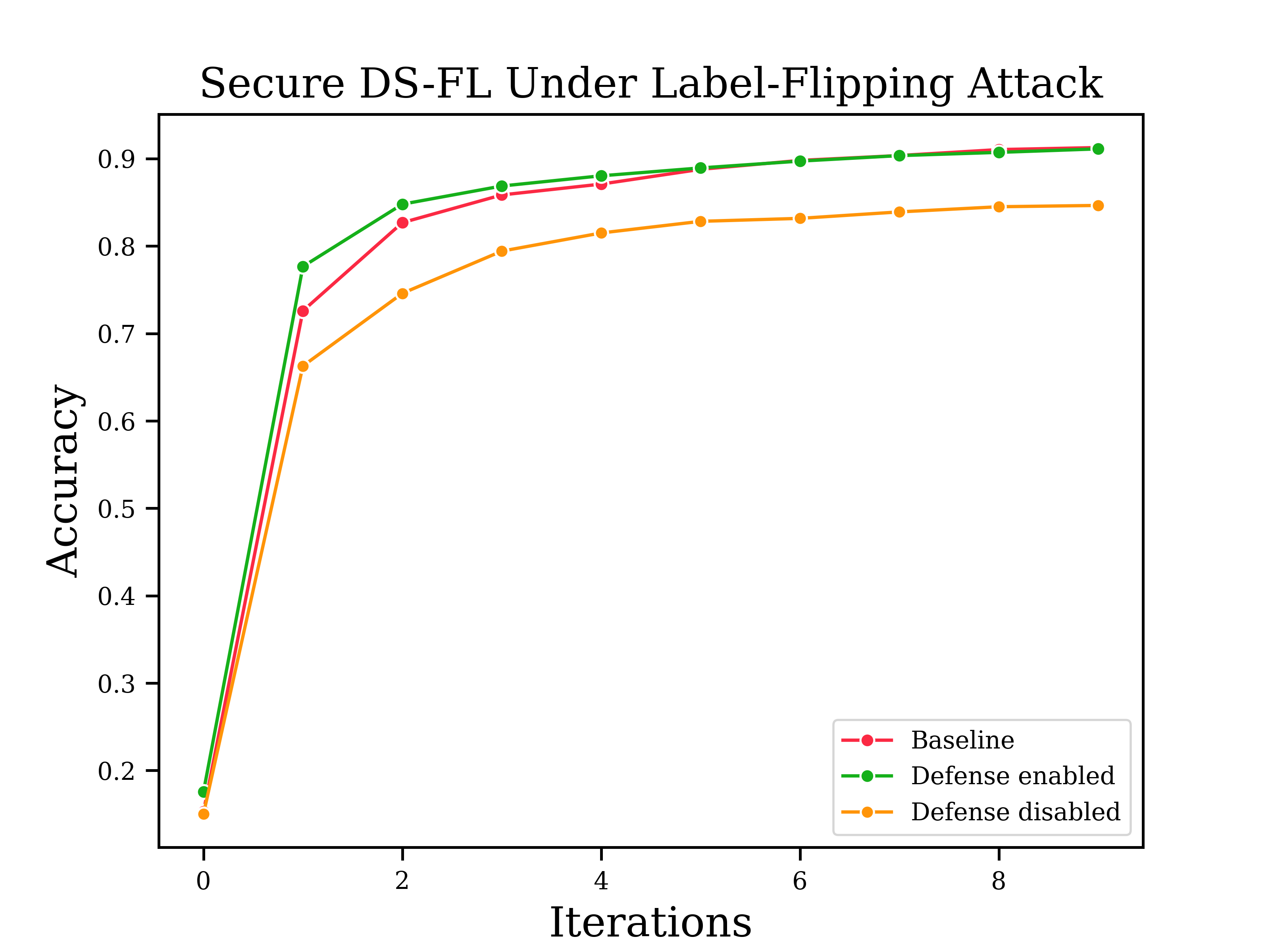}%
        \label{subfig:21}}
        \quad
        \subfloat[]{\includegraphics[width=0.3\textwidth]{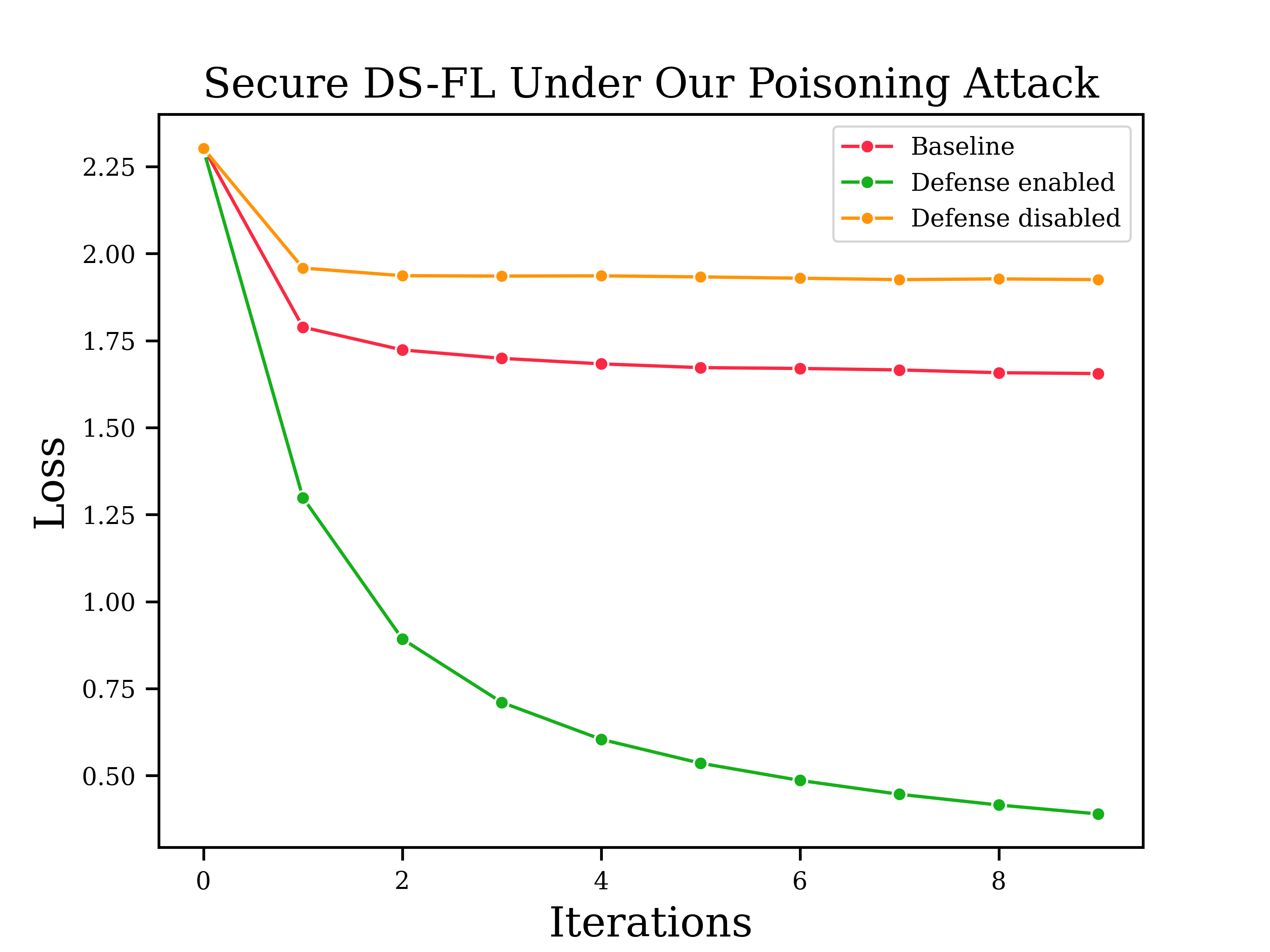}%
        \label{subfig:22}}
        \subfloat[]{\includegraphics[width=0.3\textwidth]{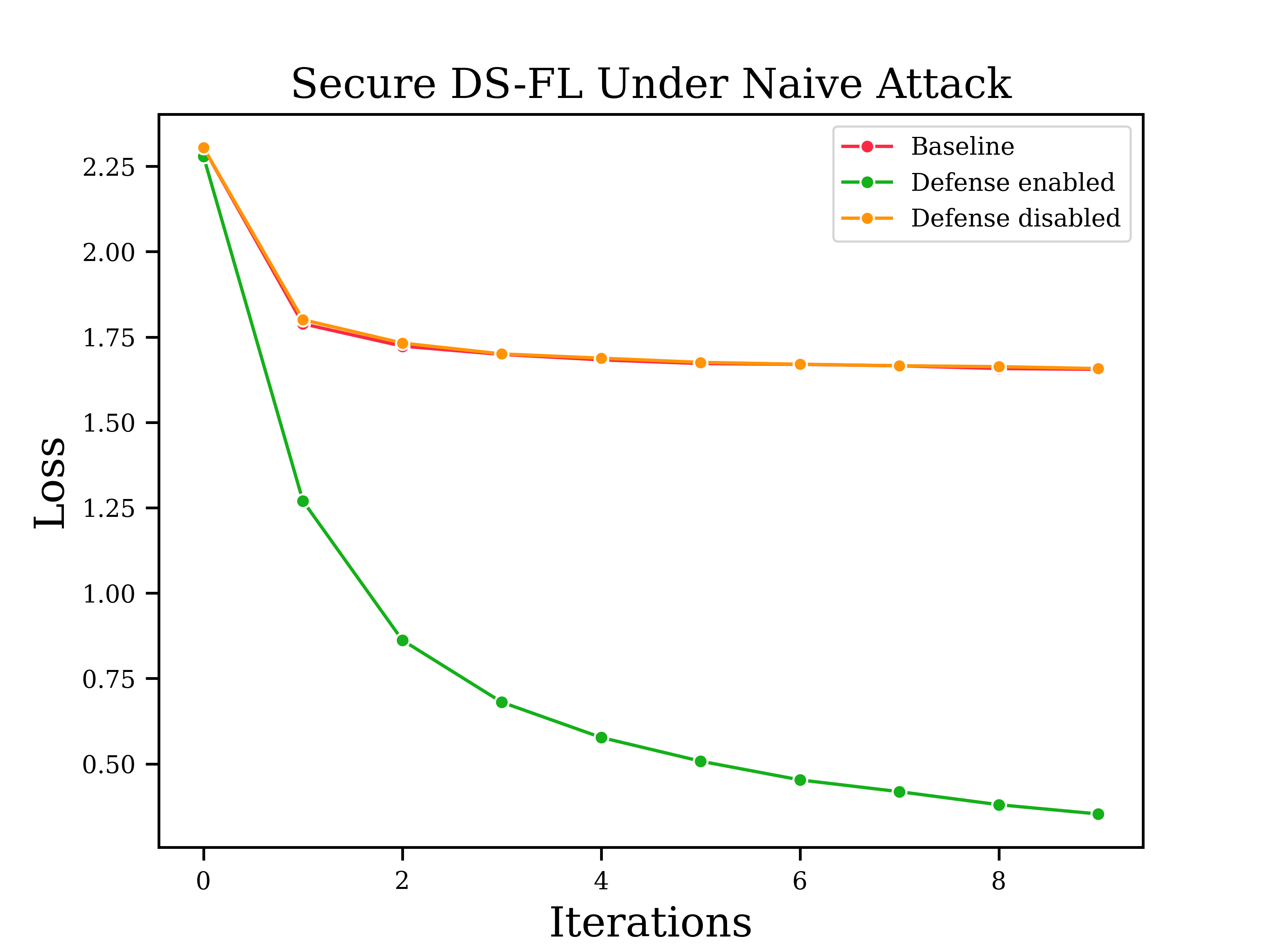}%
        \label{subfig:23}}   
        \subfloat[]{\includegraphics[width=0.3\textwidth]{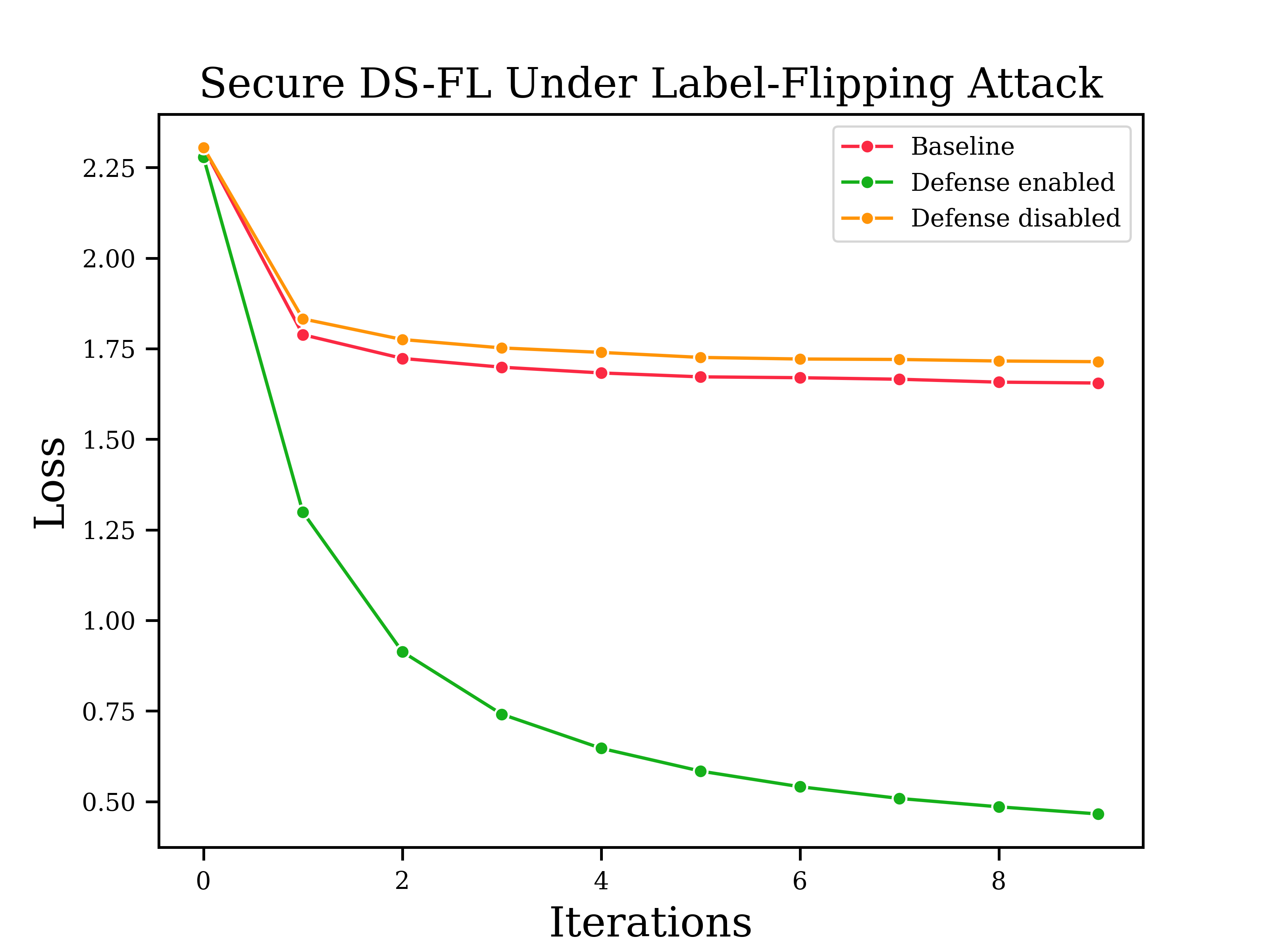}%
        \label{subfig:24}}         
    \caption{DS-FL enhanced by proposed robust aggregation algorithm}
    \label{fig:dedsfl}
\end{figure*}

\begin{figure*}
    \centering
        \subfloat[]{\includegraphics[width=0.3\textwidth]{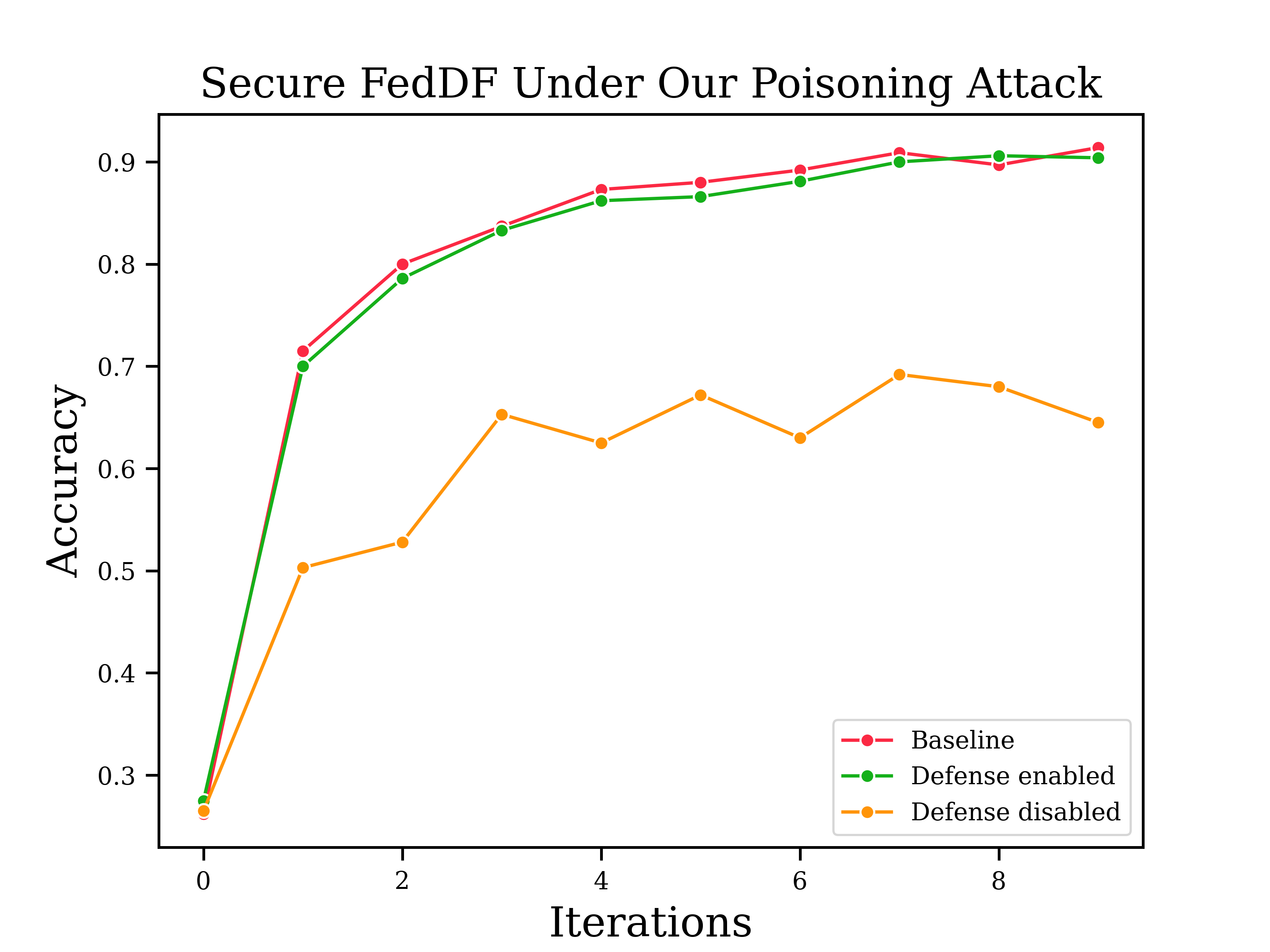}%
        \label{subfig:25}}
        \subfloat[]{\includegraphics[width=0.3\textwidth]{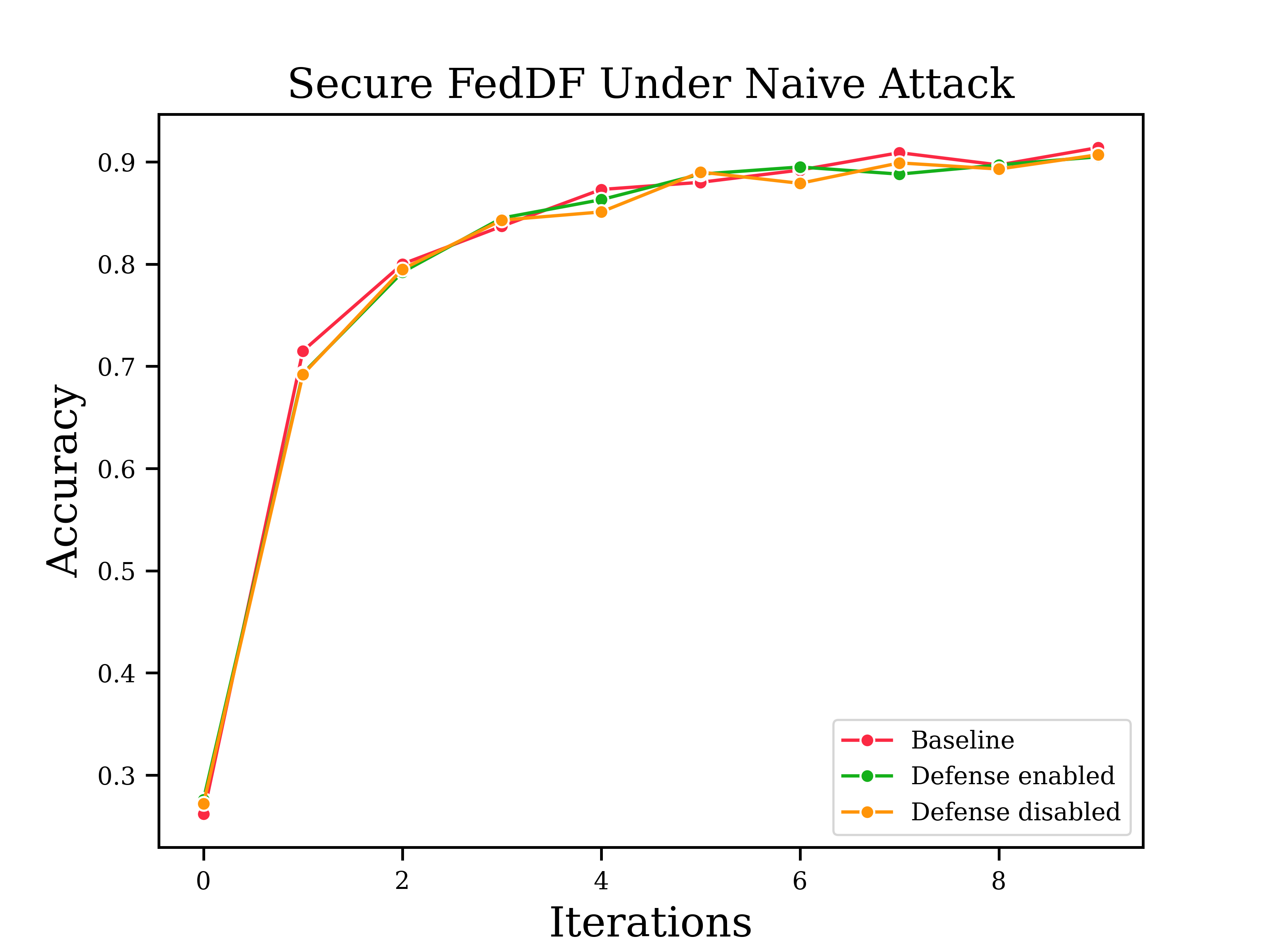}%
        \label{subfig:26}}
        \subfloat[]{\includegraphics[width=0.3\textwidth]{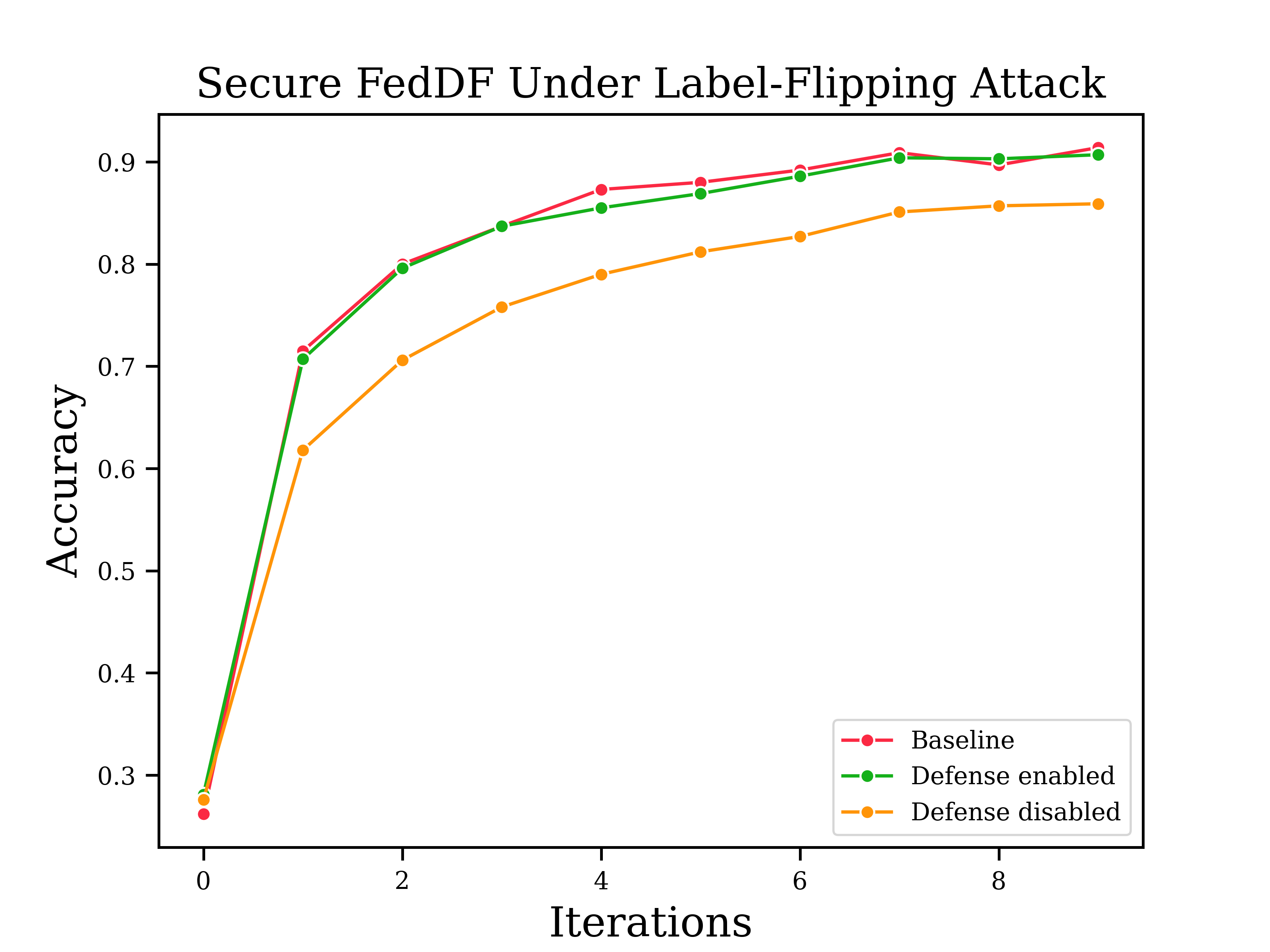}%
        \label{subfig:27}}
        \quad
        \subfloat[]{\includegraphics[width=0.3\textwidth]{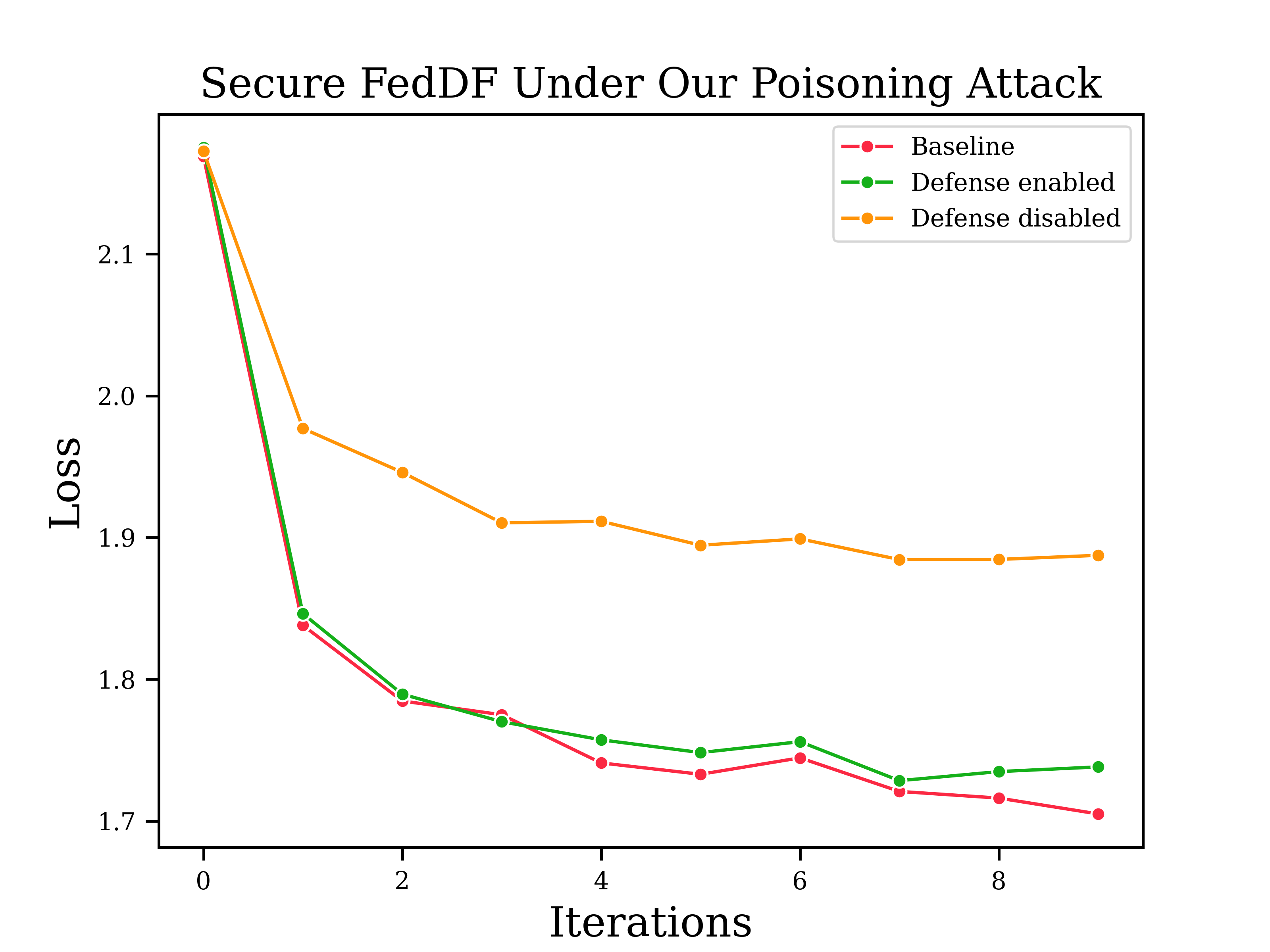}%
        \label{subfig:28}}
        \subfloat[]{\includegraphics[width=0.3\textwidth]{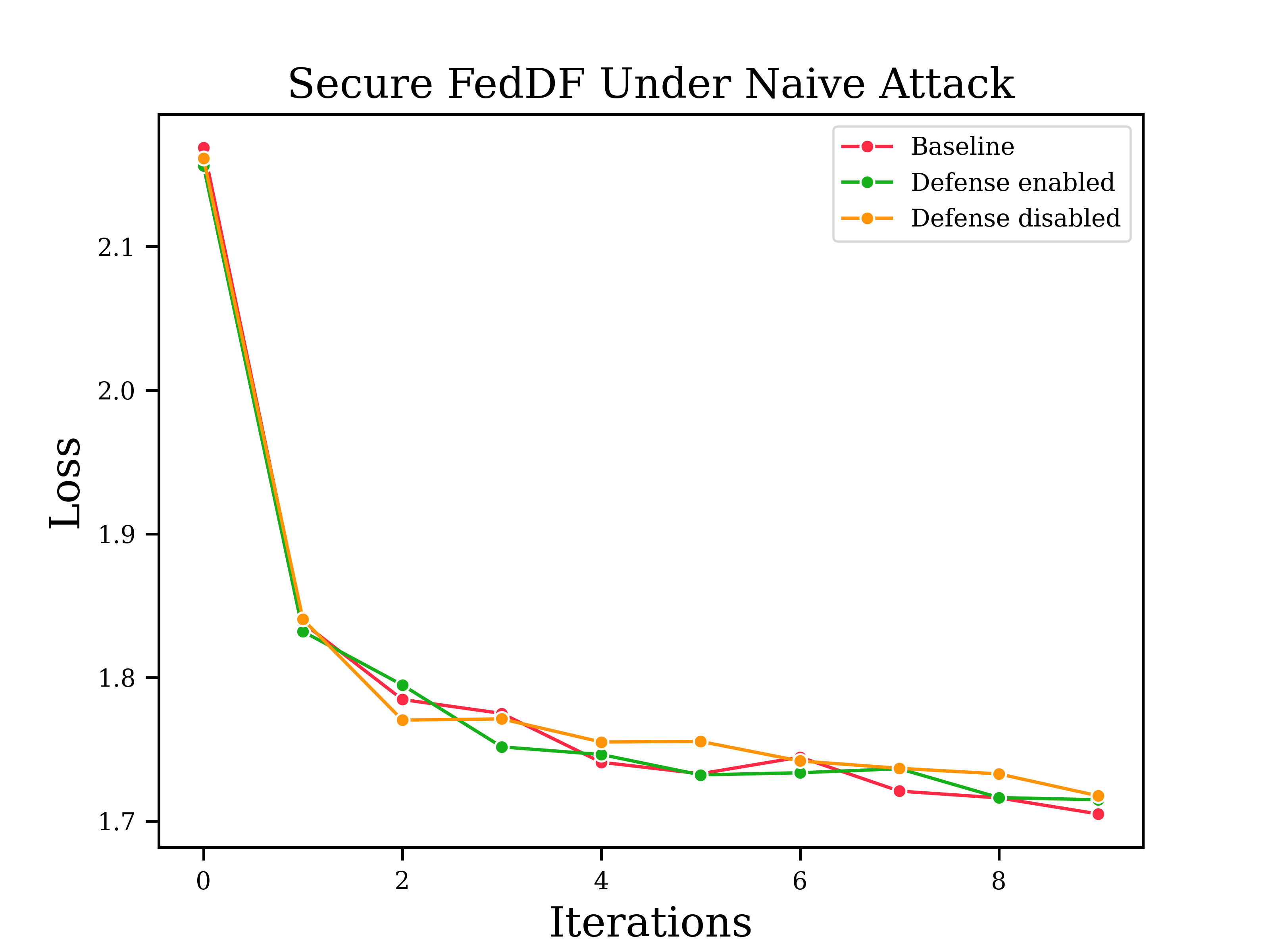}%
        \label{subfig:29}}   
        \subfloat[]{\includegraphics[width=0.3\textwidth]{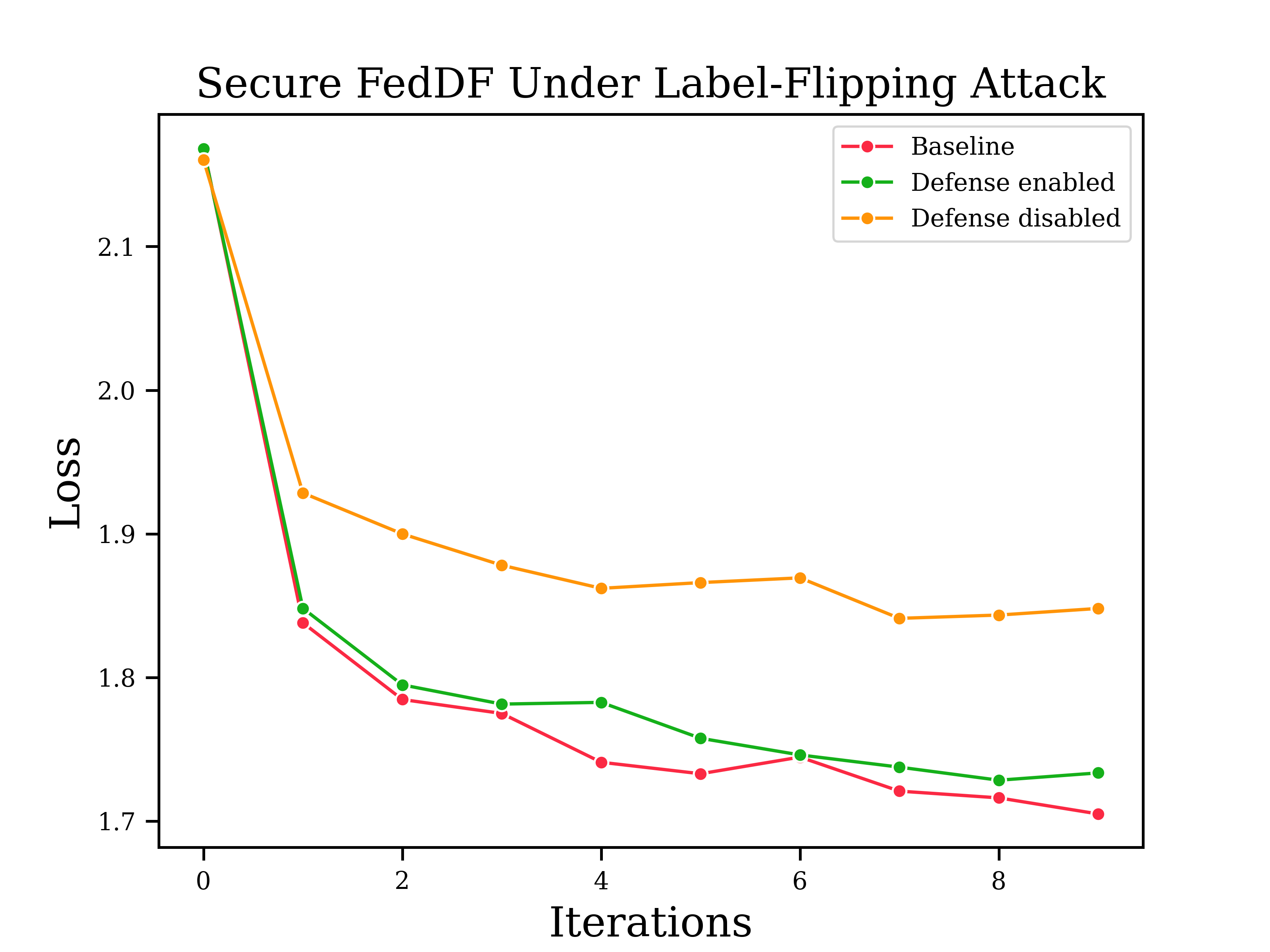}%
        \label{subfig:30}}         
    \caption{FedDF enhanced by proposed robust aggregation algorithm}
    \label{fig:defeddf}
\end{figure*}

The results of these FL schemes, fortified with our devised robust aggregation algorithm and subject to varying attack scenarios, are depicted in Fig. \ref{fig:defedmd}, Fig. \ref{fig:dedsfl}, and Fig. \ref{fig:defeddf}, respectively. Our defense scheme operates on the principle of abnormal logit vector filtering. Specifically, it detects the distribution of benign logit vectors, which are presumed to be in the majority, and assigns weights to each user. Because of this approach, our scheme is not limited to defending against our logit poisoning attack; it is also applicable to detecting and mitigating other poisoning attack schemes. We observe that, owing to the integration of our robust aggregation algorithm, all the considered schemes manifest a marked resistance to various poisoning attacks. The resultant test accuracy and loss metrics are significantly improved, aligning closely with the baseline curve. We are also encouraged to discover that for DS-FL, the test loss subsequent to the employment of our algorithm was lower than that of the baseline, while the accuracy was closely aligned with it.

\section{Conclusion}
This paper has unveiled a novel perspective on poisoning attacks within distillation-based federated learning, a complex issue that previously lacked a targeted approach. By analyzing the intrinsic properties of the logit vector, we developed an innovative two-stage scheme for executing logit poisoning attacks, successfully addressing prior limitations. Moreover, we introduced an efficient defense algorithm, estimating the benign logit vector's distribution on each class and assigning weights according to similarity to the estimated benign distribution. The extensive experimentation has underlined both the substantial threat that the newly devised logit poisoning attack can pose and the robust efficacy of our defensive mechanism. Our research thus contributes valuable insights into both offensive and defensive strategies within the realm of distillation-based federated learning, offering new avenues for further exploration and development.
%This research opens new avenues for understanding and mitigating risks in the evolving field of distillation-based federated learning.

\bibliography{references}

\begin{thebibliography}{10}
\providecommand{\url}[1]{#1}
\csname url@samestyle\endcsname
\providecommand{\newblock}{\relax}
\providecommand{\bibinfo}[2]{#2}
\providecommand{\BIBentrySTDinterwordspacing}{\spaceskip=0pt\relax}
\providecommand{\BIBentryALTinterwordstretchfactor}{4}
\providecommand{\BIBentryALTinterwordspacing}{\spaceskip=\fontdimen2\font plus
\BIBentryALTinterwordstretchfactor\fontdimen3\font minus \fontdimen4\font\relax}
\providecommand{\BIBforeignlanguage}[2]{{%
\expandafter\ifx\csname l@#1\endcsname\relax
\typeout{** WARNING: IEEEtran.bst: No hyphenation pattern has been}%
\typeout{** loaded for the language `#1'. Using the pattern for}%
\typeout{** the default language instead.}%
\else
\language=\csname l@#1\endcsname
\fi
#2}}
\providecommand{\BIBdecl}{\relax}
\BIBdecl

\bibitem{mcmahan2017communication}
B.~McMahan, E.~Moore, D.~Ramage, S.~Hampson, and B.~A. y~Arcas, ``Communication-efficient learning of deep networks from decentralized data,'' in \emph{Artificial intelligence and statistics}.\hskip 1em plus 0.5em minus 0.4em\relax PMLR, 2017, pp. 1273--1282.

\bibitem{park2019wireless}
J.~Park, S.~Samarakoon, M.~Bennis, and M.~Debbah, ``Wireless network intelligence at the edge,'' \emph{Proceedings of the IEEE}, vol. 107, no.~11, pp. 2204--2239, 2019.

\bibitem{lim2020federated}
W.~Y.~B. Lim, N.~C. Luong, D.~T. Hoang, Y.~Jiao, Y.-C. Liang, Q.~Yang, D.~Niyato, and C.~Miao, ``Federated learning in mobile edge networks: A comprehensive survey,'' \emph{IEEE Communications Surveys \& Tutorials}, vol.~22, no.~3, pp. 2031--2063, 2020.

\bibitem{kairouz2021advances}
P.~Kairouz, H.~B. McMahan, B.~Avent, A.~Bellet, M.~Bennis, A.~N. Bhagoji, K.~Bonawitz, Z.~Charles, G.~Cormode, R.~Cummings \emph{et~al.}, ``Advances and open problems in federated learning,'' \emph{Foundations and Trends{\textregistered} in Machine Learning}, vol.~14, no. 1--2, pp. 1--210, 2021.

\bibitem{zhou2022communication}
Y.~Zhou, X.~Ma, D.~Wu, and X.~Li, ``Communication-efficient and attack-resistant federated edge learning with dataset distillation,'' \emph{IEEE Transactions on Cloud Computing}, 2022.

\bibitem{zeng2023hfedms}
S.~Zeng, Z.~Li, H.~Yu, Z.~Zhang, L.~Luo, B.~Li, and D.~Niyato, ``Hfedms: Heterogeneous federated learning with memorable data semantics in industrial metaverse,'' \emph{IEEE Transactions on Cloud Computing}, 2023.

\bibitem{chang2019cronus}
H.~Chang, V.~Shejwalkar, R.~Shokri, and A.~Houmansadr, ``Cronus: Robust and heterogeneous collaborative learning with black-box knowledge transfer,'' \emph{arXiv preprint arXiv:1912.11279}, 2019.

\bibitem{li2019fedmd}
D.~Li and J.~Wang, ``Fedmd: Heterogenous federated learning via model distillation,'' \emph{arXiv preprint arXiv:1910.03581}, 2019.

\bibitem{lin2020ensemble}
T.~Lin, L.~Kong, S.~U. Stich, and M.~Jaggi, ``Ensemble distillation for robust model fusion in federated learning,'' \emph{Advances in Neural Information Processing Systems}, vol.~33, pp. 2351--2363, 2020.

\bibitem{itahara2021distillation}
S.~Itahara, T.~Nishio, Y.~Koda, M.~Morikura, and K.~Yamamoto, ``Distillation-based semi-supervised federated learning for communication-efficient collaborative training with non-iid private data,'' \emph{IEEE Transactions on Mobile Computing}, vol.~22, no.~1, pp. 191--205, 2021.

\bibitem{cheng2021fedgems}
S.~Cheng, J.~Wu, Y.~Xiao, and Y.~Liu, ``Fedgems: Federated learning of larger server models via selective knowledge fusion,'' \emph{arXiv preprint arXiv:2110.11027}, 2021.

\bibitem{fang2022robust}
X.~Fang and M.~Ye, ``Robust federated learning with noisy and heterogeneous clients,'' in \emph{Proceedings of the IEEE/CVF Conference on Computer Vision and Pattern Recognition}, 2022, pp. 10\,072--10\,081.

\bibitem{hinton2015distilling}
G.~Hinton, O.~Vinyals, and J.~Dean, ``Distilling the knowledge in a neural network,'' \emph{arXiv preprint arXiv:1503.02531}, 2015.

\bibitem{ba2014deep}
J.~Ba and R.~Caruana, ``Do deep nets really need to be deep?'' \emph{Advances in neural information processing systems}, vol.~27, 2014.

\bibitem{papernot2016semi}
N.~Papernot, M.~Abadi, U.~Erlingsson, I.~Goodfellow, and K.~Talwar, ``Semi-supervised knowledge transfer for deep learning from private training data,'' \emph{arXiv preprint arXiv:1610.05755}, 2016.

\bibitem{papernot2018scalable}
N.~Papernot, S.~Song, I.~Mironov, A.~Raghunathan, K.~Talwar, and {\'U}.~Erlingsson, ``Scalable private learning with pate,'' \emph{arXiv preprint arXiv:1802.08908}, 2018.

\bibitem{wang2018kdgan}
X.~Wang, R.~Zhang, Y.~Sun, and J.~Qi, ``Kdgan: Knowledge distillation with generative adversarial networks,'' \emph{Advances in neural information processing systems}, vol.~31, 2018.

\bibitem{anil2018large}
R.~Anil, G.~Pereyra, A.~Passos, R.~Ormandi, G.~E. Dahl, and G.~E. Hinton, ``Large scale distributed neural network training through online distillation,'' \emph{arXiv preprint arXiv:1804.03235}, 2018.

\bibitem{ejigu2023robust}
G.~F. Ejigu, S.~H. Hong, and C.~S. Hong, ``Robust federated learning with local mixed co-teaching,'' in \emph{2023 International Conference on Information Networking (ICOIN)}.\hskip 1em plus 0.5em minus 0.4em\relax IEEE, 2023, pp. 277--281.

\bibitem{jagielski2018manipulating}
M.~Jagielski, A.~Oprea, B.~Biggio, C.~Liu, C.~Nita-Rotaru, and B.~Li, ``Manipulating machine learning: Poisoning attacks and countermeasures for regression learning,'' in \emph{2018 IEEE symposium on security and privacy (SP)}.\hskip 1em plus 0.5em minus 0.4em\relax IEEE, 2018, pp. 19--35.

\bibitem{hayes2018contamination}
J.~Hayes and O.~Ohrimenko, ``Contamination attacks and mitigation in multi-party machine learning,'' \emph{Advances in neural information processing systems}, vol.~31, 2018.

\bibitem{munoz2017towards}
L.~Mu{\~n}oz-Gonz{\'a}lez, B.~Biggio, A.~Demontis, A.~Paudice, V.~Wongrassamee, E.~C. Lupu, and F.~Roli, ``Towards poisoning of deep learning algorithms with back-gradient optimization,'' in \emph{Proceedings of the 10th ACM workshop on artificial intelligence and security}, 2017, pp. 27--38.

\bibitem{baruch2019little}
G.~Baruch, M.~Baruch, and Y.~Goldberg, ``A little is enough: Circumventing defenses for distributed learning,'' \emph{Advances in Neural Information Processing Systems}, vol.~32, 2019.

\bibitem{bagdasaryan2020backdoor}
E.~Bagdasaryan, A.~Veit, Y.~Hua, D.~Estrin, and V.~Shmatikov, ``How to backdoor federated learning,'' in \emph{International Conference on Artificial Intelligence and Statistics}.\hskip 1em plus 0.5em minus 0.4em\relax PMLR, 2020, pp. 2938--2948.

\bibitem{wang2019symmetric}
Y.~Wang, X.~Ma, Z.~Chen, Y.~Luo, J.~Yi, and J.~Bailey, ``Symmetric cross entropy for robust learning with noisy labels,'' in \emph{Proceedings of the IEEE/CVF international conference on computer vision}, 2019, pp. 322--330.

\bibitem{cao2019understanding}
D.~Cao, S.~Chang, Z.~Lin, G.~Liu, and D.~Sun, ``Understanding distributed poisoning attack in federated learning,'' in \emph{2019 IEEE 25th International Conference on Parallel and Distributed Systems (ICPADS)}.\hskip 1em plus 0.5em minus 0.4em\relax IEEE, 2019, pp. 233--239.

\bibitem{lyu2023poisoning}
X.~Lyu, Y.~Han, W.~Wang, J.~Liu, B.~Wang, J.~Liu, and X.~Zhang, ``Poisoning with cerberus: stealthy and colluded backdoor attack against federated learning,'' in \emph{Thirty-Seventh AAAI Conference on Artificial Intelligence}, 2023.

\bibitem{blanchard2017machine}
P.~Blanchard, E.~M. El~Mhamdi, R.~Guerraoui, and J.~Stainer, ``Machine learning with adversaries: Byzantine tolerant gradient descent,'' \emph{Advances in neural information processing systems}, vol.~30, 2017.

\bibitem{shejwalkar2021manipulating}
V.~Shejwalkar and A.~Houmansadr, ``Manipulating the byzantine: Optimizing model poisoning attacks and defenses for federated learning,'' in \emph{NDSS}, 2021.

\bibitem{fung2020limitations}
C.~Fung, C.~J. Yoon, and I.~Beschastnikh, ``The limitations of federated learning in sybil settings.'' in \emph{RAID}, 2020, pp. 301--316.

\bibitem{sundar2022distributed}
A.~P. Sundar, F.~Li, X.~Zou, and T.~Gao, ``Distributed swift and stealthy backdoor attack on federated learning,'' in \emph{2022 IEEE International Conference on Networking, Architecture and Storage (NAS)}.\hskip 1em plus 0.5em minus 0.4em\relax IEEE, 2022, pp. 1--8.

\bibitem{lyu2022privacy}
L.~Lyu, H.~Yu, X.~Ma, C.~Chen, L.~Sun, J.~Zhao, Q.~Yang, and S.~Y. Philip, ``Privacy and robustness in federated learning: Attacks and defenses,'' \emph{IEEE transactions on neural networks and learning systems}, 2022.

\bibitem{fang2020local}
M.~Fang, X.~Cao, J.~Jia, and N.~Z. Gong, ``Local model poisoning attacks to byzantine-robust federated learning,'' in \emph{Proceedings of the 29th USENIX Conference on Security Symposium}, 2020, pp. 1623--1640.

\bibitem{yang2023model}
M.~Yang, H.~Cheng, F.~Chen, X.~Liu, M.~Wang, and X.~Li, ``Model poisoning attack in differential privacy-based federated learning,'' \emph{Information Sciences}, vol. 630, pp. 158--172, 2023.

\bibitem{wei2020federated}
K.~Wei, J.~Li, M.~Ding, C.~Ma, H.~H. Yang, F.~Farokhi, S.~Jin, T.~Q. Quek, and H.~V. Poor, ``Federated learning with differential privacy: Algorithms and performance analysis,'' \emph{IEEE Transactions on Information Forensics and Security}, vol.~15, pp. 3454--3469, 2020.

\bibitem{kullback1951information}
S.~Kullback and R.~A. Leibler, ``On information and sufficiency,'' \emph{The annals of mathematical statistics}, vol.~22, no.~1, pp. 79--86, 1951.

\bibitem{cao2022mpaf}
X.~Cao and N.~Z. Gong, ``Mpaf: Model poisoning attacks to federated learning based on fake clients,'' in \emph{Proceedings of the IEEE/CVF Conference on Computer Vision and Pattern Recognition}, 2022, pp. 3396--3404.

\bibitem{xie2020fall}
C.~Xie, O.~Koyejo, and I.~Gupta, ``Fall of empires: Breaking byzantine-tolerant sgd by inner product manipulation,'' in \emph{Uncertainty in Artificial Intelligence}.\hskip 1em plus 0.5em minus 0.4em\relax PMLR, 2020, pp. 261--270.

\bibitem{li2022learning}
H.~Li, X.~Sun, and Z.~Zheng, ``Learning to attack federated learning: A model-based reinforcement learning attack framework,'' \emph{Advances in Neural Information Processing Systems}, vol.~35, pp. 35\,007--35\,020, 2022.

\bibitem{yu2023untargeted}
Y.~Yu, Q.~Liu, L.~Wu, R.~Yu, S.~L. Yu, and Z.~Zhang, ``Untargeted attack against federated recommendation systems via poisonous item embeddings and the defense,'' in \emph{Proceedings of the AAAI Conference on Artificial Intelligence}, vol.~37, no.~4, 2023, pp. 4854--4863.

\bibitem{bhagoji2019analyzing}
A.~N. Bhagoji, S.~Chakraborty, P.~Mittal, and S.~Calo, ``Analyzing federated learning through an adversarial lens,'' in \emph{International Conference on Machine Learning}.\hskip 1em plus 0.5em minus 0.4em\relax PMLR, 2019, pp. 634--643.

\bibitem{xie2019dba}
C.~Xie, K.~Huang, P.-Y. Chen, and B.~Li, ``Dba: Distributed backdoor attacks against federated learning,'' in \emph{International conference on learning representations}, 2019.

\bibitem{suciu2018does}
O.~Suciu, R.~Marginean, Y.~Kaya, H.~Daume~III, and T.~Dumitras, ``When does machine learning $\{$FAIL$\}$? generalized transferability for evasion and poisoning attacks,'' in \emph{27th USENIX Security Symposium (USENIX Security 18)}, 2018, pp. 1299--1316.

\bibitem{biggio2013poisoning}
B.~Biggio, L.~Didaci, G.~Fumera, and F.~Roli, ``Poisoning attacks to compromise face templates,'' in \emph{2013 international conference on biometrics (ICB)}.\hskip 1em plus 0.5em minus 0.4em\relax IEEE, 2013, pp. 1--7.

\bibitem{deng2012mnist}
L.~Deng, ``The mnist database of handwritten digit images for machine learning research,'' \emph{IEEE Signal Processing Magazine}, vol.~29, no.~6, pp. 141--142, 2012.

\bibitem{lloyd1982least}
S.~Lloyd, ``Least squares quantization in pcm,'' \emph{IEEE transactions on information theory}, vol.~28, no.~2, pp. 129--137, 1982.

\bibitem{von2007tutorial}
U.~Von~Luxburg, ``A tutorial on spectral clustering,'' \emph{Statistics and computing}, vol.~17, pp. 395--416, 2007.

\bibitem{lecun1998gradient}
Y.~LeCun, L.~Bottou, Y.~Bengio, and P.~Haffner, ``Gradient-based learning applied to document recognition,'' \emph{Proceedings of the IEEE}, vol.~86, no.~11, pp. 2278--2324, 1998.

\bibitem{hinton2012improving}
G.~E. Hinton, N.~Srivastava, A.~Krizhevsky, I.~Sutskever, and R.~R. Salakhutdinov, ``Improving neural networks by preventing co-adaptation of feature detectors,'' \emph{arXiv preprint arXiv:1207.0580}, 2012.

\bibitem{kingma2014adam}
D.~P. Kingma and J.~Ba, ``Adam: A method for stochastic optimization,'' \emph{arXiv preprint arXiv:1412.6980}, 2014.

\end{thebibliography}
\bibliographystyle{IEEEtran}
% \vspace*{-1.6cm}
\begin{IEEEbiography}[{\includegraphics[width=1in,height=1.25in,clip,keepaspectratio]{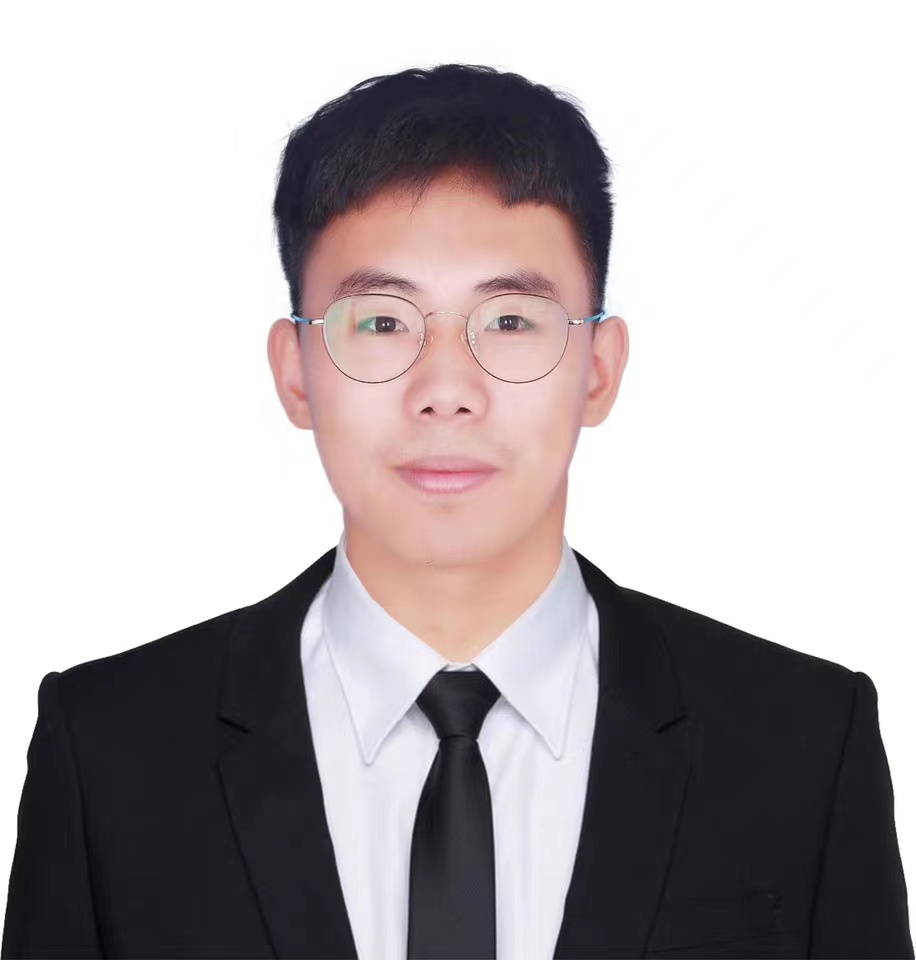}}]{Yonghao Yu}
received the B.E. degree in cyber science and engineering from Sichuan University, China, in 2023. He is currently working toward the M.E. degree at the Graduate School of Information, Production and System, Waseda University, Fukuoka, Japan. His current research interests include security in machine learning and computer vision.
\end{IEEEbiography}
% \vspace*{-1.6cm}
\begin{IEEEbiography}[{\includegraphics[width=1in,height=1.25in,clip,keepaspectratio]{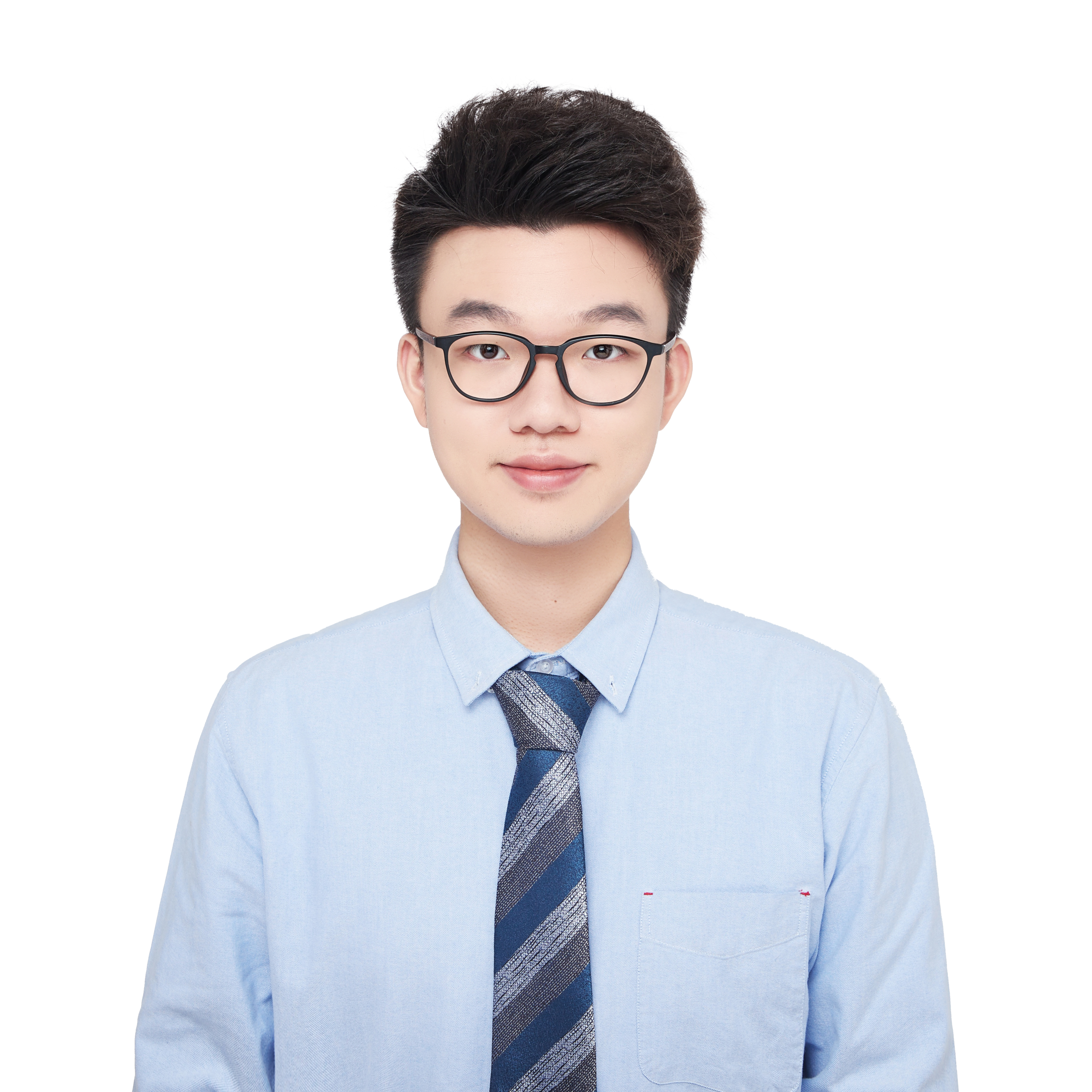}}]{Shunan Zhu}
received the B.E. degree in software engineering from Southeast University, China in 2023. He is currently working toward the M.E. degree at the Graduate School of Information, Production and System, Waseda University, Fukuoka, Japan.
His current research interests include federated learning and computer vision.
\end{IEEEbiography}
% \vspace*{-1.6cm}
\begin{IEEEbiography}[{\includegraphics[width=1in,height=1.25in,clip,keepaspectratio]{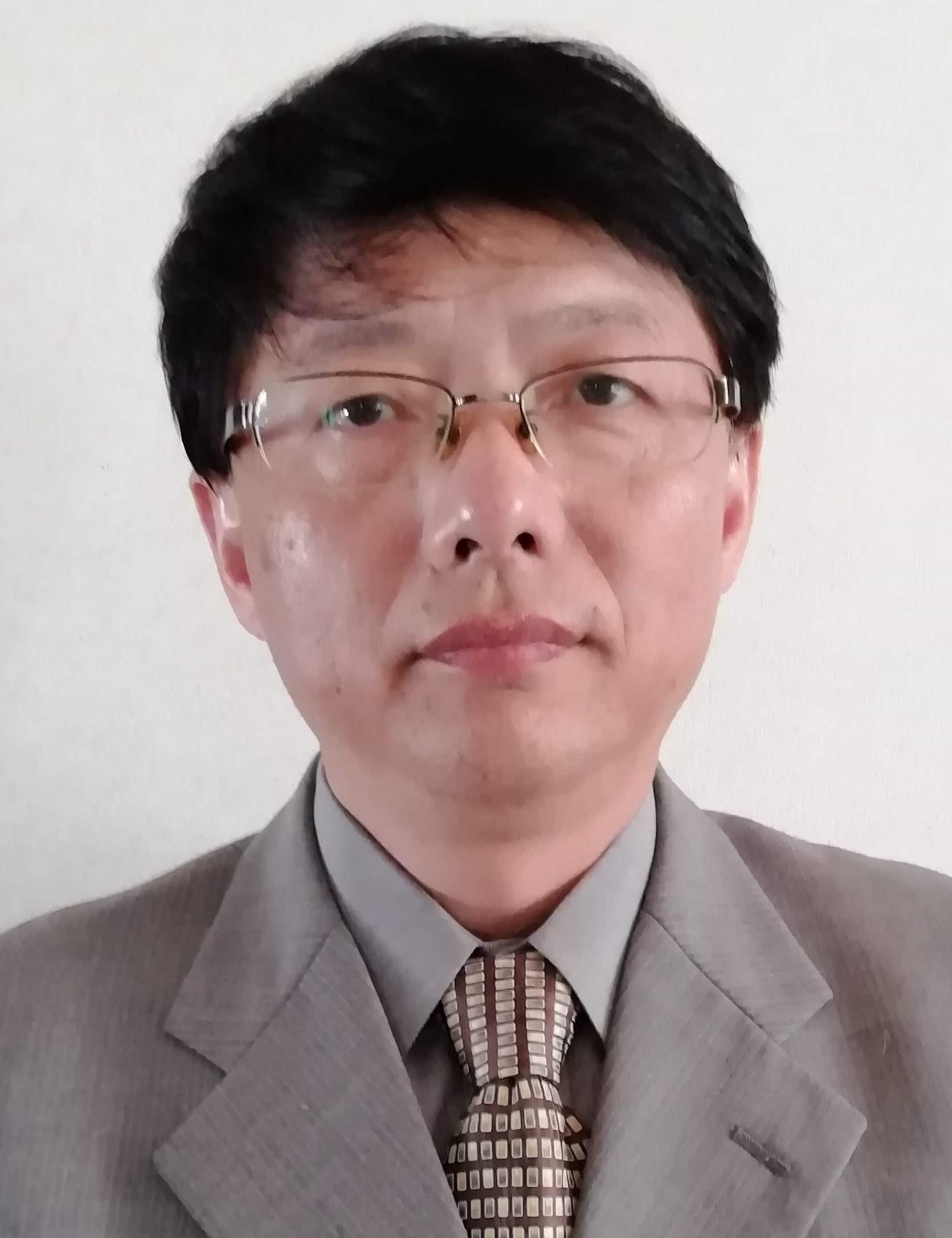}}]{Jinglu Hu}
received the M.Sci degree in Electronic Engineering from Sun Yat-Sen University, China in 1986, and a Ph.D degree in Computer Science and System Engineering from Kyushu Institute of Technology, Japan. From 1986 to 1993, he worked as a Research Associate and Lecturer at Sun Yat-Sen University. From 1997 to 2003, he worked as a Research Associate at Kyushu University. From 2003 to 2008, he worked as an Associate Professor and Since April 2008, he has been a Professor at Graduate School of Information, Production and Systems of Waseda University. His research interests include Computational Intelligence
such as neural networks and genetic algorithms, and their applications to system modeling and identification, bioinformatics, time series prediction, and so on. Dr. Hu is a member of IEEE, IEEJ, SICE and IEICE.
\end{IEEEbiography}

\end{document}